\documentclass[a4paper]{article}

\usepackage{graphicx}
\usepackage{amsmath}
\usepackage{amssymb}

\newcommand{\eop}{\sqcap\!\!\!\!\sqcup}
\newcommand{\bis}{\protect\underline{\leftrightarrow}}
\newcommand{\nbis}{\bis\!\!\!\!/\ }

\newtheorem{definition}{Definition}[section]
\newtheorem{example}{Example}[section]
\newtheorem{lemma}{Lemma}[section]
\newtheorem{proposition}{Proposition}[section]
\newtheorem{theorem}{Theorem}[section]

\date{}

\title{Behavioural equivalences for fluid stochastic Petri nets\thanks{The work was partially supported by Deutsche
Forschungsgemeinschaft (DFG) under grant BE 1267/14-1 and Russian Foundation for Basic Research (RFBR) under grant
14-01-91334.}}

\author{{\sc Igor V. Tarasyuk}\\
A.P. Ershov Institute of Informatics Systems,\\
Siberian Branch of the Russian Academy of Sciences,\\
Acad. Lavrentiev pr. 6, 630090 Novosibirsk, Russian Federation\\
{\tt itar@iis.nsk.su}
\and
{\sc Peter Buchholz}\\
Faculty of Computer Science,\\
Technical University of Dortmund,\\
Otto-Hahn-Str. 16, D-44227 Dortmund, Germany\\
{\tt peter.buchholz@cs.uni-dortmund.de}}

\begin{document}

\maketitle

\begin{abstract}
We propose fluid equivalences that allow one to compare and reduce behaviour of labeled fluid stochastic Petri nets
(LFSPNs) while preserving their discrete and continuous properties. We define a linear-time relation of fluid trace
equivalence and its branching-time counterpart, fluid bisimulation equivalence. Both fluid relations take into account
the essential features of the LFSPNs behaviour, such as {\em functional activity}, {\em stochastic timing} and {\em
fluid flow}. We consider the LFSPNs whose continuous markings have no influence to the discrete ones, i.e. every
discrete marking determines completely both the set of enabled transitions, their firing rates and the fluid flow rates
of the incoming and outgoing arcs for each continuous place. Moreover, we require that the discrete part of the LFSPNs
should be continuous time stochastic Petri nets. The underlying stochastic model for the discrete part of the LFSPNs is
continuous time Markov chains (CTMCs). The performance analysis of the continuous part of LFSPNs is accomplished via
the associated stochastic fluid models (SFMs).
%

We show that fluid trace equivalence preserves average potential fluid change volume for the transition sequences of
every certain length. We prove that
fluid bisimulation equivalence preserves the following aggregated (by such a bisimulation) probability functions:
stationary probability mass for the underlying CTMC, as well as stationary fluid buffer empty probability, fluid
density and distribution for the associated SFM. Hence, the equivalence guarantees identity of a number of discrete and
continuous performance measures. Fluid bisimulation equivalence is then used to simplify the qualitative and
quantitative analysis of LFSPNs that
is accomplished by means of quotienting (by the equivalence) the discrete reachability graph and underlying CTMC. To
describe the quotient associated SFM, the quotients of the probability functions are defined. We also characterize
logically fluid trace and bisimulation equivalences with two novel fluid modal logics $HML_{flt}$ and $HML_{flb}$,
constructed on the basis of the well-known Hennessy-Milner Logic HML. These results can
be seen as operational characterizations of the corresponding logical equivalences. The application example of a
document preparation system demonstrates the behavioural analysis via quotienting by fluid bisimulation equivalence.
\end{abstract}

\section{Introduction}
\label{introduction.sec}

An important scientific problem that has been often addressed in the last decades is the design and analysis of
parallel systems, which takes into account both qualitative (functional) and quantitative (timed, probabilistic,
stochastic) features of their behaviour. The main goal of the research on this topic is the development of models and
methods respecting performance requirements to concurrent and distributed systems with time constraints (such as
deterministic, nondeterministic and stochastic time delays) to construct, validate and optimize the performability of
realistic large-scale applications: computing systems, networks and software, controllers for industrial devices,
manufacturing lines, vehicle, aircraft and transportation engines. A fruitful approach to achieving progress in this
direction appeared to be a combined application of the theories of Petri nets,
stochastic processes and fluid flow systems to the specification and analysis of such time-dependent systems with
inherent behavioural randomicity \cite{Hav01}.

\subsection{Fluid stochastic Petri nets}

In the past, many extensions of stochastic Petri nets (SPNs) \cite{Nat80,Mol82,Mol85,Mar90,MBCDF95,Bal01,Bal07} have
been developed to specify, model, simulate and analyze some particular classes of systems, such as computer systems,
communication networks or manufacturing plants. These new formalisms have been constructed as a response to the needs
for more expressive power in describing real-world systems, and to the requirements for compact models and efficient
analysis techniques. One of the extensions are fluid stochastic Petri nets (FSPNs), capable of modeling hybrid systems
that combine continuous state variables, corresponding to the fluid levels, with discrete state variables, specifying
the token numbers. The continuous part of the FSPNs allows one to represent the fluid level in continuous places and
fluid flow along continuous arcs. This part can naturally describe
continuous variables in physical systems whose behaviour is commonly represented by differential equations.
Continuous variables may also be used to describe a macroscopic view of discrete items that appear in large
populations, e.g., packets in a computer network, molecules in a chemical reaction or people in a crowd. The discrete
part of an FSPN is essentially its underlying SPN, obtained from the FSPN by removing all the fluid-related continuous
elements. This part usually models the discrete control of the continuous process. The control may demonstrate some
stochastic behavior that captures uncertainty about the detailed system behavior.

FSPNs have been proposed in \cite{TK93,CNT97,Wol97} to model stochastic fluid flow systems \cite{GT08,GMST08}. To
analyze FSPNs, simulation, numerical and matrix-geometric methods are widely used
\cite{HKNT98,CNT99,BGGHST99,GS00,GSHB01,Gri02,GHo02,HGr02,GT07}. The major problem of FSPNs is the high complexity of
computing their solution, resulting in huge memory and time requirements while analyzing of realistic models. A
positive feature of the FSPN formalism is that it hides from a modeler the technical difficulties with solving
differential equations for the underlying stochastic processes and that it unifies in one framework the evolution
equations for the discrete and continuous parts of systems.

\subsection{Equivalences on the related models}

However, to the best of our knowledge, neither transition labeling nor behavioral equivalences have been proposed so
far for FSPNs. In \cite{TT12,TT14a,TT14b}, label equivalence and projected label equivalence have been introduced for
Fluid Process Algebra (FPA). FPA is a simple sub-algebra of Grouped PEPA (GPEPA) \cite{HB10}, which is itself a
conservative extension of Performance Evaluation Process Algebra (PEPA) \cite{Hil96}, obtained by adding fluid
semantics with an objective to simplify solving the systems of replicated ordinary differential equations. In
\cite{TT12,TT14b}, it has been proved that projected label equivalence induces a fluid lumpable partition and that both
label equivalence and projected label equivalence imply semi-isomorphism (stochastic isomorphism), in the context of a
special subclass of well-posed models. Nevertheless, the mentioned label equivalences do not respect the action names;
hence, they are not behavioral relations.

In \cite{TT14c,TT15}, the models specified with large ordinary differential equation (ODE) systems have been explored
within Fluid Extended Process Algebra (FEPA). The relations of semi-isomorphism, as well as those of ordinary and
projected label equivalence have been proposed for the sequential process components, called fluid atoms, such that
they can have a multiplicity (the number of their copies in the model specification). In addition to exact fluid
lumpability (EFL) from \cite{TT12,TT14b} that allows one to aggregate isomorphic processes with the same
multiplicities, a new notion of ordinary fluid lumpability (OFL) has been proposed. OFL does not require that the
multiplicities of the isomorphic processes coincide, but it preserves the sums of the aggregated variables instead.
Moreover, the approximate versions ($\epsilon$-variants) of semi-isomorphism,
EFL and OFL have been investigated, which abstract from small fluctuations of the parameter values in the processes
with close (similar) differential trajectories. This means that the close processes become completely symmetric
(aggregative, isomorphic) after small change (perturbation) of their parameters, resulting in the closer differential
trajectories. It has been proved that the aggregation error depends linearly in the perturbation intensity. However, as
mentioned above, the label equivalences do not respect the names of actions and therefore they are not behavioural
equivalences.

In \cite{TT16}, two notions of lumpability for the class of heterogenous systems models specified by nonlinear ODEs
have been investigated: exact lumpability (EL) \cite{TLRT97} and uniform lumpability (UL), both applied for exact
aggregation of the state variables. Unlike the EL transformations through linear mappings (in particular, those induced
by a partition of the original state space), UL considers exact symmetries of the equations due to identification of
the different variables from one partition block, which have coinciding differential trajectories (solutions) in case
of the same initial conditions. This is an extension of the ODE systems reduction technique for the formal language
$FPA$ from \cite{TT12} to arbitrary vector fields. Both the lumpability relations do not take into account the action
names and do not refer to behavioural equivalences.

In \cite{ITV15}, differential bisimulation for
FEPA has been constructed. This relation induces a partition on ODEs corresponding to the FEPA terms. Differential
bisimulation is a behavioural equivalence that is an ODE analogue of the probabilistic and stochastic bisimulations.
For each partition block, the sum of solutions of its ODEs coincides with the solution of a single aggregate ODE for
this block. In the framework of FSPNs, the ODE systems are obtained only when there is exactly one continuous place. In
the general case (more than one continuous place), the dynamics of FSPNs is described by the systems of equations with
partial derivatives of probability distribution functions (PDFs) and probability density functions
with respect to fluid levels in the continuous places. These levels are the random variables with a parameter
accounting for the work time of an FSPN, starting from the initial moment. Just for the fluid levels, the ODEs over the
time variable can be constructed in each discrete marking. However, the sojourn time in each discrete marking is a
random variable, calculated as the minimal transition delay, among all the transitions enabled in the marking. The FEPA
processes are described by the ODE systems with derivatives of the population functions that define the multiplicities
(numbers of replicas) of fluid atoms by only one variable denoting the time. Thus, the analogues of the FEPA fluid
atoms are the (mainly, continuous) places of FSPNs. Hence, the FSPN model always has a naturally embedded notion of
population, seen as a fluid in a continuous place. The systems behaviour is treated in FSPNs on a higher level of
specification using the continuous time concept and the GSPN basic model, and also on a higher analysis level using the
theories of probability and stochastic processes for constructing the underlying SMCs, CTMCs and stochastic fluid
models (SFMs). The multiplicities of the FEPA fluid atoms are the functions of time, such that their values can be
found for every particular time moment. In contrast, the
fluid levels in continuous places of FSPNs are the continuous random variables that depend on time, so that their exact
values
at a given moment of time cannot be calculated. The reason is the property of the continuous probability distributions,
stating that a continuous random variable may be equal to a concrete fixed value with zero probability only (excepting
that in FSPNs, the fluid probability mass at the boundaries may be positive). In addition, the FEPA expressivity is
rather restricted by considering only the processes, such that each of them is a parallel composition (with embedded
synchronization by the cooperation actions) of the fluid atoms denoting a large number of copies of the simple
sequential components, specified with only three operations: prefix, choice and recursive definition with constants.
Moreover, the fluid atoms in FEPA are considered uniformly, i.e. these is no difference between ``discrete'' atoms with
small multiplicities and ``continuous'' ones with large multiplicities. However, the tokens in FSPNs are jumped from
one discrete place to another instantaneously when their input of output transitions fire, whereas the fluid flow
proceeds through continuous places during all the time period when their input or output transitions are enabled. Thus,
the notion of differential bisimulation cannot be straightforwardly transferred from FEPA to FSPNs, since the two
models are different in many parts.

In \cite{CTTV15,CTTV16a,CTTV16c}, back and forth bisimulation equivalences on chemical species have been introduced for
chemical reaction networks (CRNs) with the ODE-based semantics. The forth bisimulation induces a partition where each
equivalence class is a sum of concentrations of the species from this class, and this relation guarantees the ordinary
fluid lumping on the ODEs of CRNs. The back bisimulation relates the species with the same ODE solutions at all time
points, starting from the moment for which their equal initial conditions have been defined, and this relation
characterizes the exact fluid lumping on the ODEs of CRNs. It has been noticed that the bisimulations proposed in
\cite{CTTV15} differ from the equivalences from \cite{TT12,TT14a,TT14b,TT14c,TT15}, since the former ones relate single
variables whereas the latter ones relate the sets of variables, such that each of them represents the behaviour of some
sequential process. The CRNs dynamics is described by ODEs with derivatives with respect to one variable (time), and
the CRNs behaviour is deterministic, described by differential trajectories. In \cite{CTTV16c}, an algorithm for
constructing exact aggregations for a class of ODE systems has been proposed, which computes forward and backward
bisimulation equivalences of CRNs with the time complexity $O(rs\log s)$ and space complexity $O(rs)$, where $r$ is the
number of monomials and $s$ is the number of variables in the ODEs. As mentioned above, unlike CRNs, FSPNs have a
stochastic behaviour which is influenced by the interplay of time and probabilistic factors. The FSPNs dynamics is
analyzed with (multidimensional, in general) SFMs that are solved using the differential equations with partial
derivatives with respect to several variables. In \cite{TTTV17}, back bisimulation equivalence, called there back
differential equivalence (BDE), has been used to provide an alternative characterization of emulation for CRNs,
interpreted as the systems of ODEs. Being a stricter variant of BDE, emulation requires that the ODE solutions of a
source CRN exactly overlap those of a target one at all moments of time. A genetic algorithm is presented that uses BDE
to discover emulations between~CRNs.

In \cite{CTTV16b}, back differential equivalence (BDE) and forth differential equivalence (FDE) have been explored for
a basic formalism, called Intermediate Drift Oriented Language (IDOL). IDOL has a syntax to specify drift for a class
of non-linear ODEs, for which the decidability results are known. The mentioned equivalence relations can be
transferred from IDOL to the higher-level models, such as Petri nets, process algebras and rule systems, interpreted as
ODEs. The differential equivalences embrace such notions as minimization of CTMCs based on the lumpability relation
\cite{DHS03}, bisimulations of CRNs \cite{CTTV15} and behavioural relations for process algebras with the ODE semantics
\cite{ITV15}. At the same time, the ODE class defined by the IDOL language cannot specify semantics of the systems with
stochastic continuous time delays in the discrete states, as well as many other behavioural aspects of FSPNs, including
the ones mentioned above. In \cite{CTTV17}, an application tool {\sf ERODE} has been presented for solution and
reduction of the ODE systems. The tool supports the mentioned BDE and FDE relations over the ODE variables.

In \cite{AHHBAB15}, on the product form queueing networks (QNs), the ideas of equivalent flow server and flow
equivalence have been applied to the models reduction. This has been done by aggregating server stations and their
states by the latter equivalence relation. Nevertheless, flow equivalence does not respect the names of actions, hence,
it is not a behavioural equivalence.

In \cite{Bor17}, for systems of polynomial ODEs, the questions of {\em reasoning} (detecting and proving identities
among the variables of an ODE system) and {\em reduction} (decreasing and possibly minimizing the number of variables
and equations of an ODE system while preserving all important information) have been addressed. The initial value
problem has been considered, i.e. solving the ODE systems with initial conditions. The ${\cal L}$-bisimulation
equivalence on the polynomials in the variables has been defined, which agrees with the underlying ODEs. An algorithm
has been proposed that detects all valid identities in an ODE system. This allows one to construct the reduced ODE
system with the minimal number of variables and equations so that the system is equivalent to the initial one. However,
${\cal L}$-bisimulation equivalence does not take into account the names of actions (which are not present at all in
the ODE systems specifications), therefore, the equivalence is not a behavioural relation.

\subsection{Labeled fluid stochastic Petri nets and fluid equivalences}

In this paper, we propose the behavioural relations of fluid trace and bisimulation equivalences that are useful for
the comparison and reduction of the behaviour of labeled FSPNs (LFSPNs), since these relations preserve the
functionality and performability of their discrete and continuous parts.

For every FSPN, the discrete part of its marking
is determined by the natural number of tokens contained in the discrete places. The continuous places of an FSPN are
associated with the non-negative real-valued fluid levels that determine the continuous part of the FSPN marking. Thus,
FSPNs have a hybrid (discrete-continuous) state space. The discrete part of every hybrid
marking of FSPNs is called discrete marking while the continuous part is called continuous marking.
The discrete part of each hybrid marking has an influence on the continuous part. For more general FSPNs, the reverse
dependence is possible as well. As a basic model for constructing LFSPNs, we consider only those FSPNs in which the
continuous parts of markings have no influence on the discrete ones, i.e. such that every discrete part determines
completely both the set of enabled transitions and the rates of incoming and outgoing arcs for each continuous place
\cite{Gri02,GT07}. We also require that the discrete part of LFSPNs should be labeled continuous time stochastic Petri
nets (CTSPNs) \cite{Mol82,Mar90,MBCDF95,Bal01}.

First, we define a linear-time relation of {\em fluid trace equivalence} on LFSPNs. Linear-time equivalences, unlike
branching-time ones, do not respect the points of choice among several alternative continuations of the systems
behavior. We require that fluid trace equivalence on discrete markings of two LFSPNs should be a standard (strong)
Markovian trace equivalence. Hence, for every sequence of discrete markings and transitions in the discrete
reachability graph of an LFSPN, starting from the initial discrete marking and ending in some last discrete marking
(such sequence is called path), we require a simulation of the path in the discrete reachability graph of the
equivalent LFSPN, such that the action labels of the corresponding fired transitions in the both sequences coincide.
Moreover, the average sojourn times in (or the exit rates from) the respective discrete markings should be the same.
Finally, for the two equivalent LFSPNs, the cumulative execution probabilities of all the paths corresponding to a
particular sequence of actions, together with a concrete sequence of the average sojourn times (exit rates), should be
equal. Thus, when comparing the execution probabilities, we parameterize the paths with the same extracted action
sequence by all possible sequences of the extracted average sojourn times (exit rates), i.e. we consider comparable
only the paths with the same extracted action sequence and the same value of the parameter, which is a concrete
sequence of the extracted average sojourn times (exit rates). Therefore, our definition of the trace equivalence on the
discrete markings of LFSPNs is similar to that of ordinary (that with the absolute time counter or with the countdown
timer) Markovian trace equivalence \cite{WBMC06} on transition-labeled CTMCs. Ordinary Markovian trace equivalence and
its variants from \cite{WBMC06} have been later investigated and enhanced on sequential and concurrent Markovian
process calculi SMPC and CMPC in \cite{BBo06,Bern07a,Bern07b,BBo08} and on Uniform Labeled Transition Systems (ULTraS)
in \cite{BNL13,BeTe13}.

As for the continuous markings of the two LFSPNs, we further parameterize the paths with the same extracted action
sequence and the same sequence of the extracted average sojourn times (exit rates) by counting the execution
probabilities only of those paths additionally having the same sequence of extracted fluid flow rates of the respective
continuous places (we assume that each of the two LFSPNs has exactly one continuous place) in the corresponding
discrete markings. Besides the need to respect a fluid flow in the equivalence definition, the intuition behind such a
double parameterizing by the average sojourn times and by the fluid flow rates is as follows. In each of the
corresponding discrete markings of the comparable paths we shall have the same {\em average potential fluid change
volume} in the corresponding continuous places, which is a product of the average sojourn time and the constant
(possibly zero or negative) potential fluid flow rate.

We show that fluid trace equivalence preserves average potential fluid change volume in the respective continuous
places for the transition sequences of each particular length.

Second, we propose a branching-time relation of {\em fluid bisimulation equivalence} on LFSPNs. We prove that it is
strictly stronger than fluid trace equivalence, i.e. the former relation generally makes less identifications among the
compared LFSPNs than the latter.
We require the fluid bisimulation on the discrete markings of two LFSPNs to be a standard (strong) Markovian
bisimulation. Hence, for each transition firing in an LFSPN, we require a simulation of the firing in the equivalent
LFSPN, such that the
action labels of the both fired transitions and their overall rates
coincide. Thus, our definition of the bisimulation equivalence on the discrete markings of LFSPNs is similar to that of
the performance bisimulation equivalences \cite{Buc95,Buc98} on labeled CTSPNs and labeled generalized SPNs (GSPNs)
\cite{Mar90,CMT91,MBCDF95,BPTT98,Bal01,Bal07}, as well as the strong equivalence from \cite{Hil96} on stochastic
process algebra PEPA. All these relations belong to the family of Markovian bisimulation equivalences, investigated on
sequential and concurrent Markovian process calculi SMPC and CMPC in \cite{BBo06,Bern07a,Bern07b,BBo08}, as well as on
Uniform Labeled Transition Systems (ULTraS) in \cite{BNL13,BeTe13}.

As for the continuous markings, we should fix a bijective correspondence between the sets of continuous places of the
two LFSPNs, hence, the number of their continuous places should coincide. Then each continuous place in the first LFSPN
should have exactly one corresponding continuous place in the second LFSPN and vice versa. We require that, for every
pair of the Markovian bisimilar discrete markings, the fluid flow rates of the continuous places in the first LFSPN
should coincide with those of the corresponding continuous places in the second LFSPN. Note that in our formal
definition of fluid bisimulation, we consider only LFSPNs having a single continuous place, since the definition can be
easily extended to the case of several continuous places.

We prove that the resulting fluid bisimulation equivalence of LFSPNs preserves, for the
equivalence classes of their discrete markings, the stationary
probability
distribution of the underlying continuous time Markov chain (CTMC), as well as the stationary fluid buffer empty
probability, probability distribution and density for the associated stochastic fluid model (SFM). As a consequence,
the equivalence guarantees identity of a number of discrete and hybrid performance measures, calculated for the
stationary quantitative behaviour of the LFSPNs.
The fluid bisimulation equivalence is then used to simplify the qualitative and quantitative analysis of LFSPNs, due to
diminishing the number of
discrete markings considered that are lumped into the equivalence classes, interpreted as the (aggregate) states of the
quotient discrete reachability graph and quotient underlying CTMC. We also define the quotients of the probability
functions by the equivalence, aiming at description of the quotient associated SFM. Based on the pointed equivalence, a
new quotient technique enhances and optimizes the performance evaluation of fluid systems modeled by LFSPNs.

The running example presented in the paper explains systematically the most important definitions introduced. It also
demonstrates in detail the functional and performance identity of the LFSPNs, related by fluid trace or fluid
bisimulation equivalence. The application example consists in a case study of three
LFSPNs, each of them modeling the document preparation system, and demonstrates how the LFSPNs structure and behaviour
can be reduced with respect to fluid bisimulation equivalence while preserving their functional and performance
properties.

\subsection{Logical characterization of the fluid equivalences}

A characterization of equivalences via modal logics is used to change the operational reasoning on systems behaviour by
the logical one that is more appropriate for verification. Moreover, such an interpretation elucidates the nature of
the equivalences, defined in an operational manner. It is generally accepted that the natural and nice modal
characterization of a behavioural equivalence justifies its relevance. On the other hand, we get an operational
characterization of logical equivalences. The importance of modal logical characterization for behavioural equivalences
has been explained in \cite{Ace03}, in particular, the resulting capabilities to express {\em distinguishing formulas}
for automatic verification of systems \cite{Kor92} and {\em characteristic formulas} for the equivalence classes of
processes \cite{SI94,AMFI15}, to demonstrate {\em finitariness and algebraicity} of behavioural preorders \cite{AH92},
as well as to give a {\em testing interpretation} of bisimulation equivalence \cite{Abr87}.

In the literature, several logical characterizations of stochastic and Markovian equivalences have been proposed. In
\cite{CGH99,CGHR00}, the characterization of strong equivalence has been presented with the logic PML$_{\mu}$, which is
a stochastic extension of Probabilistic Modal Logic (PML) \cite{LS91} on probabilistic transitions systems to the
stochastic process algebra PEPA \cite{Hil96}. In \cite{GHo05}, a branching time temporal logic has been described which
is an extension of Continuous Stochastic Logic (CSL) \cite{ASSB00} on CTMCs to a wide class of SFMs.
The CSL-based logical characterizations of various stochastic bisimulation equivalences have been reported in
\cite{BHHK00,BHHK03,BHKW03,SD04,BHKW05,BHKP08} on labeled CTMCs, in \cite{DP03} on labeled continuous time Markov
processes (CTMPs), in \cite{Dob05} on analytic spaces, in \cite{BHHHK13} on labeled Markov reward models (MRMs) and in
\cite{SZG14} on labeled continuous time Markov decision processes (CTMDPs). In \cite{BBo06,Bern07a}, on sequential and
concurrent Markovian process calculi SMPC (MPC) and CMPC, the logical characterizations of Markovian trace and
bisimulation equivalences have been accomplished with the modal logics $HML_{MTr}$ and $HML_{MB}$, based on
Hennessy-Milner Logic (HML) \cite{HM85}. In \cite{BBo08}, on (sequential) Markovian process calculus MPC, the logical
characterizations of Markovian trace and bisimulation equivalences have been constructed with the HML-based modal
logics $HML_{NPMTr}$ and $HML_{MB}$.

We provide fluid trace and bisimulation equivalences with the logical characterizations, accomplished via formulas of
the specially constructed novel fluid modal logics $HML_{flt}$ and $HML_{flb}$, respectively. The new logics are based
on Hennessy-Milner Logic (HML) \cite{HM85}. The logical characterizations guarantee that two LFSPNs are fluid (trace or
bisimulation) equivalent iff they satisfy the same formulas of the respective fluid modal logic, i.e. they are
logically equivalent. Thus, instead of comparing LFSPNs operationally, one may only check the corresponding
satisfaction relation. This provides one with the possibility for logical reasoning on fluid equivalences for LFSPNs.
Such an approach is often more convenient for the purpose of verification. The obtained results may be also interpreted
as operational characterizations of the corresponding logical~equivalences.

The fluid modal logic $HML_{flt}$ is used to characterize fluid trace equivalence. Therefore, the interpretation
function of the logic has an additional argument, which is the sequence of the potential fluid flow rates for the
single continuous place of an LFSPN (remember that in the {\em standard} definition of fluid trace equivalence we
compare only LFSPNs, each having exactly one continuous place). In $HML_{flt}$, one can express the properties like
``the execution probability of a sequence of actions starting from a state, with given average sojourn times and
potential fluid flow rates in the initial, intermediate and final states, is equal to a particular value''. For
example, for a production line in a food processing or a chemicals plant, we can verify the {\em probability} that the
first liquid substance fills (this is specified by the action $f_1$) the fluid reservoir with the potential flow rate
$r_1$ during the exponentially distributed time period with the average $s_1$; {\em then} the second liquid substance
fills (the action $f_2$) the reservoir with the potential flow rate $r_2$ during the exponentially distributed time
period with the average $s_2$; {\em finally}, the reservoir is emptied with the potential flow rate $r_3$ for the
exponentially distributed time period with the average $s_3$.

The fluid modal logic $HML_{flb}$ is intended to characterize fluid bisimulation equivalence. For this purpose, the
logic has a new modality, decorated with the potential fluid flow rate value for the single continuous place of an
LFSPN (again, remember that in the {\em standard} definition of fluid bisimulation equivalence consider only LFSPNs,
each having a single continuous place). The resulting formula (i.e. the new modality with the flow rate value) is used
to check whether the potential fluid flow rate in a discrete marking of an LFSPN
coincides with a certain value, the fact that
corresponds to a condition from the fluid bisimulation definition. Thus, $HML_{flb}$ is able to describe the properties
such as ``an action can be executed with a given minimal rate in a state with a given potential fluid flow rate''. For
example, for the production line mentioned above, we can verify the {\em validity} that the first liquid substance
fills (the action $f_1$) the fluid reservoir with the potential flow rate $r_1$ during the exponentially distributed
time period with the rate $\lambda_1$ {\em or} the second liquid substance fills (the action $f_2$) the reservoir with
the same potential flow rate $r_1$ during the exponentially distributed time period with the rate $\lambda_2$. Note
that disjunction in $HML_{flb}$ can be defined standardly, i.e. using conjunction and negation.

\subsection{Previous works and contributions of the paper}

The first results on this subject can be found in \cite{TB15}, where we have proposed a class of LFSPNs and defined a
novel behavioural relation of fluid bisimulation equivalence for them. We have also proven there that the equivalence
preserves aggregate fluid density and distribution, as well as discrete and continuous performance measures. The
present paper is an improved and extended version of that publication. The paper contains the following new results for
LFSPNs: fluid trace equivalence, interrelations of the fluid equivalences, quotienting by fluid bisimulation
equivalence, logical characterization of the fluid equivalences,
quotients of the probability functions
and an application example.

Thus, the main contributions of the paper are as follows.
\begin{itemize}

\item LFSPNs
extend FSPNs with the action labeling on their transitions, which allows for the functional behavioural reasoning.

\item Fluid trace and bisimulation equivalences permit to compare and reduce the qualitative and quantitative behaviour
of LFSPNs in the the linear-time and branching-time semantics, respectively.

\item The analysis of LFSPNs is simplified by quotienting their discrete reachability graphs and underlying CTMCs by
fluid bisimulation equivalence.

\item Fluid trace and bisimulation equivalences are logically characterized via two original fluid modal logics
$HML_{flt}$ and $HML_{flb}$.

\item The aggregate probability functions coincide as for the discrete part (labeled CTSPNs and their underlying
CTMCs), as for the continuous part (SFMs) of the fluid bisimulation equivalent LFSPNs.

\item Both the discrete and hybrid performance measures for LFSPNs are preserved by fluid bisimulation equivalence.

\item Application example shows in detail the functional and performance identity of the fluid bisimulation equivalent
LFSPNs specifying the document preparation system.

\end{itemize}

\subsection{Outline of the paper}

The rest of the paper is organized as follows. In Section \ref{fspnpres.sec}, we present the definition and behaviour
of LFSPNs. Section \ref{fspndisc.sec} explores the discrete part of LFSPNs, i.e. the derived labeled CTSPNs and their
underlying CTMCs. Section \ref{fspncont.sec} investigates the continuous part of LFSPNs, which is the associated SFMs.
In Section \ref{flutraeq.sec}, we construct a linear-time relation of fluid trace equivalence for LFSPNs. In Section
\ref{flubiseq.sec}, we propose a branching-time relation of fluid bisimulation equivalence for LFSPNs and compare it
with the fluid trace one. In Section \ref{reduction.sec}, we explain how to reduce discrete reachability graphs and
underlying CTMCs of LFSPNs modulo fluid bisimulation equivalence, by applying the method of the quotienting. Section
\ref{flulogs.sec} is devoted to the logical characterization of fluid trace and bisimulation equivalences with the use
of two novel fluid modal logics. Section \ref{quantbeh.sec} contains the preservation results for the quantitative
behaviour of LFSPNs modulo fluid bisimulation equivalence.
In Section \ref{funcperf.sec}, we demonstrate how fluid bisimulation equivalence preserves the functionality and
performance measures of LFSPNs. Section \ref{docprsys.sec} describes a case study of three
LFSPNs modeling the document preparation system. Finally, Section \ref{conclusion.sec} summarizes the results obtained
and outlines research perspectives in this area.

To help the reader, we have presented some important abbreviations from the paper in Table \ref{abbrfspn.tab}.

\begin{table}
\caption{Abbreviations used in the paper}
\vspace{-3mm}
\label{abbrfspn.tab}
\begin{center}
\begin{tabular}{|ll|ll|}
\hline
\multicolumn{2}{|l|}{{\bf Petri nets}} & \multicolumn{2}{|l|}{{\bf Markov chains}}\\
SPN & stochastic Petri net & CTMC & continuous time\\
CTSPN & continuous time stochastic Petri net & & Markov chain\\
GSPN & generalized stochastic Petri net & SMC & semi-Markov chain\\
FSPN & fluid stochastic Petri net & \multicolumn{2}{|l|}{{\bf Fluid models}}\\
LFSPN & labeled fluid stochastic Petri net & SFM & stochastic fluid model\\
\multicolumn{2}{|l|}{{\bf Probability functions}} & \multicolumn{2}{|l|}{{\bf Rate matrices}}\\
PMF & probability mass function & TRM & transition rate matrix\\
PDF & probability distribution function & FRM & fluid rate matrix\\
\hline
\end{tabular}
\end{center}
\end{table}

\section{Basic concepts of LFSPNs}
\label{fspnpres.sec}

Let us introduce a class of labeled fluid stochastic Petri nets (LFSPNs), whose transitions are labeled with action
names, used to specify different system activities.
Without labels, LFSPNs are essentially a subclass of FSPNs \cite{HKNT98,Gri02,GT07},
so that their discrete part describes CTSPNs \cite{Mol82,Mar90,MBCDF95,Bal01}. This means that LFSPNs have no inhibitor
arcs, priorities and immediate transitions, which are used in the standard FSPNs, which are the continuous extension of
GSPNs. However, in many practical applications, the performance analysis of GSPNs is simplified by transforming them
into CTSPNs or reducing their underlying semi-Markov chains into CTMCs (which are the underlying stochastic process of
CTSPNs) by eliminating vanishing states \cite{CMT91,MBCDF95,Bal01,Bal07}.
Transition labeling in LFSPNs is similar to the labeling, proposed for CTSPNs in \cite{Buc95}. Moreover, we suppose
that the firing rates
of transitions
and flow rates of the continuous arcs do not depend on the continuous markings (fluid levels).

Let ${\mathbb N}=\{0,1,2,\ldots\}$ be the set of {\em all natural numbers} and ${\mathbb N}_{\geq 1}=\{1,2,\ldots\}$ be
the set of {\em all positive natural numbers}. Further, let ${\mathbb R}=(-\infty ;\infty )$ be the set of {\em all
real numbers}, ${\mathbb R}_{\geq 0}=[0;\infty )$ be the set of {\em all non-negative real numbers} and ${\mathbb
R}_{>0}=(0;\infty )$ be the set of {\em all positive real numbers}. The set of {\em all row vectors of $n\in{\mathbb
N}_{\geq 1}$ elements from a set $X$} is defined as $X^n=\{(x_1,\ldots ,x_n)\mid x_i\in X\ (1\leq i\leq n)\}$. The set
of {\em all mappings from a set $X$ to a set $Y$} is defined as $Y^X=\{f\mid f:X\rightarrow Y\}$.
%
%
Let $Act=\{a,b,\ldots\}$ be the set of {\em actions}.

First, we present a formal definition of LFSPNs.

\begin{definition}
A {\em labeled fluid stochastic Petri net (LFSPN)} is a tuple\\
$N=(P_N,T_N,W_N,C_N,R_N,\Omega_N,L_N,{\cal M}_N)$, where
\begin{itemize}

\item $P_N=Pd_N\uplus Pc_N$ is a finite set of {\em discrete and continuous places} ($\uplus$ denotes disjoint union);

\item $T_N$ is a finite set of {\em transitions},
such that $P_N\cup T_N\neq\emptyset$ and $P_N\cap T_N=\emptyset$;

\item $W_N:(Pd_N\times T_N)\cup (T_N\times Pd_N)\rightarrow{\mathbb N}$ is a function providing the {\em weights of
discrete arcs} between discrete places and transitions;

\item $C_N\subseteq (Pc_N\times T_N)\cup (T_N\times Pc_N)$ is the {\em set of continuous arcs} between continuous
places and transitions;

\item $R_N:C_N\times{\mathbb N}^{|Pd_N|}\rightarrow{\mathbb R}_{\geq 0}$
is a function providing the {\em flow rates of continuous arcs}
in a given discrete marking (the markings will be defined later);

\item $\Omega_N:T_N\times{\mathbb N}^{|Pd_N|}\rightarrow{\mathbb R}_{>0}$ is the {\em transition rate}
function associating
transitions with rates
in a given discrete marking;

\item $L_N:T_N\rightarrow Act$ is the {\em transition labeling} function assigning actions to transitions;

\item ${\cal M}_N=(M_N,{\bf 0})$, where $M_N\in{\mathbb N}^{|Pd_N|}$ and ${\bf 0}$ is a row vector of $|Pc_N|$ values
$0$, is the {\em initial (discrete-continuous) marking}.

\end{itemize}
\end{definition}

Let us consider in more detail the tuple elements from the definition above. Let $N$ be an LFSPN.

Every discrete place $p_i\in Pd_N$ may contain discrete tokens, whose amount is represented by a natural number
$M_i\in{\mathbb N}\ (1\leq i\leq |Pd_N|)$. Each continuous place $q_j\in Pc_N$ may contain continuous fluid, with the
level represented by a non-negative real number $X_j\in{\mathbb R}_{\geq 0}\ (1\leq j\leq |Pc_N|)$. Then the complete
hybrid (discrete-continuous) marking of $N$ is a pair $(M,X)$,
where $M=(M_1,\ldots ,M_{|Pd_N|})$ is a
discrete marking and $X=(X_1,\ldots ,X_{|Pc_N|})$ is a
continuous marking. When needed, these vectors can also be seen as the mappings $M:Pd_N\rightarrow{\mathbb N}$ with
$M(p_i)=M_i\ (1\leq i\leq |Pd_N|)$ and $X:Pc_N\rightarrow{\mathbb R}_{\geq 0}$ with $X(q_j)=X_j\ (1\leq j\leq |Pc_N|)$.
The set of {\em all markings
(reachability set)} of $N$ is denoted by $RS(N)$. Then $DRS(N)=\{M\mid (M,X)\in RS(N)\}$ is the set of {\em all
discrete markings
(discrete reachability set)} of $N$. $DRS(N)$ will be formally defined later. Further, $CRS(N)=\{X\mid (M,X)\in
RS(N)\}\subseteq{\mathbb R}_{\geq 0}^{|Pc_N|}$ is the set of {\em all continuous markings
(continuous reachability set)} of $N$.
Every marking $(M,X)\in RS(N)$
evolves in time, hence, we can interpret it as a stochastic process $\{(M(\delta ),X(\delta ))\mid\delta\geq 0\}$.
Then the initial marking
of $N$ is that at the zero time moment, i.e. ${\cal M}_N=(M_N,{\bf 0})=(M(0),X(0))$,
where $X(0)={\bf 0}$ means that all the continuous places are initially empty.

Every
transition $t\in T_N$
has a positive real instantaneous rate $\Omega_N(t,M)\in{\mathbb R}_{>0}$ associated, which is a parameter of the
exponential distribution governing the transition delay (being a random variable), when the current discrete marking is
$M$.
Transitions are labeled with actions, each representing a sort of activity~that~they~model.

Every discrete arc $da=(p,t)$ or $da=(t,p)$, where $p\in Pd_N$ and $t\in T_N$, connects discrete places and transitions.
It has a
non-negative integer-valued weight $W_N(da)\in{\mathbb N}$ assigned, representing its multiplicity. The zero weight
indicates that the corresponding discrete arc does not exist, since its multiplicity is zero in this case. In the
discrete marking $M\in DRS(N)$,
every continuous arc $ca=(q,t)$ or $ca=(t,q)$, where $q\in Pc_N$ and $t\in T_N$,
connects continuous places and
transitions. It has a non-negative real-valued flow rate $R_N(ca,M)\in{\mathbb R}_{\geq 0}$
of fluid through $ca$, when the current discrete marking is $M$.
The zero flow rate indicates that the fluid flow along the corresponding continuous arc is stopped in some discrete
marking.

The graphical representation of LFSPNs resembles that for standard labeled Petri nets, but supplemented with the rates or
weights, written near the corresponding transitions or arcs. Discrete places are drawn with ordinary circles while
double concentric circles correspond to the continuous ones. Square boxes with the action names inside
depict
transitions and their labels.
Discrete arcs are drawn as thin lines
with arrows at the end while continuous arcs should represent pipes, so the latter are depicted by thick arrowed lines.
If the rates or the weights are not given in the picture then they are assumed to be of no importance in the
corresponding examples. The names of places and transitions are depicted near them when needed.

We now consider the behaviour of LFSPNs.

Let $N$ be an LFSPN and $M$ be a discrete marking of $N$.
%
%
A transition $t\in T_N$ is {\em enabled} in $M$ if $\forall p\in Pd_N\ W_N(p,t)\leq M(p)$.
%
%
%
%
Let $Ena(M)$ be the set of {\em all transitions enabled in $M$}.
Firings of transitions are atomic operations, and only single transitions are fired at once. Note that the enabling
condition depends only on the discrete part of $N$ and this condition is the same as for CTSPNs.
%
%
Firing of a transition $t\in Ena(M)$ changes $M$ to another discrete marking $\widetilde{M}$, such as $\forall p\in
Pd_N\ \widetilde{M}(p)=M(p)-W_N(p,t)+W_N(t,p)$,
denoted by $M\stackrel{t}{\rightarrow}_{\lambda}\widetilde{M}$, where $\lambda =\Omega_N(t,M)$.
%
%
%
We write $M\stackrel{t}{\rightarrow}\widetilde{M}$ if $\exists\lambda\
M\stackrel{t}{\rightarrow}_{\lambda}\widetilde{M}$
and $M\rightarrow\widetilde{M}$ if $\exists t\ M\stackrel{t}{\rightarrow}\widetilde{M}$.

%
%
%
Let us formally define the discrete reachability set
of $N$.
\begin{definition}
Let $N$ be an LFSPN.
%
The {\em discrete reachability set} of $N$, denoted by $DRS(N)$, is the minimal set of discrete markings such that
\begin{itemize}

\item $M_N\in DRS(N)$;

\item if $M\in DRS(N)$ and $M\rightarrow\widetilde{M}$ then $\widetilde{M}\in DRS(N)$.

\end{itemize}
%
%
\end{definition}

Let us now define the discrete reachability graph of $N$.
\begin{definition}
Let $N$ be an LFSPN. The {\em discrete reachability graph} of $N$ is a labeled transition system\\
$DRG(N)=(S_N,{\cal L}_N,{\cal T}_N,s_N)$, where
\begin{itemize}

\item the set of {\em states} is $S_N=DRS(N)$;

\item the set of {\em labels} is ${\cal L}_N=T_N\times{\mathbb R}_{>0}$;

\item the set of {\em transitions} is
${\cal T}_N=\{(M,(t,\lambda ),\widetilde{M})\mid M,\widetilde{M}\in DRS(N),\
M\stackrel{t}{\rightarrow}_{\lambda}\widetilde{M}\}$;

\item the {\em initial state} is $s_N=M_N$.

\end{itemize}
\end{definition}

%
%
%
%
%
%

\section{Discrete part of LFSPNs}
\label{fspndisc.sec}

We have restricted the class of FSPNs underlying LFSPNs to those whose discrete part is CTSPNs, since the performance
analysis of standard FSPNs with GSPNs as the discrete part is finally based on the CTMCs which are extracted from the
underlying semi-Markov chains (SMCs) of the GSPNs by removing vanishing states. Let us now consider the behaviour of
the discrete part of LFSPNs, which is labeled CTSPNs.

For an LFSPN $N$, a continuous random variable $\xi (M)$ is associated with every
discrete marking $M\in DRS(N)$.
The variable captures a residence (sojourn) time in $M$. We
adopt the {\em race} semantics, in which the fastest stochastic transition (i.e. that with the minimal exponentially
distributed firing delay) fires first. Hence, the {\em probability distribution function (PDF)} of the sojourn time in
$M$ is that of the minimal firing delay of transitions from $Ena(M)$. Since exponential distributions are closed under
minimum, the sojourn time in $M$ is (again) exponentially distributed with a parameter that is called the {\em exit
rate from the discrete marking $M$}, defined as

$$RE(M)=\sum_{t\in Ena(M)}\Omega_N(t,M).$$

Note that we may have $RE(M)=0$, meaning that there is no exit from $M$, if it is a {\em terminal discrete marking},
i.e. there are no transitions from it to different discrete markings.

Hence, the PDF of the sojourn time in $M$ (the probability of the residence time in $M$ being less than $\delta$) is
$F_{\xi (M)}(\delta )={\sf P}(\xi (M)<\delta )=1-e^{-RE(M)\delta}\ (\delta\geq 0)$. Then the {\em probability density
function}
of the residence time in $M$ (the limit probability of staying in $M$ at the time $\delta$) is $f_{\xi (M)}(\delta
)=\lim_{\Delta\to 0}\frac{F_{\xi (M)}(\delta +\Delta )-F_{\xi (M)}(\delta )}{\Delta}= \frac{dF_{\xi (M)}(\delta
)}{d\delta}=RE(M)e^{-RE(M)\delta}\ (\delta\geq 0)$.
The mean value (average, expectation) formula for the exponential distribution allows us to calculate the average
sojourn time in $M$ as ${\sf M}({\xi (M)})=\int_0^\infty\delta f_{\xi (M)}(\delta )d\delta =\frac{1}{RE(M)}$.
The variance (dispersion) formula for the exponential distribution allows us to calculate the sojourn time variance in
$M$ as ${\sf D}({\xi (M)})=\int_0^\infty (\delta -{\sf M}({\xi (M)}))^2 f_{\xi (M)}(\delta )d\delta
=\frac{1}{(RE(M))^2}$.
We are now ready to present the following two definitions.

The {\em average sojourn time in the discrete marking} $M$ is

$$SJ(M)=\frac{1}{\sum_{t\in Ena(M)}\Omega_N(t,M)}=\frac{1}{RE(M)}.$$

The {\em average sojourn time vector} of $N$, denoted by $SJ$, has the elements $SJ(M),\ M\in DRS(N)$.

Note that we may have $SJ(M)=\infty$, meaning that we stay in $M$ forever, if it is a terminal discrete marking.

The {\em sojourn time variance in the discrete marking} $M$ is

$$VAR(M)=\frac{1}{\left(\sum_{t\in Ena(M)}\Omega_N(t,M)\right)^2}=\frac{1}{RE(M)^2}.$$

The {\em sojourn time variance vector} of $N$, denoted by $VAR$, has the elements $VAR(M),\ M\in DRS(N)$.

Note that we may have $VAR(M)=\infty$, meaning that the variance of the infinite sojourn time in $M$ is infinite too,
if it is a terminal discrete marking.

To evaluate performance
with the use of the discrete part of $N$, we should investigate the stochastic process associated with it. The process
is the underlying continuous time Markov chain,
denoted by $CTMC(N)$.

%
Let $M,\widetilde{M}\in DRS(N)$. The {\em rate of moving from $M$ to $\widetilde{M}$ by firing any transition} is

$$RM(M,\widetilde{M})=\sum_{\{t\mid M\stackrel{t}{\rightarrow}\widetilde{M}\}}\Omega_N(t,M).$$

\begin{definition}
Let $N$ be an LFSPN. The {\em underlying continuous time Markov chain (CTMC)} of $N$, denoted by $CTMC(N)$, has the
state space $DRS(N)$, the initial state $M_N$ and the transitions $M\rightarrow_{\lambda}\widetilde{M}$, if
$M\rightarrow\widetilde{M}$, where $\lambda =RM(M,\widetilde{M})$.
\end{definition}
%

Isomorphism is a coincidence of systems up to renaming their components or states. Let $\simeq$ denote isomorphism
between CTMCs that binds their initial states.

Let $N$ be an LFSPN. The elements ${\cal Q}_{ij}\ (1\leq i,j\leq n=|DRS(N)|)$ of the {\em transition rate matrix
(TRM)}, also called {\em infinitesimal generator}, ${\bf Q}$ for $CTMC(N)$ are defined as

$${\cal Q}_{ij}=\left\{
\begin{array}{ll}
RM(M_i,M_j), & i\neq j;\\
-\sum_{\{k\mid 1\leq k\leq n,\ k\neq i\}}RM(M_i,M_k), & i=j.
\end{array}
\right.$$

The transient {\em probability mass function (PMF)} $\varphi (\delta )=(\varphi_1(\delta ),\ldots ,\varphi_n(\delta ))$
for $CTMC(N)$ is
calculated via matrix exponent as

$$\varphi (\delta )=\varphi (0)e^{{\bf Q}\delta},$$
where $\varphi (0)=(\varphi_1(0),\ldots ,\varphi_n(0))$ is the initial PMF, defined as

$$\varphi_i(0)=\left\{
\begin{array}{ll}
1, & M_i=M_N;\\
0, & \mbox{otherwise}.
\end{array}
\right.$$
%
%
%

The steady-state PMF $\varphi =(\varphi_1,\ldots ,\varphi_n)$
for $CTMC(N)$ is a solution of the linear equation system

$$\left\{
\begin{array}{l}
\varphi{\bf Q}={\bf 0}\\
\varphi{\bf 1}^T=1
\end{array}
\right.,$$
where ${\bf 0}$ is a row vector of $n$ values $0$ and ${\bf 1}$ is that of $n$ values $1$.

Note that the vector $\varphi$ exists and is unique, if $CTMC(N)$ is ergodic. Then
$CTMC(N)$ has a single steady state, and we have $\varphi =\lim_{\delta\to\infty}\varphi (\delta )$.

Let $N$ be an LFSPN. The following steady-state {\em discrete performance indices (measures)} can be calculated based
on the steady-state PMF $\varphi$ for $CTMC(N)$ \cite{Mol82,Mar90,CMT91,BPTT98,MBCDF95,Bal01,Bal07}.
\begin{itemize}

\item The {\em fraction (proportion) of time spent in the set of discrete markings} $S\subseteq DRS(N)$ is

$$TimeFract(S)=\sum_{\{i\mid M_i\in S\}}\varphi_i.$$

\item The {\em probability that $k\geq 0$ tokens are contained in a discrete place} $p\in Pd_N$ is

$$Tokens(p,k)=\sum_{\{i\mid M_i(p)=k,\ M_i\in DRS(N)\}}\varphi_i.$$

Then the
PMF of the number of tokens in $p$ is $Tokens(p)=(Tokens(p,0),Tokens(p,1),\ldots )$.

\item The
{\em probability of
event ${\cal A}$ defined through (a condition that holds for all discrete markings from) the set of discrete
markings} $DRS_{\cal A}(N)\subseteq DRS(N)$ is

$$Prob({\cal A})=\sum_{\{i\mid M_i\in DRS_{\cal A}(N)\}}\varphi_i.$$

\item The {\em average number of tokens in a discrete place} $p\in Pd_N$ is

$$TokensNum(p)=\sum_{k\geq 1}Tokens(p,k)\cdot k=\sum_{\{i\mid M_i(p)\geq 1,\ M_i\in DRS(N)\}}\varphi_i M_i(p).$$

\item The {\em firing frequency (throughput) of a transition} $t\in T_N$ (average number of firings per unit of time)
is

$$FiringFreq(t)=\sum_{\{i\mid t\in Ena(M_i),\ M_i\in DRS(N)\}}\varphi_i\Omega_N(t,M_i).$$

\item The {\em exit/entrance frequency of a discrete marking} $M_i\in DRS(N)\ (1\leq i\leq n)$ (average number of
exits/entrances per unit of time) is

$$ExitFreq(M_i)=\varphi_i RE(M_i)=\frac{\varphi_i}{SJ(M_i)}.$$

\item The {\em probability of the event determined by a reward function $r(M_i)=r_i\ (0\leq r_i\leq 1,\ 1\leq i\leq n)$
of the discrete markings} is

$$Prob(r)=\sum_{\{i\mid M_i\in DRS(N)\}}\varphi_i r_i.$$

\item The {\em traversal frequency of the move from a discrete marking} $M_i$ {\em to a discrete marking} $M_j\in
DRS(N)\ (1\leq i,j\leq n)$ (average number of traversals per unit of time) is

$$TravFreq(M_i,M_j)=\varphi_i RM(M_i,M_j).$$

\item Let $TravTokens$ be the average number of tokens traversing a subnet of $N$ and $Rate$ be the average input
(output) token rate into (out of) the subnet. The {\em average delay of a token} traversing the subnet
is

$$Delay=\frac{TravTokens}{Rate}.$$

\end{itemize}

\section{Continuous part of LFSPNs}
\label{fspncont.sec}

We now consider the impact the discrete part of LFSPNs has on their continuous part, which is stochastic fluid models
(SFMs). We investigate LFSPNs with a single continuous place, since the definitions and our subsequent results on the
fluid bisimulation can be transferred
straightforwardly to the case of several continuous places, where multidimensional SFMs
have to be explored.

%
Let $N$ be an LFSPN such that $Pc_N=\{q\}$ and $M(\delta )\in DRS(N)$
be its
discrete marking at the time $\delta\geq 0$. Every continuous arc $ca=(q,t)$ or $ca=(t,q)$, where
$t\in T_N$, changes the fluid level in the continuous place $q$ at the time $\delta$ with the flow rate $R_N(ca,M(\delta
))$.
This means that in the discrete marking $M(\delta )$
fluid can leave $q$ along the continuous arc $(q,t)$ with the rate $R_N((q,t),M(\delta ))$
and can enter $q$ along the continuous arc $(t,q)$ with the rate $R_N((t,q),M(\delta ))$
for every
transition $t\in Ena(M(\delta ))$.

The {\em potential rate of the fluid level change (fluid flow rate) for the continuous place} $q$ in the discrete
marking $M(\delta )$
is

$$RP(M(\delta ))=\sum_{\{t\in Ena(M(\delta ))\mid (t,q)\in C_N\}}R_N((t,q),M(\delta ))-
\sum_{\{t\in Ena(M(\delta ))\mid (q,t)\in C_N\}}R_N((q,t),M(\delta )).$$

Let $X(\delta )$ be the fluid level in $q$ at the time $\delta$. It is clear that the fluid level in a continuous place
can never be negative. Therefore, $X(\delta )$ satisfies the following ordinary differential equation describing the
{\em actual fluid flow rate for the continuous place} $q$ in the marking $(M(\delta ),X(\delta ))$:

$$RA(M(\delta ),X(\delta ))=\frac{dX(\delta )}{d\delta }=\left\{
\begin{array}{ll}
\max\{RP(M(\delta )),0\}, & X(\delta )=0;\\
RP(M(\delta )), & (X(\delta )>0)\wedge (RP(M(\delta^-))RP(M(\delta^+))\geq 0);\\
0, & (X(\delta )>0)\wedge (RP(M(\delta^-))RP(M(\delta^+))<0).
\end{array}
\right.$$

In the first case considered in the definition above, we have $X(\delta )=0$. In this case, if $RP(M(\delta ))\geq 0$
then the fluid level is growing and the derivative is equal to the potential rate. Otherwise, if $RP(M(\delta ))<0$
then we should prevent the fluid level from crossing the lower boundary (zero) by stopping the fluid flow. For an
explanation of the more complex second and third cases please refer to \cite{EM94,HKNT98,Gri02,GT07}. Note that
$\frac{dX(\delta )}{d\delta }$ is a piecewise constant function of $X(\delta )$; hence, for each different ``constant''
segment we have $\frac{dX(\delta )}{d\delta }=RP(M(\delta ))$ or $\frac{dX(\delta )}{d\delta }=0$ and, therefore, we
can suppose that within each such segment $RP(M(\delta ))$ or $0$ are the {\em actual} fluid flow rates for the
continuous place $q$ in the marking $(M(\delta ),X(\delta ))$. While constructing differential equations that describe
the behaviour of
SFMs associated with LFSPNs, we are interested only in the segments where $\frac{dX(\delta )}{d\delta }=RP(M(\delta
))$. The SFMs behaviour within the remaining segments, where $\frac{dX(\delta )}{d\delta}=0$, is completely comprised
by the buffer empty probability function that collects the probability mass at the lower boundary.

The elements ${\cal R}_{ij}\ (1\leq i,j\leq n=|DRS(N)|)$
of the {\em fluid rate matrix (FRM)} ${\bf R}$
for the continuous place $q$ are defined as

$${\cal R}_{ij}=\left\{
\begin{array}{ll}
RP(M_i), & i=j;\\
0, & i\neq j.
\end{array}
\right.$$

According to \cite{Gri02,GT07}, the underlying SFMs
of LFSPNs are the first order, infinite buffer, homogeneous Markov fluid models. The discrete part of the SFM derived
from an LFSPN $N$ is the CTMC $CTMC(N)$ with the TRM ${\bf Q}$.
The evolution of the continuous part of the SFM (the fluid flow drift) is described by the FRM ${\bf R}$.
%

Let us consider the {\em transient behaviour} of the SFM associated with an LFSPN $N$. We introduce the following
transient probability functions.
\begin{itemize}

\item $\varphi_i(\delta )={\sf P}(M(\delta )=M_i)$ is the {\em discrete marking probability};

\item $\ell_i(\delta )={\sf P}(X(\delta )=0,\ M(\delta )=M_i)$ is the {\em buffer empty probability (probability mass
at the lower boundary)};

\item $F_i(\delta ,x)={\sf P}(X(\delta )<x,\ M(\delta )=M_i)$ is the {\em fluid probability distribution function};

\item $f_i(\delta ,x)=\frac{\partial F_i(\delta ,x)}{\partial x}=\lim_{h\to 0}\frac{F_i(\delta ,x+h)-
F_i(\delta ,x)}{h}=\lim_{h\to 0}\frac{{\sf P}(x<X(\delta )<x+h,\ M(\delta )=M_i)}{h}$ is the {\em fluid probability
density function}.

\end{itemize}

The initial conditions are:

$$\ell_i(0)=\left\{
\begin{array}{ll}
1, & M_i=M_N;\\
0, & \mbox{otherwise};
\end{array}
\right.$$

$$F_i(0,x)=\left\{
\begin{array}{ll}
1, & (M_i=M_N)\wedge (x\geq 0);\\
0, & \mbox{otherwise};
\end{array}
\right.$$

$$f_i(0,x)=0\ \forall (M_i,x)\in RS(N).$$

Let $\varphi (\delta ),\ell (\delta ),F(\delta ,x),f(\delta ,x)$ be the row vectors with the elements $\varphi_i(\delta
),\ell_i(\delta ),F_i(\delta ,x),f_i(\delta ,x)$, respectively $(1\leq i\leq n)$.

By the total probability law, we have

$$\ell (\delta )+\int_{0+}^{\infty}f(\delta ,x)dx=\varphi (\delta ).$$

The partial differential equations describing the transient behaviour are

$$\frac{\partial F(\delta ,x)}{\partial\delta}+\frac{\partial F(\delta ,x)}{\partial x}{\bf R}=F(\delta ,x){\bf Q},\
x>0;$$

$$\frac{\partial f(\delta ,x)}{\partial\delta}+\frac{\partial f(\delta ,x)}{\partial x}{\bf R}=f(\delta ,x){\bf Q},\
x>0.$$

Note that we have $\frac{\partial F(\delta ,x)}{\partial x}=f(\delta ,x),\ F(\delta ,0)=\ell (\delta ),\ F(\delta
,\infty )=\varphi (\delta )$.

The partial differential equation for the buffer empty probabilities (lower boundary conditions) are

$$\frac{d\ell(\delta)}{d\delta}+f(\delta ,0){\bf R}=\ell(\delta ){\bf Q}.$$

The lower boundary constraint is: if ${\cal R}_{ii}=RP(M_i)>0$ then $\ell_i(\delta )=F_i(\delta ,0)=0\ (1\leq i\leq
n)$.

The normalizing condition is

$$\ell(\delta ){\bf 1}^T+\int_{0+}^{\infty}f(\delta ,x)dx{\bf 1}^T=1,$$
where ${\bf 1}$ is a row vector of $n$ values $1$.

Let us now consider the {\em stationary behaviour} of the SFM associated with an LFSPN $N$.
We do not discuss here in detail the conditions under which the steady state for the associated SFM {\em exists} and is
{\em unique}, since this topic has been extensively explored in \cite{HKNT98,Gri02,GT07}. Particularly, according to
\cite{HKNT98,GT07}, the steady-state PDF {\em exists} (i.e. the transient functions approach their stationary values,
as the time parameter $\delta$ tends to infinity in the transient equations), when the associated SFM is a Markov fluid
model, whose fluid flow drift (described by the matrix ${\bf R}$) and transition rates (described by the matrix ${\bf
Q}$) are fluid level independent, and the following {\em stability condition} holds:

$$FluidFlow(q)=\sum_{i=1}^n\varphi_i RP(M_i)=\varphi{\bf R}{\bf 1}^T<0,$$
stating that the steady-state {\em mean potential fluid flow rate for the continuous place} $q$ is negative. Stable
infinite buffer models usually converge, hence, the existing steady-state PDF is also {\em unique} in this case.

We introduce the following steady-state probability functions, obtained from the transient ones by taking the limit
$\delta\to\infty$.
\begin{itemize}

\item $\varphi_i=\lim_{\delta\to\infty}{\sf P}(M(\delta )=M_i)$ is the {\em steady-state discrete marking probability};

\item $\ell_i=\lim_{\delta\to\infty}{\sf P}(X(\delta )=0,\ M(\delta )=M_i)$ is the {\em steady-state buffer empty
probability
(probability mass at the lower boundary)};

\item $F_i(x)=\lim_{\delta\to\infty}{\sf P}(X(\delta )<x,\ M(\delta )=M_i)$ is the {\em steady-state fluid probability
distribution function};

\item $f_i(x)=\frac{dF_i(x)}{dx}=\lim_{h\to 0}\frac{F_i(x+h)-F_i(x)}{h}=
\lim_{\delta\to\infty}\lim_{h\to 0}\frac{{\sf P}(x<X(\delta )<x+h,\ M(\delta )=M_i)}{h}$ is the {\em steady-state fluid
probability density function}.

\end{itemize}

Let $\varphi ,\ell ,F(x),f(x)$ be the row vectors with the elements $\varphi_i,\ell_i,F_i(x),f_i(x)$, respectively
$(1\leq i\leq n)$.

By the total probability law for the stationary behaviour, we have

$$\ell +\int_{0+}^{\infty}f(x)dx=\varphi .$$

The ordinary differential equations describing the stationary behaviour are

$$\frac{dF(x)}{dx}{\bf R}=F(x){\bf Q},\ x>0;$$

$$\frac{df(x)}{dx}{\bf R}=f(x){\bf Q},\ x>0.$$

Note that we have $\frac{dF(x)}{dx}=f(x),\ F(0)=\ell ,\ F(\infty )=\varphi$.

The ordinary differential equation for the steady-state buffer empty probabilities (stationary lower boundary
conditions) are

$$f(0){\bf R}=\ell{\bf Q}.$$

The stationary lower boundary constraint is: if ${\cal R}_{ii}=RP(M_i)>0$ then $F_i(0)=\ell_i=0\ (1\leq i\leq n)$.

The stationary normalizing condition is

$$\ell{\bf 1}^T+\int_{0+}^{\infty}f(x)dx{\bf 1}^T=1,$$
where ${\bf 1}$ is a row vector of $n$ values $1$.

The solutions of the equations for $F(x)$ and $f(x)$ in the form of {\em matrix exponent} are $F(x)=\ell e^{x{\bf
Q}{\bf R}^{-1}}$ and $f(x)=\ell{\bf Q}{\bf R}^{-1}e^{x{\bf Q}{\bf R}^{-1}}$, respectively. Since the steady-state
existence implies boundedness of the SFM associated with an LFSPN and we do not have a finite upper fluid level bound,
the positive eigenvalues of ${\bf Q}{\bf R}^{-1}$ must be excluded. Moreover, ${\bf R}^{-1}$ does not exist if for some
$i\ (1\leq i\leq n)$ we have ${\cal R}_{ii}=0$. These difficulties are avoided in the alternative solution method for
$F(x)$, called {\em spectral decomposition} \cite{TK93,HKNT98,Gri02,GT07,GMST08}, which we
outline below.

Let us define the sets of {\em negative discrete markings} of $N$ as $DRS^-(N)=\{M\in DRS(N)\mid RP(M)<0\}$, {\em zero
discrete markings} of $N$ as $DRS^0(N)=\{M\in DRS(N)\mid RP(M)=0\}$ and {\em positive discrete markings} of $N$ as
$DRS^+(N)=\{M\in DRS(N)\mid RP(M)>0\}$. The spectral decomposition is
$F(x)=\sum_{j=1}^m a_j e^{\gamma_j x}v_j$, where $a_j$ are some scalar coefficients, $\gamma_j$ are the eigenvalues and
$v_j=(v_{j1},\ldots ,v_{jn})$ are the eigenvectors of ${\bf Q}{\bf R}^{-1}$. Thus, each $v_j$ is the solution of the
equation $v_j({\bf Q}{\bf R}^{-1}-\gamma_j{\bf I})=0$, where ${\bf I}$ is the identity matrix of the order $n$, hence,
it holds $v_j({\bf Q}-\gamma_j{\bf R})=0$.

Since for each non-zero $v_j$ we must have $|{\bf Q}-\gamma_j{\bf R}|=0$,
the number of solutions $\gamma_1 ,\ldots ,\gamma_m$ is the number of non-zero elements among ${\cal R}_{ii}=RP(M_i)\
(1\leq i\leq n)$, i.e. $m=|DRS^-(N)|+|DRS^+(N)|$. We have $1$ zero eigenvalue, $|DRS^+(N)|$ eigenvalues with a negative
real part and $|DRS^-(N)|-1$ eigenvalues with a positive real part. Let us reorder all the eigenvalues according to the
sign of their real part (first, with a zero real part; then with a negative one; at last, with a positive one). The
boundedness of $F(x)$ requires $a_j=0$ if $Re({\gamma_j})>0\ (1\leq j\leq m)$. Further, for the zero eigenvalue
$\gamma_1=0$ we have $a_1 e^{\gamma_1 x}v_1=a_1 v_1$, and for the corresponding eigenvector it holds $v_1{\bf Q}=0$.
Then $F(x)=a_1 v_1+\sum_{k=2}^{|DRS^+(N)|+1}a_k e^{\gamma_k x}v_k$, where $Re({\gamma_k})<0\ (2\leq k\leq
|DRS^+(N)|+1)$. Remember that $\varphi =F(\infty )=a_1 v_1$, hence, $F(x)=\varphi +\sum_{k=2}^{|DRS^+(N)|+1}a_k
e^{\gamma_k x}v_k$.

It remains to find $|DRS^+(N)|$ coefficients $a_k$ corresponding to the eigenvalues $\gamma_k\ (2\leq k\leq
|DRS^+(N)|+1)$. Remember the stationary lower boundary constraint: if ${\cal R}_{ll}=RP(M_l)>0$ then $F_l(0)=\ell_l=0$.
Then for each positive discrete marking $M_l\in DRS^+(N)$ we have $F_l(0)=\varphi_l+\sum_{k=2}^{|DRS^+(N)|+1}a_k
v_{kl}=0$.
We obtain a system of $|DRS^+(N)|$ independent linear equations with $|DRS^+(N)|$ unknowns, for which a unique solution
exists.

Then, using $F(x)$, we can find $f(x)=\frac{dF(x)}{dx}$ and $\ell =F(0)$.

Let $N$ be an LFSPN. The following steady-state {\em hybrid (discrete-continuous) performance indices (measures)} can
be calculated based on the steady-state fluid probability density function $f(x)$ for the SFM of $N$
\cite{BGGHST99,GS00,GSHB01,GHo02,Gri02,HGr02}. Note that the hybrid performance indices that do not depend on the fluid
level coincide with the corresponding discrete performance measures.
\begin{itemize}

\item The {\em fraction (proportion) of time spent in the set of discrete markings} $S\subseteq DRS(N)$ is

$$TimeFract(S)=\sum_{\{i\mid M_i\in S\}}\left(\ell_i+\int_{0+}^\infty f_i(x)dx\right)=
\sum_{\{i\mid M_i\in S\}}\varphi_i.$$

\item The {\em probability that $k\geq 0$ tokens are contained in a discrete place} $p\in Pd_N$ is

$$Tokens(p,k)=\sum_{\{i\mid M_i(p)=k,\ M_i\in DRS(N)\}}\left(\ell_i+\int_{0+}^\infty f_i(x)dx\right)=
\sum_{\{i\mid M_i(p)=k,\ M_i\in DRS(N)\}}\varphi_i.$$

Then the
PMF of the number of tokens in $p$ is $Tokens(p)=(Tokens(p,0),Tokens(p,1),\ldots )$.

\item The {\em probability of the event ${\cal A}$ defined through (a condition that holds for all discrete markings
from) the set of discrete markings} $DRS_{\cal A}(N)\subseteq DRS(N)$ is

$$Prob({\cal A})=\sum_{\{i\mid M_i\in DRS_{\cal A}(N)\}}\left(\ell_i+\int_{0+}^\infty f_i(x)dx\right)=
\sum_{\{i\mid M_i\in DRS_{\cal A}(N)\}}\varphi_i.$$

\item The {\em average number of tokens in a discrete place} $p\in Pd_N$ is

$$\begin{array}{c}
TokensNum(p)=\sum_{k\geq 1}Tokens(p,k)\cdot k=
\displaystyle\sum_{\{i\mid M_i(p)\geq 1,\ M_i\in DRS(N)\}}\left(\ell_i+\int_{0+}^\infty f_i(x)dx\right)M_i(p)=\\
\displaystyle\sum_{\{i\mid M_i(p)\geq 1,\ M_i\in DRS(N)\}}\varphi_i M_i(p).
\end{array}$$

\item The {\em firing frequency (throughput) of a transition} $t\in T_N$ (average number of firings per unit of time)
is

$$\begin{array}{c}
FiringFreq(t)=\displaystyle\sum_{\{i\mid t\in Ena(M_i),\ M_i\in DRS(N)\}}\left(\ell_i+\int_{0+}^\infty f_i(x)dx\right)
\Omega_N(t,M_i)=\\
\displaystyle\sum_{\{i\mid t\in Ena(M_i),\ M_i\in DRS(N)\}}\varphi_i\Omega_N(t,M_i).
\end{array}$$

\item The {\em exit/entrance frequency of a discrete marking} $M_i\in DRS(N)\ (1\leq i\leq n)$ (average number of
exits/entrances per unit of time) is

$$ExitFreq(M_i)=\left(\ell_i+\int_{0+}^\infty f_i(x)dx\right)\frac{1}{SJ(M_i)}=\frac{\varphi_i}{SJ(M_i)}.$$

\item The {\em mean potential fluid flow rate for the continuous place} $q\in Pc_N$ is

$$FluidFlow(q)=\sum_{\{i\mid M_i\in DRS(N)\}}\left(\ell_i+\int_{0+}^\infty f_i(x)dx\right)RP(M_i)=
\sum_{\{i\mid M_i\in DRS(N)\}}\varphi_i RP(M_i).$$

\item The {\em probability of the event determined by a reward function $r(M_i)=r_i\ (0\leq r_i\leq 1,\\
1\leq i\leq n)$ of the discrete markings} is

$$Prob(r)=\sum_{\{i\mid M_i\in DRS(N)\}}\left(\ell_i+\int_{0+}^\infty f_i(x)dx\right)r_i=
\sum_{\{i\mid M_i\in DRS(N)\}}\varphi_i r_i.$$

\item The {\em traversal frequency of the move from a discrete marking} $M_i$ {\em to a discrete marking} $M_j\in
DRS(N)\ (1\leq i,j\leq n)$ (average number of traversals per unit of time) is

$$TravFreq(M_i,M_j)=\left(\ell_i+\int_{0+}^\infty f_i(x)dx\right)RM(M_i,M_j)=\varphi_i RM(M_i,M_j).$$

\item The {\em probability
of
a positive fluid level in a continuous place} $q\in Pc_N$ is

$$\begin{array}{c}
FluidLevel(q)=\displaystyle\sum_{\{i\mid M_i\in DRS(N)\}}\left(\ell_i\cdot 0+\int_{0+}^\infty f_i(x)\cdot 1dx\right)=
\displaystyle\sum_{\{i\mid M_i\in DRS(N)\}}\int_{0+}^\infty f_i(x)dx=\\
\displaystyle\sum_{\{i\mid M_i\in DRS(N)\}}(\varphi_i-\ell_i)=1-\displaystyle\sum_{\{i\mid M_i\in DRS(N)\}}\ell_i.
\end{array}$$

\item The {\em probability
that
the fluid level in a continuous place} $q\in Pc_N$ {\em does not lie below the value} $v\in{\mathbb R}_{>0}$ is

$$\begin{array}{c}
FluidLevel(q,v)=\displaystyle\sum_{\{i\mid M_i\in DRS(N)\}}\left(\ell_i\cdot 0+\int_{0+}^v f_i(x)\cdot 0dx+
\int_v^\infty f_i(x)\cdot 1dx\right)=\\
\displaystyle\sum_{\{i\mid M_i\in DRS(N)\}}\int_v^\infty f_i(x)dx=
\displaystyle\sum_{\{i\mid M_i\in DRS(N)\}}(\varphi_i-F_i(v))=1-\displaystyle\sum_{\{i\mid M_i\in DRS(N)\}}F_i(v).
\end{array}$$

\item The {\em mean proportional flow rate across a continuous arc} $(q,t),\ q\in Pc_N, t\in T_N$, is

$$FluidFlow(q,t)=\sum_{\{i\mid t\in Ena(M_i),\ M_i\in DRS(N)\}}\left(\ell_i R_N^*((q,t),(M_i,0))+
\int_{0+}^\infty f_i(x)R_N^*((q,t),(M_i,x))dx\right),$$
where $R_N^*((q,t),(M,x))$ is the fluid level dependent {\em proportional flow rate function in the marking}
$(M,x)\in RS(N)$, defined as

$$R_N^*((q,t),(M,x))=\left\{
\begin{array}{ll}
R_N((q,t),M), & x>0;\\
R_N((q,t),M)\cdot\frac{\sum_{u\in Ena(M)}R_N((u,q),M)}{\sum_{v\in Ena(M)}R_N((q,v),M)}, & x=0.
\end{array}
\right.$$

Thus,

$$\begin{array}{c}
FluidFlow(q,t)=\\
\displaystyle\sum_{\{i\mid t\in Ena(M_i),\ M_i\in DRS(N)\}}\left(\ell_i\cdot\frac{\sum_{u\in Ena(M)}R_N((u,q),M)}
{\sum_{v\in Ena(M)}R_N((q,v),M)}+\int_{0+}^\infty f_i(x)dx\right)R_N((q,t),M)=\\
\displaystyle\sum_{\{i\mid t\in Ena(M_i),\ M_i\in DRS(N)\}}
\left(\ell_i\left(\frac{\sum_{u\in Ena(M)}R_N((u,q),M)}{\sum_{v\in Ena(M)}R_N((q,v),M)}-1\right)+
\varphi_i\right)R_N((q,t),M).
\end{array}$$

\item The {\em mean proportional flow rate across a continuous arc} $(t,q),\ t\in T_N,\ q\in Pc_N$, is

$$FluidFlow(t,q)=\sum_{\{i\mid t\in Ena(M_i),\ M_i\in DRS(N)\}}\left(\ell_i R_N^*((t,q),(M_i,0))+
\int_{0+}^\infty f_i(x)R_N^*((t,q),(M_i,x))dx\right),$$
where $R_N^*((t,q),(M,x))$ is the fluid level dependent {\em proportional flow rate function in the marking}
$(M,x)\in RS(N)$, defined as

$$R_N^*((t,q),(M,x))=\left\{
\begin{array}{ll}
R_N((t,q),M), & x>0;\\
R_N((t,q),M)\cdot\frac{\sum_{u\in Ena(M)}R_N((q,u),M)}{\sum_{v\in Ena(M)}R_N((v,q),M)}, & x=0.
\end{array}
\right.$$

Thus,

$$\begin{array}{c}
FluidFlow(t,q)=\\
\displaystyle\sum_{\{i\mid t\in Ena(M_i),\ M_i\in DRS(N)\}}\left(\ell_i\cdot\frac{\sum_{u\in Ena(M)}R_N((q,u),M)}
{\sum_{v\in Ena(M)}R_N((v,q),M)}+\int_{0+}^\infty f_i(x)dx\right)R_N((t,q),M)=\\
\displaystyle\sum_{\{i\mid t\in Ena(M_i),\ M_i\in DRS(N)\}}
\left(\ell_i\left(\frac{\sum_{u\in Ena(M)}R_N((q,u),M)}{\sum_{v\in Ena(M)}R_N((v,q),M)}-1\right)+
\varphi_i\right)R_N((t,q),M).
\end{array}$$

\item The {\em probability of the event determined by a hybrid reward function $r(M_i,x)=r_i(x)\ (0\leq r_i(x)\leq 1,\
1\leq i\leq n)$ of the markings} is

$$Prob(r)=\sum_{\{i\mid M_i\in DRS(N)\}}\left(\ell_i r_i(0)+\int_{0+}^\infty f_i(x)r_i(x)dx\right).$$

%
%
\end{itemize}

\section{Fluid trace equivalence}
\label{flutraeq.sec}

Trace equivalences are the least discriminating ones. In the trace semantics, the behavior of a system is associated
with the set of all possible sequences of actions, i.e. the protocols of work or computations. Thus, the points of
choice of an external observer between several extensions of a particular computation are not taken into account.

The formal definition of fluid trace equivalence resembles that of ordinary Markovian trace equivalence, proposed on
transition-labeled CTMCs in \cite{WBMC06}, on sequential and concurrent Markovian process calculi SMPC and CMPC in
\cite{BBo06,Bern07a,Bern07b,BBo08} and on Uniform Labeled Transition Systems (ULTraS) in \cite{BNL13,BeTe13}. While
defining fluid trace equivalence, we additionally have to take into account the fluid flow rates in the corresponding
discrete markings of two compared LFSPNs. Hence, in order to construct fluid trace equivalence, we should determine how
to calculate the cumulative execution probabilities of all the specific (selected) paths. A {\em path} in the discrete
reachability graph of an LFSPN is a sequence of its discrete markings and transitions that is generated by some firing
sequence in the LFSPN.

First, we should {\em multiply the transition firing probabilities} for all the transitions along the paths starting in
the initial discrete marking of the LFSPN. The resulting product will be the {\em execution probability of the path}.
Second, we should {\em sum the path execution probabilities} for all the selected paths corresponding to the same {\em
sequence of actions},
moreover, to the same {\em sequence of the average sojourn times} and the same {\em sequence of the fluid flow rates}
in all the discrete markings participating the paths. We suppose that each LFSPN has exactly one continuous place. The
resulting sum will be the {\em cumulative execution probability of the selected paths} corresponding to some fluid
stochastic trace. A {\em fluid stochastic trace} is a pair with the first element being the triple of the correlated
sequences of actions, average sojourn times and fluid flow rates, and the second element being the execution
probability of the triple. Each element of the triple guarantees that fluid trace equivalence respects the following
important aspects of the LFSPNs behaviour: {\em functional activity}, {\em stochastic timing} and {\em fluid flow}.

It is also possible to define fluid trace equivalence between LFSPNs with more than one continuous place, if they have
the same number of the {\em corresponding} continuous places. Then one should consider the sequences of the {\em
vectors} of the average sojourn times and {\em vectors} of the fluid flow rates. The elements of each such a vector
will be the average sojourn times or fluid flow rates, respectively, for all continuous places in a particular discrete
marking.

Note that $CTMC(N)$ can be interpreted as a semi-Markov chain (SMC) \cite{Kul09}, denoted by $SMC(N)$, which is
analyzed by extracting from it the embedded (absorbing) discrete time Markov chain (EDTMC) corresponding to $N$,
denoted by $EDTMC(N)$. The construction of the latter is analogous to that applied in the context of GSPNs in
\cite{Mar90,MBCDF95,Bal01,Bal07}. $EDTMC(N)$ only describes the state changes of $SMC(N)$ while ignoring its time
characteristics. Thus, to construct the EDTMC, we should abstract from all time aspects of behaviour of the SMC, i.e.
from the sojourn time in its states. It is well-known that every SMC is fully described by the EDTMC and the state
sojourn time distributions (the latter can be specified by the vector of PDFs of residence time in the states)
\cite{Hav01}.

We first propose some helpful definitions of the probability functions for the transition firings and discrete marking
changes. Let $N$ be an LFSPN, $M,\widetilde{M}\in DRS(N)$ be its discrete markings and $t\in Ena(M)$.

The (time-abstract) {\em probability that the transition $t$ fires in $M$} is

$$PT(t,M)=\frac{\Omega_N(t,M)}{\sum_{u\in Ena(M)}\Omega_N(u,M)}=\frac{\Omega_N(t,M)}{RE(M)}=SJ(M)\Omega_N(t,M).$$

We have $\forall M\in{\mathbb N}^{|Pd_N|}\ \sum_{t\in Ena(M)}PT(t,M)=\sum_{t\in Ena(M)}\frac{\Omega_N(t,M)}{\sum_{u\in
Ena(M)}\Omega_N(u,M)}=\frac{\sum_{t\in Ena(M)}\Omega_N(t,M)}{\sum_{u\in Ena(M)}\Omega_N(u,M)}=1$, i.e. $PT(t,M)$
defines a probability distribution.

The {\em probability to move from $M$ to $\widetilde{M}$ by firing any transition} is

$$PM(M,\widetilde{M})=\sum_{\{t\mid M\stackrel{t}{\rightarrow}\widetilde{M}\}}PT(t,M)=
\frac{\sum_{\{t\mid M\stackrel{t}{\rightarrow}\widetilde{M}\}}\Omega_N(t)}{RE(M)}=
SJ(M)\cdot\sum_{\{t\mid M\stackrel{t}{\rightarrow}\widetilde{M}\}}\Omega_N(t).$$

We write $M\rightarrow_{\cal P}\widetilde{M}$, if $M\rightarrow\widetilde{M}$, where ${\cal P}=PM(M,\widetilde{M})$. We
have $\forall M\in{\mathbb N}^{|Pd_N|}\ \sum_{\{\widetilde{M}\mid M\rightarrow\widetilde{M}\}}PM(M,\widetilde{M})=
\sum_{\{\widetilde{M}\mid M\rightarrow\widetilde{M}\}}\sum_{\{t\mid M\stackrel{t}{\rightarrow}\widetilde{M}\}}PT(t,M)=
\sum_{t\in Ena(M)}PT(t,M)=1$, i.e. $PM(M,\widetilde{M})$ defines a probability distribution.

\begin{definition}
Let $N$ be an LFSPN. The {\em embedded (absorbing) discrete time Markov chain (EDTMC)} of $N$, denoted by $EDTMC(N)$,
has the state space $DRS(N)$, the initial state $M_N$ and the transitions $M\rightarrow_{\cal P}\widetilde{M}$, if
$M\rightarrow\widetilde{M}$,
where ${\cal P}=PM(M,\widetilde{M})$.

The {\em underlying SMC} of $N$, denoted by $SMC(N)$, has the EDTMC $EDTMC(N)$ and the sojourn time in every $M\in
DRS(N)$ is exponentially distributed with the parameter $RE(M)$.
\end{definition}

Since the sojourn time in every $M\in DRS(N)$ is exponentially distributed, we have $SMC(N)=CTMC(N)$.

Let $N$ be an LFSPN. The elements ${\cal P}_{ij}\ (1\leq i,j\leq n=|DRS(N)|)$ of the {\em (one-step) transition
probability matrix (TPM)} ${\bf P}$ for $EDTMC(N)$ are defined as

$${\cal P}_{ij}=\left\{
\begin{array}{ll}
PM(M_i,M_j), & M_i\rightarrow M_j;\\
0, & \mbox{otherwise}.
\end{array}
\right.$$

Let $X$ be a set, $n\in{\mathbb N}_{\geq 1}$ and $x_i\in X\ (1\leq i\leq n)$. Then $\chi =x_1\cdots x_n$ is a finite
sequence over $X$ of {\em length} $|\chi |=n$.
When $X$ is a set on numbers, we usually write $\chi =x_1\circ\cdots\circ x_n$, to avoid confusion because of mixing up
the operations of concatenation of sequences ($\circ$) and multiplication of numbers ($\cdot$). The {\em empty
sequence} $\varepsilon$ of length $|\varepsilon |=0$ is an extra case. Let $X^*$ denote the {\em set of all finite
sequences} (including the empty one) over $X$.

Let $M_N=M_0\stackrel{t_1}{\rightarrow}M_1\stackrel{t_2}{\rightarrow}\cdots\stackrel{t_n}{\rightarrow}M_n\
(n\in{\mathbb N})$ be a finite sequence of transition firings starting in the initial discrete marking $M_N$ and called
{\em firing sequence} in $N$. The firing sequence generates the {\em path} $M_0 t_1 M_1 t_2\cdots t_n M_n$ in the
discrete reachability graph $DRG(N)$. Since the {\em first discrete marking} $M_N=M_0$ of the path is fixed, one can
see that the (finite) {\em transition sequence} $\vartheta =t_1\cdots t_n$ in $N$ uniquely determines the {\em discrete
marking sequence} $M_0\cdots M_n$, ending with the {\em last discrete marking} $M_n$ of the mentioned path in $DRG(N)$.
Hence, to refer the paths, one can simply use the transition sequences extracted from them as shown above. The {\em
empty transition sequence} $\varepsilon$ refers to the path $M_0$, consisting just of one discrete marking (which is
the first and last one of the path in such a case).

Let $N$ be an LFSPN. The set of {\em all (finite) transition sequences} in $N$ is defined as

$$TranSeq(N)=\{\vartheta\mid\vartheta =\varepsilon\mbox{ or }\vartheta =t_1\cdots t_n,\ M_N=M_0
\stackrel{t_1}{\rightarrow}M_1\stackrel{t_2}{\rightarrow}\cdots\stackrel{t_n}{\rightarrow}M_n\}.$$

Let $\vartheta =t_1\cdots t_n\in TranSeq(N)$ and $M_N=M_0\stackrel{t_1}{\rightarrow}M_1\stackrel{t_2}{\rightarrow}\cdots
\stackrel{t_n}{\rightarrow}M_n$. The {\em probability to execute the transition sequence $\vartheta$} is

$$PT(\vartheta )=\prod_{i=1}^n PT(t_i,M_{i-1}).$$

For $\vartheta =\varepsilon$ we define $PT(\varepsilon )=1$.
Let us prove that $\forall n\in{\mathbb N}\ \sum_{\{\vartheta\in TranSeq(N)\mid |\vartheta |=n\}}PT(\vartheta )=1$,
i.e. $PT(\vartheta )$ defines a probability distribution.

\begin{lemma}
Let $N$ be an LFSPN. Then $\forall n\in{\mathbb N}$

$$\sum_{\{\vartheta\in TranSeq(N)\mid |\vartheta |=n\}}PT(\vartheta )=1.$$

\label{sumprtrseq.lem}
\end{lemma}
{\em Proof.} We prove by induction on the transition sequences length $n$.
\begin{itemize}

\item $n=0$

By definition, $\sum_{\{\vartheta\in TranSeq(N)\mid |\vartheta |=0\}}PT(\vartheta )=PT(\varepsilon )=1$.

\item $n\rightarrow n+1$

By distributivity law for multiplication and addition, and since $\forall M\in{\mathbb N}^{|Pd_N|}\
\sum_{t\in Ena(M)}PT(t,M)=1$,\\
$\sum_{\{\vartheta\in TranSeq(N)\mid |\vartheta |=n+1\}}PT(\vartheta )=\sum_{\{t_1,\ldots ,t_n,t_{n+1}\mid
M_N=M_0\stackrel{t_1}{\rightarrow}M_1\stackrel{t_2}{\rightarrow}\cdots\stackrel{t_n}{\rightarrow}M_n
\stackrel{t_{n+1}}{\rightarrow}M_{n+1}\}}\prod_{i=1}^{n+1}PT(t_i,M_{i-1})=\\
\sum_{\{t_1,\ldots ,t_n\mid M_N=M_0\stackrel{t_1}{\rightarrow}M_1\stackrel{t_2}{\rightarrow}\cdots
\stackrel{t_n}{\rightarrow}M_n\}}\sum_{\{t_{n+1}\mid M_n\stackrel{t_{n+1}}{\rightarrow}M_{n+1}\}}\prod_{i=1}^n
PT(t_i,M_{i-1})PT(t_{n+1},M_n)=\\
\sum_{\{t_1,\ldots ,t_n\mid M_N=M_0\stackrel{t_1}{\rightarrow}M_1\stackrel{t_2}{\rightarrow}\cdots
\stackrel{t_n}{\rightarrow}M_n\}}\left(\prod_{i=1}^n PT(t_i,M_{i-1})\sum_{\{t_{n+1}\mid
M_n\stackrel{t_{n+1}}{\rightarrow}M_{n+1}\}}PT(t_{n+1},M_n)\right)=\\
\sum_{\{t_1,\ldots ,t_n\mid M_N=M_0\stackrel{t_1}{\rightarrow}M_1\stackrel{t_2}{\rightarrow}\cdots
\stackrel{t_n}{\rightarrow}M_n\}}\prod_{i=1}^n PT(t_i,M_{i-1})\cdot 1=1$. \hfill $\eop$

\end{itemize}

Let $\vartheta =t_1\cdots t_n\in TranSeq(N)$ be a transition sequence in $N$. The {\em action sequence} of $\vartheta$
is $L_N(\vartheta )=a_1\cdots a_n\in Act^*$, where $L_N(t_i)=a_i\ (1\leq i\leq n)$, i.e. it is the sequence of actions
which label the transitions of that transition sequence. For $\vartheta =\varepsilon$ we define $L_N(\varepsilon
)=\varepsilon$. Further, the {\em average sojourn time sequence} of $\vartheta$ is
$SJ(\vartheta )=SJ(M_0)\circ\cdots\circ SJ(M_n)\in{\mathbb R}_{>0}^*$, i.e. it is the sequence of average sojourn times
in the discrete markings of the path to which $\vartheta$ refers. For $\vartheta =\varepsilon$ we define
$SJ(\varepsilon )=SJ(M_0)$. Similarly, the {\em (potential) fluid flow rate sequence} of $\vartheta$ is
$RP(\vartheta )=RP(M_0)\circ\cdots\circ RP(M_n)\in{\mathbb R}^*$, i.e. it is the sequence of (potential) fluid flow
rates in the discrete markings of the path to which $\vartheta$ refers. For $\vartheta =\varepsilon$ we define
$RP(\varepsilon )=RP(M_0)$.

Let $N$ be an LFSPN and $(\sigma ,\varsigma ,\varrho )\in Act^*\times{\mathbb R}_{>0}^*\times{\mathbb R}^*$. The set of
{\em $(\sigma ,\varsigma ,\varrho )$-selected (finite) transition sequences} in $N$ is defined as

$$TranSeq(N,\sigma ,\varsigma ,\varrho )=\{\vartheta\in TranSeq(N)\mid L_N(\vartheta )=\sigma ,\ SJ(\vartheta
)=\varsigma ,\ RP(\vartheta )=\varrho\}.$$

Let $TranSeq(N,\sigma ,\varsigma ,\varrho )\neq\emptyset$. Then the triple $(\sigma ,\varsigma ,\varrho )$, together
with its execution probability, which is the cumulative execution probability of all the paths from which the triple is
extracted (as described above), constitute a {\em fluid stochastic trace} of the LFSPN $N$. Fluid stochastic traces are
formally introduced below, followed by the (first) definition of fluid stochastic trace equivalence.

\begin{definition}
A {\em (finite) fluid stochastic trace} of an LFSPN $N$ is a pair $((\sigma ,\varsigma ,\varrho ),PT(\sigma ,\varsigma
,\varrho ))$, where\\
$TranSeq(N,\sigma ,\varsigma ,\varrho )\neq\emptyset$ and the (cumulative) {\em probability to execute $(\sigma
,\varsigma ,\varrho )$-selected transition sequences} is

$$PT(\sigma ,\varsigma ,\varrho )=\sum_{\vartheta\in TranSeq(N,\sigma ,\varsigma ,\varrho )}PT(\vartheta ).$$

We denote the set of {\em all fluid stochastic traces} of an LFSPN $N$ by $FluStochTraces(N)$. Two LFSPNs $N$ and $N'$
are {\em fluid trace equivalent}, denoted by $N\equiv_{fl}N'$, if

$$FluStochTraces(N)=FluStochTraces(N').$$

\label{flteq1.def}
\end{definition}

By Lemma \ref{sumprtrseq.lem}, we have $\forall n\in{\mathbb N}\ \sum_{\{(\sigma ,\varsigma ,\varrho )\mid |\sigma
|=n\}}PT(\sigma ,\varsigma ,\varrho )=\sum_{\{(\sigma ,\varsigma ,\varrho )\mid |\sigma |=n\}}\sum_{\vartheta\in
TranSeq(N,\sigma ,\varsigma ,\varrho )}PT(\vartheta )=\\
\sum_{(\sigma ,\varsigma ,\varrho )}\sum_{\{\vartheta\in TranSeq(N,\sigma ,\varsigma ,\varrho )\mid |\vartheta
|=n\}}PT(\vartheta )=\sum_{\{\vartheta\in TranSeq(N)\mid |\vartheta |=n\}}PT(\vartheta )=1$, i.e. $PT(\sigma ,\varsigma
,\varrho )$ defines a probability distribution.

The following (second) definition of fluid stochastic trace equivalence does not use fluid stochastic traces.

\begin{definition}
Two LFSPNs $N$ and $N'$ are {\em fluid trace equivalent}, denoted by $N\equiv_{fl}N'$, if $\forall (\sigma ,\varsigma
,\varrho )\in Act^*\times{\mathbb R}_{>0}^*\times{\mathbb R}^*$ we have

$$\sum_{\vartheta\in TranSeq(N,\sigma ,\varsigma ,\varrho )}PT(\vartheta )=
\sum_{\vartheta '\in TranSeq(N',\sigma ,\varsigma ,\varrho )}PT(\vartheta ').$$

\label{flteq2.def}
\end{definition}

Note that in
Definition \ref{flteq2.def}, for $\vartheta =t_1\cdots t_n\in TranSeq(N,\sigma ,\varsigma ,\varrho )$ with
$M_N=M_0\stackrel{t_1}{\rightarrow}M_1\stackrel{t_2}{\rightarrow}\cdots\stackrel{t_n}{\rightarrow}M_n$ and $\vartheta
'=t_1'\cdots t_n'\in TranSeq(N',\sigma ,\varsigma ,\varrho )$ with
$M_{N'}=M_0'\stackrel{t_1'}{\rightarrow}M_1'\stackrel{t_2'}{\rightarrow}\cdots\stackrel{t_n'}{\rightarrow}M_n'$, we
have $PT(\vartheta )=\prod_{i=1}^n PT(t_i,M_{i-1})=\prod_{i=1}^n SJ(M_{i-1})\Omega_N(t_i,M_{i-1})$ and $PT(\vartheta
')=\prod_{i=1}^n PT(t_i',M_{i-1}')=\\
\prod_{i=1}^n SJ(M_{i-1}')\Omega_N(t_i',M_{i-1}')$. Then the equality $SJ(M_0)\circ\cdots\circ SJ(M_n)=SJ(\vartheta
)=\varsigma =SJ(\vartheta ')=SJ(M_0')\circ\cdots\circ SJ(M_n')$ implies that $\prod_{i=1}^n SJ(M_{i-1})=\prod_{i=1}^n
SJ(M_{i-1}')$. Hence, $PT(\vartheta )=PT(\vartheta ')$ iff $\prod_{i=1}^n\Omega_N(t_i,M_{i-1})=
\prod_{i=1}^n\Omega_N(t_i',M_{i-1}')$. This alternative equality results in the following (third) definition of fluid
trace equivalence.

\begin{definition}
Two LFSPNs $N$ and $N'$ are {\em fluid trace equivalent}, denoted by $N\equiv_{fl}N'$, if $\forall (\sigma ,\varsigma
,\varrho )\in Act^*\times{\mathbb R}_{>0}^*\times{\mathbb R}^*$ we have

$$\begin{array}{c}
\sum_{\{t_1\cdots t_n\in TranSeq(N,\sigma ,\varsigma ,\varrho )\mid M_N=M_0\stackrel{t_1}{\rightarrow}M_1
\stackrel{t_2}{\rightarrow}\cdots\stackrel{t_n}{\rightarrow}M_n\}}\prod_{i=1}^n\Omega_N(t_i,M_{i-1})=\\
\sum_{\{t_1'\cdots t_n'\in TranSeq(N',\sigma ,\varsigma ,\varrho )\mid M_{N'}=M_0'\stackrel{t_1'}{\rightarrow}M_1'
\stackrel{t_2'}{\rightarrow}\cdots\stackrel{t_n'}{\rightarrow}M_n'\}}\prod_{i=1}^n\Omega_N(t_i',M_{i-1}').
\end{array}$$

\label{flteq3.def}
\end{definition}

Note that in the definition of $TranSeq(N,\sigma ,\varsigma ,\varrho )$, as well as in
Definitions \ref{flteq1.def}, \ref{flteq2.def} and \ref{flteq3.def}, for $\vartheta\in T_N^*$, we may use the {\em exit
rate sequences} $RE(\vartheta )=RE(M_0)\circ\cdots\circ RE(M_n)\in{\mathbb R}_{\geq 0}^*$ instead of average sojourn
time sequences $\varsigma =SJ(\vartheta )=SJ(M_0)\circ\cdots\circ SJ(M_n)\in{\mathbb R}_{>0}^*$, since we have $\forall
M\in DRS(N)\ SJ(M)=\frac{1}{RE(M)}$ and $\forall M\in DRS(N)\ \forall M'\in DRS(N')\ SJ(M)=SJ(M')\ \Leftrightarrow\
RE(M)=RE(M')$.

Let $N$ and $N'$ be LFSPNs such that $Pc_N=\{q\}$ and $Pc_{N'}=\{q'\}$. In this case the continuous place $q'$ of $N$
{\em corresponds} to $q$ of $N$, in other words, $q$ and $q'$ are the {\em respective} continuous places. Then for
$M\in DRS(N)$ (or for $M'\in DRS(N')$) we denote by $RP(M)$ (or by $RP(M')$) the fluid level change rate for the
continuous place $q$ (or for the corresponding one $q'$), i.e. the argument discrete marking determines for which of
the two continuous places, $q$ or $q'$, the flow rate function $RP$ is taken.

Let $N$ be an LFSPN. The {\em average potential fluid change volume} in a continuous place $q\in Pc_N$ in the discrete
marking $M\in DRS(N)$ is

$$FluidChange(q,M)=SJ(M)RP(M).$$

In order to define the probability function $PT(\sigma ,\varsigma ,\varrho )$, the transition sequences corresponding
to a particular action sequence are also selected according to the specific average sojourn times and fluid flow rates
in the discrete markings of the paths to which those transition sequences refer. One of several intuitions behind such
an additional selection is as follows. The average potential fluid change volume in a continuous place $q$ in the
discrete marking $M$ is a product of the average sojourn time and the constant (possibly zero or negative) potential
fluid flow rate in $M$. In each of the corresponding discrete markings $M$ and $M'$ of the paths to which the
corresponding transition sequences $\vartheta\in TranSeq(N,\sigma ,\varsigma ,\varrho )$ and $\vartheta '\in
TranSeq(N',\sigma ,\varsigma ,\varrho )$ refer, we shall have the same average potential fluid change volume in the
respective continuous places $q$ and $q'$, i.e. $FluidChange(q,M)=SJ(M)RP(M)=SJ(M')RP(M')=FluidChange(q',M')$. Note
that the average {\em actual} and {\em potential} fluid change volumes coincide unless the lower boundary of fluid in
some continuous place is reached, setting hereupon the actual fluid flow rate in it equal to zero till the end of the
sojourn time in the current discrete marking.

Note that our notion of fluid trace equivalence is based rather on that of Markovian trace equivalence from
\cite{WBMC06}, since there the average sojourn times in the states ``surrounding'' the actions of the corresponding
traces of the equivalent processes should {\em coincide} while in the definition of the mentioned equivalence from
\cite{BBo06,Bern07a,Bern07b,BBo08}, the shorter average sojourn time may simulate the longer one. If we would adopt
such a simulation then the smaller average potential fluid change volume
would model the bigger one, since the potential fluid flow rate remains constant while residing in a discrete marking.
Since we observe no intuition behind that modeling, we do not use it.

Let $\vartheta =t_1\cdots t_n\in TranSeq(N)$ and $M_N=M_0\stackrel{t_1}{\rightarrow}M_1\stackrel{t_2}{\rightarrow}
\cdots\stackrel{t_n}{\rightarrow}M_n$. The {\em average potential fluid change volume for the transition sequence
$\vartheta$} in a continuous place $q\in Pc_N$ is

$$FluidChange(q,\vartheta )=\sum_{i=0}^n FluidChange(q,M_i).$$

In \cite{BNL13,BeTe13}, the following two types of Markovian trace equivalence have been proposed. The {\em
state-to-state} Markovian trace equivalence requires coincidence of average sojourn times in all corresponding discrete
markings of the paths. The {\em end-to-end} Markovian trace equivalence demands that only the sums of average sojourn
times for all corresponding discrete markings of the paths should be equal. As a basis for constructing fluid trace
equivalence, we have taken the state-to-state relation, since the constant potential fluid flow rate in the discrete
markings may differ with their change (moreover, the actual fluid flow rate function may become discontinuous when the
lower fluid boundary for a continuous place is reached in some discrete marking). Therefore, while summing the
potential fluid flow rates for all discrete markings of a path, an important information is lost. The information is
needed to calculate the average potential fluid change volume for a transition sequence that refers to the path. The
mentioned value is a sum of the average potential fluid change volumes for all corresponding discrete markings of the
path. It coincides for the corresponding transition sequences $\vartheta\in TranSeq(N,\sigma ,\varsigma ,\varrho )$ and
$\vartheta '\in TranSeq(N',\sigma ,\varsigma ,\varrho )$, i.e. $FluidChange(q,\vartheta )=FluidChange(q',\vartheta ')$
for the respective continuous places $q$ and $q'$. Again, note that the average {\em actual} and {\em potential} fluid
change volumes for a transition sequence may differ, due to discontinuity of the actual fluid flow rate functions for
some discrete markings of the path to which the transition sequence refers.

Let $TranSeq(N,\sigma ,\varsigma ,\varrho )\neq\emptyset$. The {\em average potential fluid change volume for the
$(\sigma ,\varsigma ,\varrho )$-selected (finite) transition sequences} in a continuous place $q\in Pc_N$ is

$$FluidChange(q,(\sigma ,\varsigma ,\varrho ))=FluidChange(q,\vartheta )\ \forall\vartheta\in TranSeq(N,\sigma
,\varsigma ,\varrho ).$$

Then, as mentioned above, for the respective continuous places $q$ and $q'$ of the LFSPNs $N$ and $N'$, such that
$TranSeq(N,\sigma ,\varsigma ,\varrho )\neq\emptyset\neq TranSeq(N,\sigma ,\varsigma ,\varrho )$, we have
$FluidChange(q,(\sigma ,\varsigma ,\varrho ))=FluidChange(q',(\sigma ,\varsigma ,\varrho ))$.

Let $n\in{\mathbb N}$. The {\em average potential fluid change volume for the transition sequences of length $n$} in a
continuous place $q\in Pc_N$ is

$$FluidChange(q,n)=\sum_{\{\vartheta\in TranSeq(N)\mid |\vartheta |=n\}}FluidChange(q,\vartheta )PT(\vartheta ).$$

Note that we have $FluidChange(q,n)=\sum_{\{\vartheta\in TranSeq(N)\mid |\vartheta |=n\}}FluidChange(q,\vartheta
)PT(\vartheta )=\\
\sum_{\{(\sigma ,\varsigma ,\varrho )\mid TranSeq(N,\sigma ,\varsigma ,\varrho )\neq\emptyset\wedge |\sigma
|=n\}}FluidChange(q,(\sigma ,\varsigma ,\varrho ))PT(\sigma ,\varsigma ,\varrho )$. For the respective continuous
places $q$ and $q'$ of the LFSPNs $N$ and $N'$ with $N\equiv_{fl}N'$, we have $\forall n\in{\mathbb N}\
FluidChange(q,n)=FluidChange(q',n)$. Thus, fluid trace equivalence preserves average potential fluid change volume for
the transition sequences of every certain length in the respective continuous places.

\begin{figure}
\begin{center}
\includegraphics{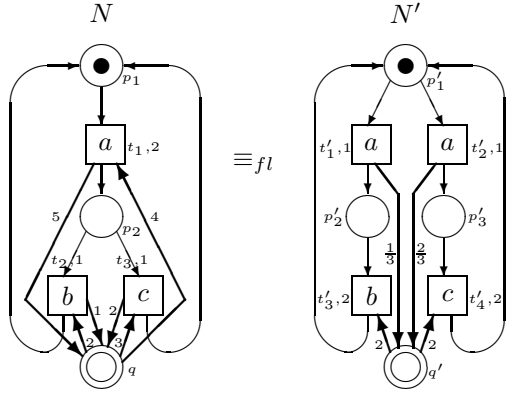}
\end{center}
\vspace{-7mm}
\caption{Fluid trace equivalent LFSPNs}
\label{fltlfspn.fig}
\end{figure}

\begin{figure}
\begin{center}
\includegraphics{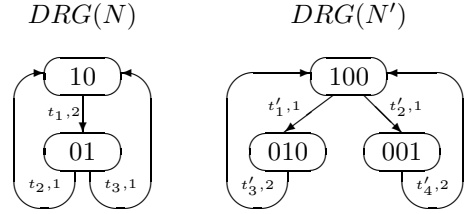}
\end{center}
\vspace{-7mm}
\caption{The discrete reachability graphs of the fluid trace equivalent LFSPNs}
\label{fltdrg.fig}
\end{figure}

\begin{figure}
\begin{center}
\includegraphics{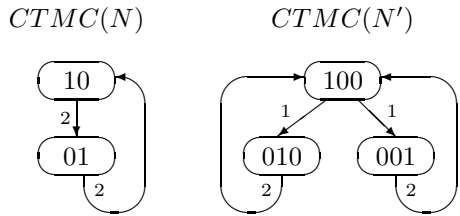}
\end{center}
\vspace{-7mm}
\caption{The underlying CTMCs of the fluid trace equivalent LFSPNs}
\label{fltctmc.fig}
\end{figure}

\begin{figure}
\begin{center}
\includegraphics{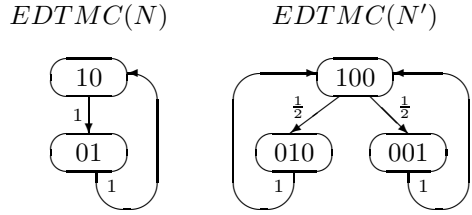}
\end{center}
\vspace{-7mm}
\caption{The EDTMCs of the fluid trace equivalent LFSPNs}
\label{fltedtmc.fig}
\end{figure}

\begin{figure}
\begin{center}
\includegraphics{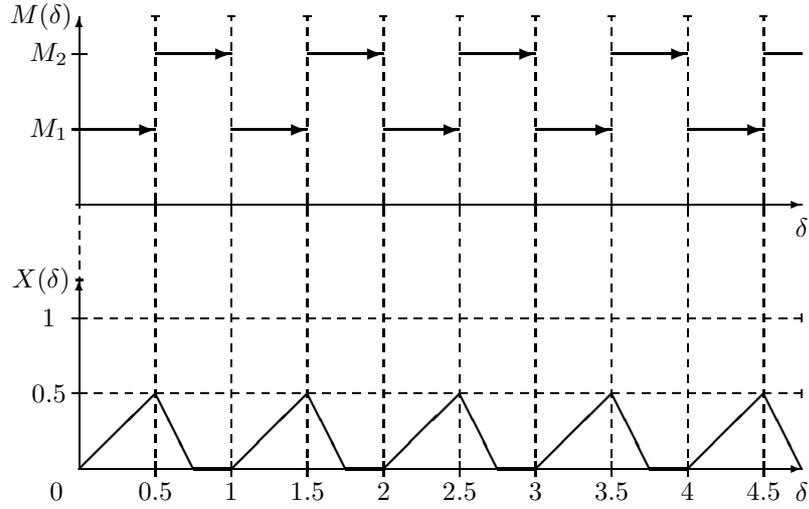}
\end{center}
\vspace{-7mm}
\caption{The ideal evolution of the actual fluid level in the first of two fluid trace equivalent LFSPNs}
\label{actflulevi.fig}
\end{figure}

\begin{figure}
\begin{center}
\includegraphics{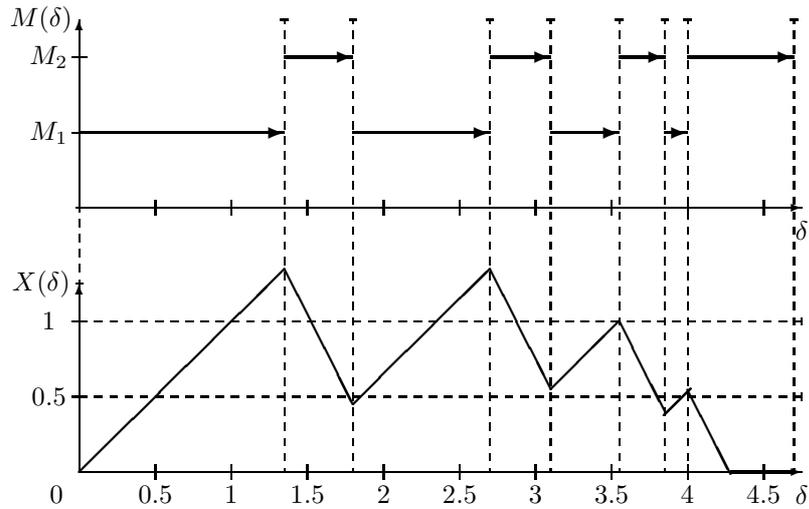}
\end{center}
\vspace{-7mm}
\caption{Possible evolution of the actual fluid level in the first of two fluid trace equivalent LFSPNs}
\label{actflulevp.fig}
\end{figure}

\begin{example}
\label{flutraeq.exm}
In Figure \ref{fltlfspn.fig}, the LFSPNs $N$ and $N'$ are presented, such that $N\equiv_{fl}N'$. We have
$DRS(N)=\{M_1,M_2\}$, where $M_1=(1,0),\ M_2=(0,1)$, and $DRS(N')=\{M_1',M_2',M_3'\}$, where $M_1'=(1,0,0),\
M_2'=(0,1,0),\ M_3'=(0,0,1)$.

In Figure \ref{fltdrg.fig}, the discrete reachability graphs $DRG(N)$ and $DRG(N')$ are depicted. In Figure
\ref{fltctmc.fig}, the underlying CTMCs $CTMC(N)$ and $CTMC(N')$ are drawn. In Figure \ref{fltedtmc.fig}, the EDTMCs
$EDTMC(N)$ and $EDTMC(N')$ are presented.

The sojourn time average and variance vectors of $N$ are

$$SJ=\left(\frac{1}{2},\frac{1}{2}\right),\ VAR=\left(\frac{1}{4},\frac{1}{4}\right).$$

The TRM ${\bf Q}$ for $CTMC(N)$, TPM ${\bf P}$ for $EDTMC(N)$ and FRM ${\bf R}$ for the SFM of $N$ are

$$\begin{array}{ccc}
{\bf Q}=\left(\begin{array}{cc}
-2 & 2\\
2 & -2
\end{array}\right),
&
{\bf P}=\left(\begin{array}{cc}
0 & 1\\
1 & 0
\end{array}\right),
&
{\bf R}=\left(\begin{array}{cc}
1 & 0\\
0 & -2
\end{array}\right).
\end{array}$$

The sojourn time average and variance vectors of $N'$ are

$$SJ'=\left(\frac{1}{2},\frac{1}{2},\frac{1}{2}\right),\ VAR'=\left(\frac{1}{4},\frac{1}{4},\frac{1}{4}\right).$$

The TRM ${\bf Q}'$ for $CTMC(N')$, TPM ${\bf P}'$ for $EDTMC(N')$ and FRM ${\bf R}'$ for the SFM of $N'$ are

$$\begin{array}{ccc}
{\bf Q}'=\left(\begin{array}{ccc}
-2 & 1 & 1\\
2 & -2 & 0\\
2 & 0 & -2
\end{array}\right),
&
{\bf P}'=\left(\begin{array}{ccc}
0 & \frac{1}{2} & \frac{1}{2}\\
1 & 0 & 0\\
1 & 0 & 0
\end{array}\right),
&
{\bf R}'=\left(\begin{array}{ccc}
1 & 0 & 0\\
0 & -2 & 0\\
0 & 0 & -2
\end{array}\right).
\end{array}$$

We have $t_1 t_2\in TranSeq\left(N,ab,\frac{1}{2}\circ\frac{1}{2}\circ\frac{1}{2},1\circ (-2)\circ 1\right)$ and $t_1
t_3\in TranSeq\left(N,ac,\frac{1}{2}\circ\frac{1}{2}\circ\frac{1}{2},1\circ (-2)\circ 1\right)$, hence,
$FluidChange(q,t_1 t_2)=FluidChange(q,t_1 t_3)=\frac{1}{2}\cdot 1+\frac{1}{2}\cdot (-2)+\frac{1}{2}\cdot 1=0$.

We have $t_1't_3'\in TranSeq\left(N',ab,\frac{1}{2}\circ\frac{1}{2}\circ\frac{1}{2},1\circ (-2)\circ 1\right)$ and
$t_2't_4'\in TranSeq\left(N',ac,\frac{1}{2}\circ\frac{1}{2}\circ\frac{1}{2},1\circ (-2)\circ 1\right)$, hence,
$FluidChange(q',t_1't_3')=FluidChange(q',t_2't_4')=\frac{1}{2}\cdot 1+\frac{1}{2}\cdot (-2)+\frac{1}{2}\cdot 1=0$.

It holds $PT(t_1 t_2)=PT(t_1 t_3)=1\cdot\frac{1}{2}=\frac{1}{2}$ and $PT(t_1't_3')=PT(t_2't_4')=\frac{1}{2}\cdot
1=\frac{1}{2}$.

We get $FluStochTraces(N)=\{\left(\left(\varepsilon ,\frac{1}{2},1\right),1\right),
\left(\left(a,\frac{1}{2}\circ\frac{1}{2},1\circ (-2)\right),1\right),
\left(\left(ab,\frac{1}{2}\circ\frac{1}{2}\circ\frac{1}{2},1\circ (-2)\circ 1\right),\frac{1}{2}\right),\\
\left(\left(ac,\frac{1}{2}\circ\frac{1}{2}\circ\frac{1}{2},1\circ (-2)\circ 1\right),\frac{1}{2}\right),\ldots\}=
FluStochTraces(N')$.

It holds $FluidChange\left(q,\left(a,\frac{1}{2}\circ\frac{1}{2},1\circ (-2)\right)\right)=
FluidChange\left(q',\left(a,\frac{1}{2}\circ\frac{1}{2},1\circ (-2)\right)\right)=\frac{1}{2}\cdot 1+\frac{1}{2}\cdot
(-2)=-\frac{1}{2}$.

We then get $FluidChange(q,1)=FluidChange(q,t_1)PT(t_1)=(-\frac{1}{2})\cdot 1=-\frac{1}{2}=
(-\frac{1}{2})\cdot\frac{1}{2}+(-\frac{1}{2})\cdot\frac{1}{2}=FluidChange(q',t_1')PT(t_1')+
FluidChange(q',t_2')PT(t_2')=FluidChange(q',1)$.

In Figure \ref{actflulevi.fig}, the ideal (since we have a stochastic process here, the actual and average sojourn
times may differ) evolution of the actual fluid level for the continuous place $q$ of the LFSPN $N$ is depicted. One
can see that $X(0.75)=0$, i.e. at the time moment $\delta =0.75$, the fluid level $X(\delta )$ reaches the zero low
boundary while $N$ resides in the discrete marking $M(\delta )=M_2$ for all $\delta\in [0.5;1)$. Then the actual fluid
flow rate function $RA(M(\delta ),X(\delta ))$ has a discontinuity at that point, where the function value is changed
instantly from $-2$ to $0$. If it would exist no lower boundary, the average potential and actual fluid change volumes
for the transition sequences of length $1$ in the continuous place $q$ would coincide and be equal to
$FluidChange(q,1)=-0.5=0.5-1=X(1)$.

In Figure \ref{actflulevp.fig}, possible evolution of the actual fluid level for the continuous place $q$ of the LFSPN
$N$ is presented, where the actual and average sojourn times in the discrete markings demonstrate substantial
differences.

\end{example}

\section{Fluid bisimulation equivalence}
\label{flubiseq.sec}

Bisimulation equivalences respect particular points of choice in the behavior of a system. To define fluid bisimulation
equivalence, we have to consider a bisimulation being an {\em equivalence} relation that partitions the states of the
{\em union} of the discrete reachability graphs $DRG(N)$ and $DRG(N')$ of the LFSPNs $N$ and $N'$. For $N$ and $N'$ to
be bisimulation equivalent the initial states $M_N$ and $M_{N'}$ of their discrete reachability graphs should be
related by a bisimulation having the following transfer property: if two states are related then in each of them the
same action can occur, leading with the identical overall rate from each of the two states to {\em the same equivalence
class} for every such action.

The definition of fluid bisimulation should be given at the level of LFSPNs, but it must use the transition rates of
the extracted CTMC. These rates cannot be easily (i.e. with a simple
expression) defined at the level of more general LFSPNs, whose discrete part is labeled GSPNs. In addition, the action
labels of immediate transitions are lost and their individual probabilities are redistributed while GSPNs are
transformed into CTSPNs. The individual probabilities of immediate transitions are ``dissolved'' in the total
transition rates between tangible states when vanishing states are eliminated from SMCs while reducing them to CTMCs.
Therefore, to make the definition of fluid bisimulation less intricate and complex, we have decided to consider only
LFSPNs with labeled CTSPNs as their discrete part. Then the underlying stochastic process of the discrete part of
LFSPNs will be that of CTSPNs, i.e. CTMCs.

The novelty of the fluid bisimulation definition with respect to that of the Markovian bisimulations from
\cite{Buc95,Hil96,BBo06,Bern07a,Bern07b,BBo08,BNL13,BeTe13} is that, for each pair of bisimilar discrete markings of
$N$ and $N'$, we require coincidence of the fluid flow rates of the {\em corresponding} (i.e. related by a {\em
correspondence bijection}) continuous places of $N$ and $N'$ in these two discrete markings. Thus, fluid bisimulation
equivalence takes into account {\em functional activity}, {\em stochastic timing} and {\em fluid flow}, like fluid
trace equivalence does.

We first propose some helpful extensions of the rate functions for the discrete marking changes and for the fluid flow
in continuous places. Let $N$ be an LFSPN and ${\cal H}\subseteq DRS(N)$. Then, for each $M\in DRS(N)$ and $a\in Act$,
we write $M\stackrel{a}{\rightarrow}_{\lambda}{\cal H}$, where $\lambda =RM_a(M,{\cal H})$ is the {\em overall rate to
move from $M$ into the set of discrete markings ${\cal H}$ by action $a$}, defined as

$$RM_a(M,{\cal H})=\sum_{\{t\mid\exists\widetilde{M}\in{\cal H}\ M\stackrel{t}{\rightarrow}\widetilde{M},\
L_N(t)=a\}}\Omega_N(t,M).$$

We write $M\stackrel{a}{\rightarrow}{\cal H}$ if $\exists\lambda\ M\stackrel{a}{\rightarrow}_{\lambda}{\cal H}$.
Further, we write $M\rightarrow_{\lambda}{\cal H}$ if $\exists a\ M\stackrel{a}{\rightarrow}{\cal H}$, where
$\lambda=RM(M,{\cal H})$ is the {\em overall rate to move from $M$ into the set of discrete markings ${\cal H}$ by
any actions}, defined as

$$RM(M,{\cal H})=\sum_{\{t\mid\exists\widetilde{M}\in{\cal H}\ M\stackrel{t}{\rightarrow}\widetilde{M}\}}
\Omega_N(t,M).$$

To construct a fluid bisimulation between LFSPNs $N$ and $N'$, we should consider the ``composite'' set of their
discrete markings $DRS(N)\cup DRS(N')$, since we have to identify the rates to come from any two equivalent discrete
markings into the same ``composite'' equivalence class (with respect to the fluid bisimulation). Note that, for $N\neq
N'$, transitions starting from the discrete markings of $DRS(N)$ (or $DRS(N')$) always lead to those from the same set,
since $DRS(N)\cap DRS(N')=\emptyset$, and this allows us to ``mix'' the sets of discrete markings in the definition of
fluid bisimulation.

Let $Pc_N=\{q\}$ and $Pc_{N'}=\{q'\}$. In this case the continuous place $q'$ of $N$ {\em corresponds} to $q$ of $N$,
according to a trivial {\em correspondence bijection} $\beta :Pc_N\rightarrow Pc_{N'}$ such that $\beta (q)=q'$. Then
for $M\in DRS(N)$ (or for $M'\in DRS(N')$) we denote by $RP(M)$ (or by $RP(M')$) the fluid level change rate for the
continuous place $q$ (or for the corresponding one $q'$),
i.e. the argument discrete marking determines for which of the two continuous places, $q$ or $q'$, the flow rate
function $RP$ is taken.

Note that if $N$ and $N'$ have more than one continuous place and there exists a {\em correspondence bijection} $\beta
:Pc_N\rightarrow Pc_{N'}$ then we should consider several flow rate functions $RP_i\ (1\leq i\leq l=|Pc_N|=|Pc_{N'}|)$
in the same manner, i.e. each $RP_i$ is used for the pair of the corresponding continuous places $q_i\in Pc_N$ and
$\beta (q_i)=q_i'\in Pc_{N'}$. In other words, we require that the vectors $(RP_1(M),\ldots ,RP_l(M))$ and
$(RP_1(M'),\ldots ,RP_l(M'))$ coincide for each pair of fluid bisimilar discrete markings $M$ and $M'$ in such a case.

\begin{definition}
Let $N$ and $N'$ be LFSPNs such that $Pc_N=\{q\},\ Pc_{N'}=\{q'\}$ and $q'$ corresponds to $q$. An {\em equivalence}
relation ${\cal R}\subseteq (DRS(N)\cup DRS(N'))^2$ is a {\em fluid bisimulation} between $N$ and $N'$, denoted by
${\cal R}:N\bis_{fl}N'$, if:
\begin{enumerate}

\item $(M_N,M_{N'})\in{\cal R}$.

\item $(M_1,M_2)\in{\cal R}\ \Rightarrow\ RP(M_1)=RP(M_2),\ \forall{\cal H}\in (DRS(N)\cup DRS(N'))/_{\cal R},\
\forall a\in Act$

$$M_1\stackrel{a}{\rightarrow}_{\lambda}{\cal H}\ \Leftrightarrow\ M_2\stackrel{a}{\rightarrow}_{\lambda}{\cal H}.$$

\end{enumerate}
Two LFSPNs $N$ and $N'$ are {\em fluid bisimulation equivalent}, denoted by $N\bis_{fl}N'$, if $\exists{\cal
R}:N\bis_{fl}N'$.
\end{definition}

Let ${\cal R}_{fl}(N,N')=\bigcup\{{\cal R}\mid{\cal R}:N\bis_{fl}N'\}$ be the {\em union of all fluid bisimulations}
between $N$ and $N'$. The following proposition proves that ${\cal R}_{fl}(N,N')$ is also an {\em equivalence} and
${\cal R}_{fl}(N,N'):N\bis_{fl}N'$.

\begin{proposition}
Let $N$ and $N'$ be LFSPNs and $N\bis_{fl}N'$. Then ${\cal R}_{fl}(N,N')$ is the largest fluid bisimulation between $N$
and $N'$.
\label{largestflbis.pro}
\end{proposition}
{\em Proof.}
Analogous to that of Proposition 8.2.1 from \cite{Hil96}, which establishes the result for strong equivalence. \hfill
$\eop$

Let $N,N'$ be LFSPNs with ${\cal R}:N\bis_{fl}N'$ and ${\cal H}\in (DRS(N)\cup DRS(N'))/_{\cal R}$. We now present a
number of remarks on the important equalities and helpful notations based on the rate functions $RM_a,\ RM,\ RP$ and
sojourn time characteristics $SJ,\ VAR$.

\noindent{\em Remark 1.} We have $\forall M_1,M_2\in{\cal H}\ \forall\widetilde{\cal H}\in (DRS(N)\cup DRS(N'))/_{\cal
R}\ \forall a\in Act\ M_1\stackrel{a}{\rightarrow}_{\lambda}\widetilde{\cal H}\ \Leftrightarrow\
M_2\stackrel{a}{\rightarrow}_{\lambda}\widetilde{\cal H}$. Since the previous equality is valid for all
$M_1,M_2\in{\cal H}$, we can rewrite it as ${\cal H}\stackrel{a}{\rightarrow}_{\lambda}\widetilde{\cal H}$, where
$\lambda =RM_a({\cal H},\widetilde{\cal H})=RM_a(M_1,\widetilde{\cal H})=RM_a(M_2,\widetilde{\cal H})=RM_a({\cal H}\cap
DRS(N),\widetilde{\cal H})=RM_a({\cal H}\cap DRS(N'),\widetilde{\cal H})$. Then we write ${\cal
H}\stackrel{a}{\rightarrow}\widetilde{\cal H}$ if $\exists{\lambda}\ {\cal H}\stackrel{a}{\rightarrow}_{\lambda}
\widetilde{\cal H}$ and ${\cal H}\rightarrow\widetilde{\cal H}$ if $\exists a\ {\cal H}\stackrel{a}{\rightarrow}
\widetilde{\cal H}$.

Since the transitions from the discrete markings of $DRS(N)$ always lead to those from the same set, we have
$\forall M\in DRS(N)\ \forall a\in Act\ RM_a(M,\widetilde{\cal H})=RM_a(M,\widetilde{\cal H}\cap DRS(N))$. Hence,
$\forall M\in{\cal H}\cap DRS(N)\ \forall a\in Act\ RM_a({\cal H},\widetilde{\cal H})=RM_a(M,\widetilde{\cal
H})=RM_a(M,\widetilde{\cal H}\cap DRS(N))=RM_a({\cal H}\cap DRS(N),\widetilde{\cal H}\cap DRS(N))$. The same is
true for $DRS(N')$. Thus, $\forall\widetilde{\cal H}\in (DRS(N)\cup DRS(N'))/_{\cal R}$

$$RM_a({\cal H}\cap DRS(N),\widetilde{\cal H}\cap DRS(N))=RM_a({\cal H},\widetilde{\cal H})=
RM_a({\cal H}\cap DRS(N'),\widetilde{\cal H}\cap DRS(N')).$$

\noindent{\em Remark 2.} We have $\forall M_1,M_2\in{\cal H}\ \forall\widetilde{\cal H}\in (DRS(N)\cup DRS(N'))/_{\cal
R}\ RM(M_1,\widetilde{\cal H})=\\
\sum_{\{t\mid\exists\widetilde{M}_1\in\widetilde{\cal H}\
M_1\stackrel{t}{\rightarrow}\widetilde{M}_1\}}\Omega_N(t,M_1)=
\sum_{a\in Act}\sum_{\{t\mid\exists\widetilde{M}_1\in\widetilde{\cal H}\ M_1\stackrel{t}{\rightarrow}\widetilde{M}_1,\
L_N(t)=a\}}\Omega_N(t,M_1)=\sum_{a\in Act}RM_a(M_1,\widetilde{\cal H})=\\
\sum_{a\in Act}RM_a(M_2,\widetilde{\cal H})=\sum_{a\in Act}\sum_{\{t\mid\exists\widetilde{M}_2\in\widetilde{\cal H}\
M_2\stackrel{t}{\rightarrow}\widetilde{M}_2,\ L_N(t)=a\}}\Omega_N(t,M_2)=
\sum_{\{t\mid\exists\widetilde{M}_2\in\widetilde{\cal H}\
M_2\stackrel{t}{\rightarrow}\widetilde{M}_2\}}\Omega_N(t,M_2)=\\
RM(M_2,\widetilde{\cal H})$. Since the previous equality is valid for all $M_1,M_2\in{\cal H}$, we can denote $RM({\cal
H},\widetilde{\cal H})=RM(M_1,\widetilde{\cal H})= RM(M_2,\widetilde{\cal H})$. Then we write ${\cal
H}\rightarrow_{\lambda}\widetilde{\cal H}$, where $\lambda =RM({\cal H},\widetilde{\cal H})=RM(M_1,\widetilde{\cal
H})=RM(M_2,\widetilde{\cal H})$.

Since the transitions from the discrete markings of $DRS(N)$ always lead to those from the same set, we have
$\forall M\in DRS(N)\ RM(M,\widetilde{\cal H})=RM(M,\widetilde{\cal H}\cap DRS(N))$. Hence, $\forall M\in{\cal
H}\cap DRS(N)\ RM({\cal H},\widetilde{\cal H})=RM(M,\widetilde{\cal H})=RM(M,\widetilde{\cal H}\cap
DRS(N))=RM({\cal H}\cap DRS(N),\widetilde{\cal H}\cap DRS(N))$. The same is true for $DRS(N')$. Thus,
$\forall\widetilde{\cal H}\in (DRS(N)\cup DRS(N'))/_{\cal R}$

$$RM({\cal H}\cap DRS(N),\widetilde{\cal H}\cap DRS(N))=RM({\cal H},\widetilde{\cal H})=
RM({\cal H}\cap DRS(N'),\widetilde{\cal H}\cap DRS(N')).$$

\noindent{\em Remark 3.} We have $\forall M_1,M_2\in{\cal H}\ RP(M_1)=RP(M_2)$. Since the previous equality is valid
for all $M_1,M_2\in{\cal H}$, we can denote $RP({\cal H})=RP(M_1)=RP(M_2)$.

Since any argument discrete marking $M\in DRS(N)\cup DRS(N')$ completely determines for which continuous place the
flow rate function $RP(M)$ is taken (either for $q$ if $M\in DRS(N)$ or for $q'$ if $M\in DRS(N')$), we have
$\forall M\in{\cal H}\cap DRS(N)\ RP({\cal H})=RP(M)=RP({\cal H}\cap DRS(N))$. The same is true for $DRS(N')$.
Thus,

$$RP({\cal H}\cap DRS(N))=RP({\cal H})=RP({\cal H}\cap DRS(N')).$$

\noindent{\em Remark 4.} We have $\forall M_1,M_2\in{\cal H}\ SJ(M_1)=
\frac{1}{\sum_{t\in Ena(M_1)}\Omega_N(t,M_1)}=\\
\frac{1}{\sum_{\widetilde{\cal H}\in (DRS(N)\cup DRS(N'))/_{\cal R}}\sum_{\{t\mid\exists\widetilde{M}_1\in
\widetilde{\cal H}\ M_1\stackrel{t}{\rightarrow}\widetilde{M}_1\}}\Omega_N(t,M_1)}=
\frac{1}{\sum_{\widetilde{\cal H}\in (DRS(N)\cup DRS(N'))/_{\cal R}}RM(M_1,\widetilde{\cal H})}=\\
\frac{1}{\sum_{\widetilde{\cal H}\in (DRS(N)\cup DRS(N'))/_{\cal R}}RM({\cal H},\widetilde{\cal H})}=
\frac{1}{\sum_{\widetilde{\cal H}\in (DRS(N)\cup DRS(N'))/_{\cal R}}RM(M_2,\widetilde{\cal H})}=\\
\frac{1}{\sum_{\widetilde{\cal H}\in (DRS(N)\cup DRS(N'))/_{\cal R}}\sum_{\{t\mid\exists\widetilde{M}_2\in
\widetilde{\cal H}\ M_2\stackrel{t}{\rightarrow}\widetilde{M}_2\}}\Omega_N(t,M_2)}=
\frac{1}{\sum_{t\in Ena(M_2)}\Omega_N(t,M_2)}=SJ(M_2)$.\\
Since the previous equality is valid for all $M_1,M_2\in{\cal H}$, we can denote $SJ_{\cal R}({\cal
H})=SJ(M_1)=SJ(M_2)$.

Since any argument discrete marking $M\in DRS(N)\cup DRS(N')$ completely determines, for which LFSPN the average
sojourn time function $SJ(M)$ is considered (either for $N$ if $M\in DRS(N)$, or for $N'$ if $M\in DRS(N')$), we
have $\forall M\in{\cal H}\cap DRS(N)\ SJ({\cal H})=SJ(M)=SJ({\cal H}\cap DRS(N))$. The same is true for $DRS(N')$.
Thus,

$$SJ({\cal H}\cap DRS(N))=SJ({\cal H})=SJ({\cal H}\cap DRS(N')).$$

\noindent{\em Remark 5.} We have $\forall M_1,M_2\in{\cal H}\ VAR(M_1)=
\frac{1}{(\sum_{t\in Ena(M_1)}\Omega_N(t,M_1))^2}=\\
\frac{1}{(\sum_{\widetilde{\cal H}\in (DRS(N)\cup DRS(N'))/_{\cal R}}\sum_{\{t\mid\exists\widetilde{M}_1\in
\widetilde{\cal H}\ M_1\stackrel{t}{\rightarrow}\widetilde{M}_1\}}\Omega_N(t,M_1))^2}=
\frac{1}{(\sum_{\widetilde{\cal H}\in (DRS(N)\cup DRS(N'))/_{\cal R}}RM(M_1,\widetilde{\cal H}))^2}=\\
\frac{1}{(\sum_{\widetilde{\cal H}\in (DRS(N)\cup DRS(N'))/_{\cal R}}RM({\cal H},\widetilde{\cal H}))^2}=
\frac{1}{(\sum_{\widetilde{\cal H}\in (DRS(N)\cup DRS(N'))/_{\cal R}}RM(M_2,\widetilde{\cal H}))^2}=\\
\frac{1}{(\sum_{\widetilde{\cal H}\in (DRS(N)\cup DRS(N'))/_{\cal R}}\sum_{\{t\mid\exists\widetilde{M}_2\in
\widetilde{\cal H}\ M_2\stackrel{t}{\rightarrow}\widetilde{M}_2\}}\Omega_N(t,M_2))^2}=
\frac{1}{(\sum_{t\in Ena(M_2)}\Omega_N(t,M_2))^2}=VAR(M_2)$.\\
Since the previous equality is valid for all $M_1,M_2\in{\cal H}$, we can denote $VAR_{\cal R}({\cal
H})=VAR(M_1)=VAR(M_2)$.

Since any argument discrete marking $M\in DRS(N)\cup DRS(N')$ completely determines, for which LFSPN the sojourn
time variance function $VAR(M)$ is considered (either for $N$ if $M\in DRS(N)$, or for $N'$ if $M\in DRS(N')$), we
have $\forall M\in{\cal H}\cap DRS(N)\ VAR({\cal H})=VAR(M)=VAR({\cal H}\cap DRS(N))$. The same is true for $DRS(N')$.
Thus,

$$VAR({\cal H}\cap DRS(N))=VAR({\cal H})=VAR({\cal H}\cap DRS(N')).$$

\begin{figure}
\begin{center}
\includegraphics{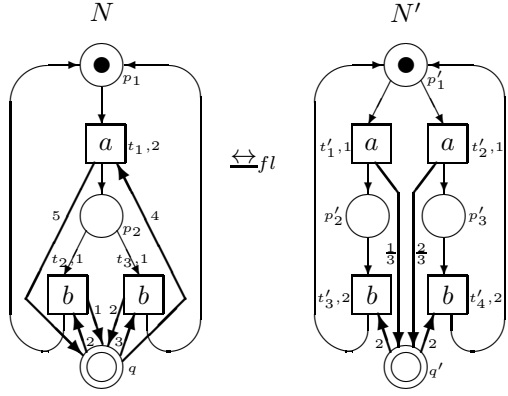}
\end{center}
\vspace{-7mm}
\caption{Fluid bisimulation equivalent LFSPNs}
\label{flblfspn.fig}
\end{figure}

\begin{example}
\label{flubiseq.exm}
In Figure \ref{flblfspn.fig}, the LFSPNs $N$ and $N'$ are presented, such that $N\bis_{fl}N'$. The only difference
between the {\em respective} LFSPNs in Figure \ref{fltlfspn.fig} and those in Figure \ref{flblfspn.fig} is that the
transitions $t_3$ and $t_4'$ are labeled with action $c$ in the former, instead of action $b$ in the latter.

Therefore, the following notions coincide for the {\em respective} LFSPNs in Figure \ref{fltlfspn.fig} and those in
Figure \ref{flblfspn.fig}: the discrete reachability sets $DRS(N)$ and $DRS(N')$, the discrete reachability graphs
$DRG(N)$ and $DRG(N')$, the underlying CTMCs $CTMC(N)$ and $CTMC(N')$, the sojourn time average vectors $SJ$ and $SJ'$
of $N$ and $N'$, the variance vectors $VAR$ and $VAR'$ of $N$ and $N'$, the TRMs ${\bf Q}$ and ${\bf Q}'$ for $CTMC(N)$
and $CTMC(N')$, the TPMs ${\bf P}$ and ${\bf P}'$ for $EDTMC(N)$ and $EDTMC(N')$, the FRMs ${\bf R}$ and ${\bf R}'$ for
the SFMs of $N$ and $N'$.

We have $DRS(N)/_{{\cal R}_{fl}(N)}=\{{\cal K}_1,{\cal K}_2\}$, where ${\cal K}_1=\{M_1\},\ {\cal K}_2=\{M_2\}$, and
$DRS(N')/_{{\cal R}_{fl}(N')}=\{{\cal K}_1',{\cal K}_2'\}$, where ${\cal K}_1'=\{M_1'\},\ {\cal K}_2'=\{M_2',M_3'\}$.

%
%
%
%
%
%
%
%
%
\end{example}

We now intend to compare the introduced fluid equivalences to discover their interrelations. The following proposition
demonstrates that fluid bisimulation equivalence
implies fluid trace one.

\begin{proposition}
For LFSPNs $N$ and $N'$ the following holds:

$$N\bis_{fl}N'\ \Rightarrow\ N\equiv_{fl}N'.$$

\label{intfltbe.pro}
\end{proposition}
{\em Proof.} Let ${\cal R}:N\bis_{fl}N',\ {\cal H}\in (DRS(N)\cup DRS(N'))/_{\cal R}$ and $M_1,M_2\in{\cal H}$. We have
$RP(M_1)=RP(M_2)$ and $\forall\widetilde{\cal H}\in (DRS(N)\cup DRS(N'))/_{\cal R}\ \forall a\in Act\
M_1\stackrel{a}{\rightarrow}_{\lambda}\widetilde{\cal H}\ \Leftrightarrow\
M_2\stackrel{a}{\rightarrow}_{\lambda}\widetilde{\cal H}$.
Note that transitions from the discrete markings of $DRS(N)$ always lead to those from the same set, hence, $\forall
M\in DRS(N)\ RM_a(M,\widetilde{\cal H})=RM_a(M,\widetilde{\cal H}\cap DRS(N))$. The same is true for $DRS(N')$.

By Remark 1
from Section \ref{flubiseq.sec}, we can write ${\cal H}\stackrel{a}{\rightarrow}_{\lambda}\widetilde{\cal H}$ and
denote $\lambda =RM_a(M_1,\widetilde{\cal H})=RM_a(M_2,\widetilde{\cal H})=RM_a({\cal H},\widetilde{\cal H})=RM_a({\cal
H}\cap DRS(N),\widetilde{\cal H}\cap DRS(N))=RM_a({\cal H}\cap DRS(N'),\widetilde{\cal H}\cap DRS(N'))$.

Further, by Remark 4
from Section \ref{flubiseq.sec}, we can denote
$SJ(M_1)=SJ(M_2)=SJ({\cal H})=SJ({\cal H}\cap DRS(N))=SJ({\cal H}\cap DRS(N'))$.

At last, by Remark 3
from Section \ref{flubiseq.sec}, we can denote
$RP(M_1)=RP(M_2)=RP({\cal H})=RP({\cal H}\cap DRS(N))=RP({\cal H}\cap DRS(N'))$.

%
Let $TranSeq(N,\sigma ,\varsigma ,\varrho )\neq\emptyset$ and $\sigma =a_1\cdots a_n\in Act^*,\ \varsigma =s_0\circ
\cdots\circ s_n\in{\mathbb R}_{>0}^*,\ \varrho =r_0\circ\cdots\circ r_n\in{\mathbb R}^*$.
Taking into account the notes above and ${\cal R}:N\bis_{fl}N'$, we have $SJ(M_N)=SJ(M_{N'})=s_0,\
RP(M_N)=RP(M_{N'})=r_0$ and for all ${\cal H}_1,\ldots ,{\cal H}_n\in (DRS(N)\cup DRS(N'))/_{\cal R}$, such that
$SJ({\cal H}_i)=s_i,\ RP({\cal H}_i)=r_i\ (1\leq i\leq n)$, it holds $M_N\stackrel{a_1}{\rightarrow}_{\lambda_1}{\cal
H}_1\stackrel{a_2}{\rightarrow}_{\lambda_2}\cdots \stackrel{a_n}{\rightarrow}_{\lambda_n}{\cal H}_n\ \Leftrightarrow\
M_{N'}\stackrel{a_1}{\rightarrow}_{\lambda_1}{\cal H}_1\stackrel{a_2}{\rightarrow}_{\lambda_2}\cdots
\stackrel{a_n}{\rightarrow}_{\lambda_n}{\cal H}_n$.
Then we have $TranSeq(N',\sigma ,\varsigma ,\varrho )\neq\emptyset$. Thus, $TranSeq(N,\sigma ,\varsigma ,\varrho
)\neq\emptyset$ implies $TranSeq(N',\sigma ,\varsigma ,\varrho )\neq\emptyset$.

We now intend to prove that the sum of the transition rates products for all the paths starting in $M_N=M_0$ and going
through the discrete markings from ${\cal H}_1,\ldots ,{\cal H}_n$
is equal to the product of $\lambda_1,\ldots ,\lambda_n$, which is essentially the transition rates product for the
``composite'' path starting in ${\cal H}_0=[M_0]_{\cal R}$ and going through the equivalence classes ${\cal H}_1,\ldots
,{\cal H}_n$ in $DRG(N)$:

$$\sum_{\{t_1,\ldots ,t_n\mid M_N=M_0\stackrel{t_1}{\rightarrow}\cdots\stackrel{t_n}{\rightarrow}M_n,\ L_N(t_i)=a_i,\
M_i\in{\cal H}_i\ (1\leq i\leq n)\}}\prod_{i=1}^n\Omega_N(t_i,M_{i-1})=\prod_{i=1}^n
RM_{a_i}({\cal H}_{i-1},{\cal H}_i).$$

We prove this equality by induction on the ``composite'' path length $n$.
\begin{itemize}

\item $n=1$

$\sum_{\{t_1\mid M_N=M_0\stackrel{t_1}{\rightarrow}M_1,\ L_N(t_1)=a_1,\ M_1\in{\cal H}_1\}}\Omega_N(t_1,M_0)=
RM_{a_1}(M_0,{\cal H}_1)=RM_{a_1}({\cal H}_0,{\cal H}_1)$.

\item $n\rightarrow n+1$

$\sum_{\{t_1,\ldots ,t_n,t_{n+1}\mid M_N=M_0\stackrel{t_1}{\rightarrow}\cdots
\stackrel{t_n}{\rightarrow}M_n\stackrel{t_{n+1}}{\rightarrow}M_{n+1},\ L_N(t_i)=a_i,\
M_i\in{\cal H}_i\ (1\leq i\leq n+1)\}}\prod_{i=1}^{n+1}\Omega_N(t_i,M_{i-1})=\\
\sum_{\{t_1,\ldots ,t_n\mid M_N=M_0\stackrel{t_1}{\rightarrow}\cdots\stackrel{t_n}{\rightarrow}M_n,\ L_N(t_i)=a_i,\
M_i\in{\cal H}_i\ (1\leq i\leq n)\}}\\
\sum_{\{t_{n+1}\mid M_n\stackrel{t_{n+1}}{\rightarrow}M_{n+1},\ L_N(t_{n+1})=a_{n+1},\ M_n\in{\cal
H}_n,\ M_{n+1}\in{\cal H}_{n+1}\}}\prod_{i=1}^n\Omega_N(t_i,M_{i-1})\Omega_N(t_{n+1},M_n)=\\
\sum_{\{t_1,\ldots ,t_n\mid M_N=M_0\stackrel{t_1}{\rightarrow}\cdots\stackrel{t_n}{\rightarrow}M_n,\ L_N(t_i)=a_i,\
M_i\in{\cal H}_i\ (1\leq i\leq n)\}}\\
\left[\prod_{i=1}^n\Omega_N(t_i,M_{i-1})\sum_{\{t_{n+1}\mid M_n\stackrel{t_{n+1}}{\rightarrow}M_{n+1},\
L_N(t_{n+1})=a_{n+1},\ M_n\in{\cal H}_n,\ M_{n+1}\in{\cal H}_{n+1}\}}\Omega_N(t_{n+1},M_n)\right]=\\
\sum_{\{t_1,\ldots ,t_n\mid M_N=M_0\stackrel{t_1}{\rightarrow}\cdots
\stackrel{t_n}{\rightarrow}M_n,\ L_N(t_i)=a_i,\ M_i\in{\cal H}_i\ (1\leq i\leq n)\}}\prod_{i=1}^n
\Omega_N(t_i,M_{i-1})RM_{a_{n+1}}(M_n,{\cal H}_{n+1})=\\
\sum_{\{t_1,\ldots ,t_n\mid M_N=M_0\stackrel{t_1}{\rightarrow}\cdots
\stackrel{t_n}{\rightarrow}M_n,\ L_N(t_i)=a_i,\ M_i\in{\cal H}_i\ (1\leq i\leq n)\}}\prod_{i=1}^n
\Omega_N(t_i,M_{i-1})RM_{a_{n+1}}({\cal H}_n,{\cal H}_{n+1})=\\
RM_{a_{n+1}}({\cal H}_n,{\cal H}_{n+1})\sum_{\{t_1,\ldots ,t_n\mid
M_N=M_0\stackrel{t_1}{\rightarrow}\cdots\stackrel{t_n}{\rightarrow}M_n,\ L_N(t_i)=a_i,\
M_i\in{\cal H}_i\ (1\leq i\leq n)\}}\prod_{i=1}^n\Omega_N(t_i,M_{i-1})=\\
RM_{a_{n+1}}({\cal H}_n,{\cal H}_{n+1})\prod_{i=1}^n RM_{a_i}({\cal H}_{i-1},{\cal H}_i)=\prod_{i=1}^{n+1}
RM_{a_i}({\cal H}_{i-1},{\cal H}_i)$.

\end{itemize}

Note that the equality that we have just proved can also be applied to $N'$.

One can see that the summation over {\em all $(\sigma ,\varsigma ,\varrho )$-selected transition sequences} is the same
as the summation over {\em all accordingly selected equivalence classes}:
$\sum_{t_1\cdots t_n\in TranSeq(N,\sigma ,\varsigma ,\varrho )}\prod_{i=1}^n\Omega_N(t_i,M_{i-1})=\\
\sum_{\{t_1,\ldots ,t_n\mid M_N=M_0\stackrel{t_1}{\rightarrow}\cdots\stackrel{t_n}{\rightarrow}M_n,\ L_N(t_i)=a_i,\
SJ(M_i)=s_i,\ RP(M_j)=r_i\ (1\leq i\leq n)\}}\prod_{i=1}^n\Omega_N(t_i,M_{i-1})=\\
\sum_{\{{\cal H}_1,\ldots ,{\cal H}_n\mid SJ({\cal H}_i)=s_i, RP({\cal H}_i)=r_i\ (1\leq i\leq n)\}}\!\!
\sum_{\{t_1,\ldots ,t_n\mid M_N=M_0\stackrel{t_1}{\rightarrow}\cdots\stackrel{t_n}{\rightarrow}M_n,\ L_N(t_i)=a_i,
M_i\in{\cal H}_i\ (1\leq i\leq n)\}}\!\!\prod_{i=1}^n\Omega_N(t_i,M_{i-1})\!\!=\\
\sum_{\{{\cal H}_1,\ldots ,{\cal H}_n\mid SJ({\cal H}_i)=s_i,\ RP({\cal H}_i)=r_i\ (1\leq i\leq n)\}}\prod_{i=1}^n
RM_{a_i}({\cal H}_{i-1},{\cal H}_i)=\\
\sum_{\{{\cal H}_1,\ldots ,{\cal H}_n\mid SJ({\cal H}_i)=s_i, RP({\cal H}_i)=r_i\ (1\leq i\leq n)\}}\!\!
\sum_{\{t_1',\ldots ,t_n'\mid M_{N'}=M_0'\stackrel{t_1'}{\rightarrow}\cdots\stackrel{t_n'}{\rightarrow}M_n',\
L_N(t_i')=a_i, M_i'\in{\cal H}_i\ (1\leq i\leq n)\}}\!\!\!\!\prod_{i=1}^n\Omega_{N'}(t_i',M_{i-1}')\!\!=\\
\sum_{\{t_1',\ldots ,t_n'\mid M_{N'}=M_0'\stackrel{t_1'}{\rightarrow}\cdots\stackrel{t_n'}{\rightarrow}M_n',\
L_{N'}(t_i')=a_i,\ SJ(M_i')=s_i,\ RP(M_j')=r_i\ (1\leq i\leq n)\}}\prod_{i=1}^n\Omega_{N'}(t_i',M_{i-1}')=\\
\sum_{t_1'\cdots t_n'\in TranSeq(N',\sigma ,\varsigma ,\varrho )}\prod_{i=1}^n\Omega_N(t_i',M_{i-1}')$. By the remark
before
Definition \ref{flteq3.def}, the probabilities to execute $(\sigma ,\varsigma ,\varrho )$-selected transition sequences
in $N$ and $N'$ coincide.

We conclude that for all triples $(\sigma ,\varsigma ,\varrho )\in Act^*\times{\mathbb R}_{>0}^*\times{\mathbb R}^*$,
it holds that $TranSeq(N,\sigma ,\varsigma ,\varrho )\neq\emptyset$ implies $TranSeq(N',\sigma ,\varsigma ,\varrho
)\neq\emptyset$ and the execution probabilities of $(\sigma ,\varsigma ,\varrho )$ in $N$
and $N'$ are equal. The reverse implication is proved by symmetry of fluid bisimulation. \hfill $\eop$

%
%
\begin{figure}
\begin{center}
\includegraphics{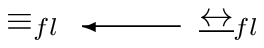}
\end{center}
\vspace{-7mm}
\caption{Interrelations of fluid equivalences}
\label{intfleq.fig}
\end{figure}

The following theorem compares discriminating power of the introduced fluid equivalences.

\begin{theorem}
For LFSPNs $N$ and $N'$ the following {\em strict} implication holds that is also depicted in Figure \ref{intfleq.fig}:

$$N\bis_{fl}N'\ \Rightarrow\ N\equiv_{fl}N'.$$

\label{intfleq.the}
\end{theorem}
{\em Proof.} Let us check the validity of the implication.
\begin{itemize}

\item The implication $\bis_{fl}\rightarrow\equiv_{fl}$ is valid by Proposition \ref{intfltbe.pro}.

\end{itemize}

Let us see that the implication is strict, i.e. the reverse one does not work, by the following counterexample.
\begin{itemize}

\item In Figure \ref{fltlfspn.fig}, $N\equiv_{fl}N'$, but $N\nbis_{fl}N'$, since only in the LFSPN $N'$ an action $a$
can be executed so (by firing the transition $t_2'$) that no action $b$ can occur afterwards. \hfill $\eop$

\end{itemize}

\section{Reduction of the behaviour}
\label{reduction.sec}

Fluid bisimulation equivalence can be used to reduce the discrete reachability graphs and underlying CTMCs of LFSPNs.
Reductions of graph-based models, like transition systems (whose instances are reachability graphs and CTMCs), result
in those with less states (the graph nodes). The goal of the reduction is to decrease the number of states in the
semantic representation of the modeled system while preserving its important qualitative and quantitative properties.
Thus, the reduction allows one to simplify the behavioural and performance analysis of systems.

An {\em autobisimulation} is a bisimulation between an LFSPN and itself. Let $N$ be an LFSPN with ${\cal
R}:N\bis_{fl}N$ and ${\cal K}\in DRS(N)/_{\cal R}$.

%
Then Remarks 2, 4 and 5
from Section \ref{flubiseq.sec} allow us to present the following definitions.

The {\em average sojourn time in the equivalence class (with respect to ${\cal R}$) of discrete markings ${\cal K}$} is

$$SJ_{\cal R}({\cal K})=\frac{1}{\sum_{\widetilde{\cal K}\in DRS(N)/_{\cal R}}RM({\cal K},\widetilde{\cal K})}=
SJ(M)\ \forall M\in{\cal K}.$$

The {\em average sojourn time vector for the equivalence classes (with respect to ${\cal R}$) of discrete markings} of
$N$, denoted by $SJ_{\cal R}$, has the elements $SJ_{\cal R}({\cal K}),\ {\cal K}\in DRS(N)/_{\cal R}$.

The {\em sojourn time variance in the equivalence class (with respect to ${\cal R}$) of discrete markings ${\cal K}$} is

$$VAR_{\cal R}({\cal K})=\frac{1}{\left(\sum_{\widetilde{\cal K}\in DRS(N)/_{\cal R}}RM({\cal K},
\widetilde{\cal K})\right)^2}=VAR(M)\ \forall M\in{\cal K}.$$

The {\em sojourn time variance vector for the equivalence classes (with respect to ${\cal R}$) of discrete markings} of
$N$, denoted by $VAR_{\cal R}$, has the elements $VAR_{\cal R}({\cal K}),\ {\cal K}\in DRS(N)/_{\cal R}$.

Let ${\cal R}_{fl}(N)=\bigcup\{{\cal R}\mid{\cal R}:N\bis_{fl}N\}$ be the {\em union of all fluid autobisimulations} on
$N$. By Proposition \ref{largestflbis.pro}, ${\cal R}_{fl}(N)$ is the largest fluid autobisimulation on $N$. Based on
the equivalence classes with respect to ${\cal R}_{fl}(N)$, the quotient (by $\bis_{fl}$) discrete reachability graphs
and quotient (by $\bis_{fl}$) underlying CTMCs of LFSPNs can be defined. The mentioned equivalence classes become the
quotient states. The average and variance for the sojourn time in a quotient state are those in the corresponding
equivalence class, respectively. Every quotient transition between two such composite states represents all transitions
(having the same action label in case of the discrete reachability graph quotient) from the first state to the second
one.

\begin{definition}
Let $N$ be an LFSPN. The {\em quotient (by $\bis_{fl}$) discrete reachability graph} of $N$ is a labeled transition
system $DRG_{\bis_{fl}}(N)=(S_{\bis_{fl}},{\cal L}_{\bis_{fl}},{\cal T}_{\bis_{fl}},s_{\bis_{fl}})$, where
\begin{itemize}

\item $S_{\bis_{fl}}=DRS(N)/_{{\cal R}_{fl}(N)}$;

\item ${\cal L}_{\bis_{fl}}=Act\times{\mathbb R}_{>0}$;

\item ${\cal T}_{\bis_{fl}}=\{({\cal K},(a,RM_a({\cal K},\widetilde{\cal K})),\widetilde{\cal K})\mid{\cal K},
\widetilde{\cal K}\in DRS(N)/_{{\cal R}_{fl}(N)},\ {\cal K}\stackrel{a}{\rightarrow}\widetilde{\cal K}\}$;

\item $s_{\bis_{fl}}=[M_N]_{{\cal R}_{fl}(N)}$.

\end{itemize}
\end{definition}

The transition $({\cal K},(a,\lambda ),\widetilde{\cal K})\in{\cal T}_{\bis_{fl}}$ will be written as ${\cal
K}\stackrel{a}{\rightarrow}_\lambda\widetilde{\cal K}$.

Let $\simeq$ denote isomorphism between the quotient discrete reachability graphs that binds their initial states.

The {\em quotient (by $\bis_{fl}$) average sojourn time vector} of $N$ is defined as $SJ_{\bis_{fl}}=SJ_{{\cal
R}_{fl}(N)}$. The {\em quotient (by $\bis_{fl}$) sojourn time variance vector} of $N$ is defined as
$VAR_{\bis_{fl}}=VAR_{{\cal R}_{fl}(N)}$.

\begin{definition}
Let $N$ be an LFSPN. The {\em quotient (by $\bis_{fl}$) underlying CTMC} of $N$, denoted by\\
$CTMC_{\bis_{fl}}(N)$, has the state space $DRS(N)/_{{\cal R}_{fl}(N)}$, the initial state $[M_N]_{{\cal R}_{fl}(N)}$
and the transitions ${\cal K}\rightarrow_{\lambda}\widetilde{\cal K}$ if ${\cal K}\rightarrow\widetilde{\cal K}$, where
$\lambda =RM({\cal K},\widetilde{\cal K})$.
\end{definition}

The steady-state PMF $\varphi_{\bis_{fl}}$ for $CTMC_{\bis_{fl}}(N)$ is defined like the corresponding notion $\varphi$
for $CTMC(N)$.

The quotients of both discrete reachability graphs and underlying CTMCs are the minimal reductions of the mentioned
objects modulo fluid bisimulation. The quotients can be used to simplify analysis of system properties which are
preserved by $\bis_{fl}$, since less states should be examined for it. Such a reduction method resembles that from
\cite{AS92} based on place bisimulation equivalence for Petri nets, excepting that the former method merges states,
while the latter one merges places.

Let $N$ be an LFSPN. We shall now demonstrate how to construct the quotients (by $\bis_{fl}$) of the TRM for $CTMC(N)$,
FRM for the associated SFM of $N$, average sojourn time vector and sojourn time variance vector of $N$, using special
collector and distributor matrices. The quotient TRMs and FRMs will be later applied to describe the quotient
associated SFMs of LFSPNs. Let $DRS(N)=\{M_1,\ldots ,M_n\}$ and $DRS(N)/_{{\cal R}_{fl}(N)}=\{{\cal K}_1,\ldots ,{\cal
K}_l\}$.

The elements $({\cal Q}_{\bis_{fl}})_{rs}\ (1\leq r,s\leq l)$ of the
TRM ${\bf Q}_{\bis_{fl}}$ for $CTMC_{\bis_{fl}}(N)$ are defined as

$$({\cal Q}_{\bis_{fl}})_{rs}=\left\{
\begin{array}{ll}
RM({\cal K}_r,{\cal K}_s), & r\neq s;\\
-\sum_{\{k\mid 1\leq k\leq l,\ k\neq r\}}RM({\cal K}_r,{\cal K}_k), & r=s.
\end{array}
\right.$$

Like it has been done for strong performance bisimulation on labeled CTSPNs in \cite{Buc95}, the $l\times l$ TRM ${\bf
Q}_{\bis_{fl}}$ for $CTMC_{\bis_{fl}}(N)$ can be constructed from the $n\times n$ TRM ${\bf Q}$ for $CTMC(N)$ using the
$n\times l$ {\em collector matrix} ${\bf V}$ for the largest fluid autobisimulation ${\cal R}_{fl}(N)$ on $N$ and the
$l\times n$ {\em distributor matrix} ${\bf W}$ for ${\bf V}$. Then ${\bf W}$ should be a non-negative matrix (i.e. all
its elements must be non-negative) with the elements of each its row summed to one, such that ${\bf W}{\bf V}={\bf I}$,
where ${\bf I}$ is the identity matrix of order $l$, i.e. ${\bf W}$ is a {\em left-inverse matrix} for ${\bf V}$. It is
known that for each collector matrix there is at least one distributor matrix, in particular, the matrix obtained by
transposing ${\bf V}$ and subsequent normalizing its rows, to guarantee that the elements of each row of the transposed
matrix are summed to one. We now present the formal definitions.

The elements ${\cal V}_{ir}\ (1\leq i\leq n,\ 1\leq r\leq l)$ of the {\em collector matrix} ${\bf V}$ for the largest
fluid autobisimulation ${\cal R}_{fl}(N)$ on $N$ are defined as

$${\cal V}_{ir}=\left\{
\begin{array}{ll}
1, & M_i\in{\cal K}_r;\\
0, & \mbox{otherwise}.
\end{array}
\right.$$

Thus, all the elements of ${\cal V}$ are non-negative, as required. The row elements of ${\cal V}$ are summed to one,
since for each $M_i\ (1\leq i\leq n)$ there exists exactly one ${\cal K}_r\ (1\leq r\leq l)$ such that $M_i\in{\cal
K}_r$. Hence,

$${\bf V}{\bf 1}^T={\bf 1}^T,$$
where ${\bf 1}$ on the left side is the row vector of $l$ values $1$ while ${\bf 1}$ on the right side is the row
vector of $n$ values $1$.

For a vector $v=(v_1,\ldots ,v_l)$, let $Diag(v)$ be a
diagonal matrix with the elements $Diag_{rs}(v)\ (1\leq r,s\leq l)$ defined as

$$Diag_{rs}(v)=\left\{
\begin{array}{ll}
v_r, & r=s;\\
0, & \mbox{otherwise}.
\end{array}
\right.$$

The {\em distributor matrix} ${\bf W}$ for the collector matrix ${\bf V}$ is defined as

$${\bf W}=(Diag({\bf V}^T{\bf 1}^T))^{-1}{\bf V}^T,$$
where ${\bf 1}$ is the row vector of $n$ values $1$. One can check that ${\bf W}{\bf V}={\bf I}$, where ${\bf I}$ is
the identity matrix of order $l$.

The elements $({\cal Q}{\cal V})_{is}\ (1\leq i\leq n,\ 1\leq s\leq l)$ of the matrix ${\bf Q}{\bf V}$ are

$$({\cal Q}{\cal V})_{is}=\sum_{j=1}^n{\cal Q}_{ij}{\cal V}_{js}=\sum_{\{j\mid 1\leq j\leq n,\
M_j\in{\cal K}_s\}}RM(M_i,M_j)=RM(M_i,{\cal K}_s).$$

As we know, for each $M_i\ (1\leq i\leq n)$ there exists exactly one ${\cal K}_r\ (1\leq r\leq l)$ such that
$M_i\in{\cal K}_r$. By Remark 2
from Section \ref{flubiseq.sec}, for all $M_i\in{\cal K}_r$ we have $RM({\cal K}_r,{\cal K}_s)=RM(M_i,{\cal K}_s)\
(1\leq i\leq n,\ 1\leq r,s\leq l)$. Then the elements $({\cal V}{\cal Q}_{\bis_{fl}})_{is}\ (1\leq i\leq n,\ 1\leq
s\leq l)$ of the matrix ${\bf V}{\bf Q}_{\bis_{fl}}$ are

$$({\cal V}{\cal Q}_{\bis_{fl}})_{is}=\sum_{r=1}^l{\cal V}_{ir}({\cal Q}_{\bis_{fl}})_{rs}=\sum_{\{r\mid 1\leq r\leq
l,\ M_i\in{\cal K}_r\}}RM({\cal K}_r,{\cal K}_s)=RM(M_i,{\cal K}_s).$$

Therefore, we have

$${\bf Q}{\bf V}={\bf V}{\bf Q}_{\bis_{fl}},\ {\bf W}{\bf Q}{\bf V}={\bf Q}_{\bis_{fl}}.$$
%
%
%
%
%
%

The elements $({\cal R}_{\bis_{fl}})_{rs}\ (1\leq r,s\leq l)$ of the
FRM ${\bf R}_{\bis_{fl}}$ of the quotient (by $\bis_{fl}$) SFM of $N$ for the continuous place $q$ are defined as

$$({\cal R}_{\bis_{fl}})_{rs}=\left\{
\begin{array}{ll}
RP({\cal K}_r), & r=s;\\
0, & r\neq s.
\end{array}
\right.$$

Let ${\bf R}$ be the
FRM of the SFM of $N$ for the continuous place $q$. The elements $({\cal R}{\cal V})_{is}\ (1\leq i\leq n,\ 1\leq s\leq
l)$ of the matrix ${\bf R}{\bf V}$ are

$$({\cal R}{\cal V})_{is}=\sum_{j=1}^n{\cal R}_{ij}{\cal V}_{js}=RP(M_i){\cal V}_{is}=\left\{
\begin{array}{ll}
RP(M_i), & M_i\in{\cal K}_s;\\
0, & \mbox{otherwise}.
\end{array}
\right.$$

By Remark 2
from Section \ref{flubiseq.sec}, for all $M_i\in{\cal K}_s$ we have $RP({\cal K}_s)=RP(M_i)\ (1\leq i\leq n,\ 1\leq
s\leq l)$. Then the elements $({\cal V}{\cal R}_{\bis_{fl}})_{is}\ (1\leq i\leq n,\ 1\leq s\leq l)$ of the matrix ${\bf
V}{\bf R}_{\bis_{fl}}$ are

$$({\cal V}{\cal R}_{\bis_{fl}})_{is}=\sum_{r=1}^l{\cal V}_{ir}({\cal R}_{\bis_{fl}})_{rs}={\cal V}_{is}RP({\cal K}_s)=
\left\{
\begin{array}{ll}
RP({\cal K}_s)=RP(M_i), & M_i\in{\cal K}_s;\\
0, & \mbox{otherwise}.
\end{array}
\right.$$

Therefore, we also have

$${\bf R}{\bf V}={\bf V}{\bf R}_{\bis_{fl}},\ {\bf W}{\bf R}{\bf V}={\bf R}_{\bis_{fl}}.$$

Let us consider the matrices $Diag(SJ)$ and $Diag(SJ_{\bis_{fl}})$. By analogy with the proven above for ${\bf R}$ and
${\bf R}_{\bis_{fl}}$, we can deduce $Diag(SJ){\bf V}={\bf V}Diag(SJ_{\bis_{fl}})$ and ${\bf W}Diag(SJ){\bf
V}=Diag(SJ_{\bis_{fl}})$. Therefore, we have

$${\bf 1}{\bf W}Diag(SJ){\bf V}={\bf 1}Diag(SJ_{\bis_{fl}})=SJ_{\bis_{fl}},$$
where ${\bf 1}$ is the row vector of $l$ values $1$. In a similar way, we obtain

$${\bf 1}{\bf W}Diag(VAR){\bf V}={\bf 1}Diag(VAR_{\bis_{fl}})=VAR_{\bis_{fl}},$$
where ${\bf 1}$ is the row vector of $l$ values $1$.

\begin{figure}
\begin{center}
\includegraphics{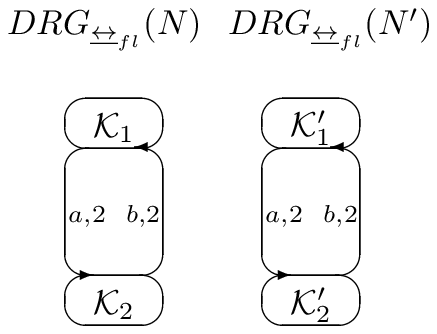}
\end{center}
\vspace{-7mm}
\caption{The quotient discrete reachability graphs of the fluid bisimulation equivalent LFSPNs}
\label{flbqdrg.fig}
\end{figure}

\begin{figure}
\begin{center}
\includegraphics{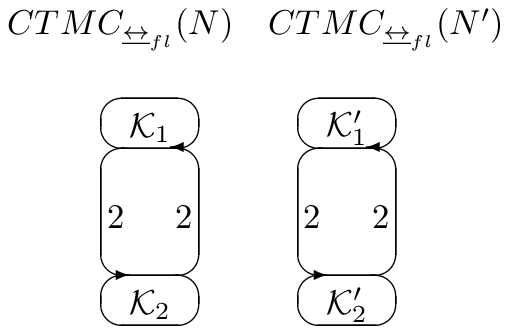}
\end{center}
\vspace{-7mm}
\caption{The quotient underlying CTMCs of the fluid bisimulation equivalent LFSPNs}
\label{flbqctmc.fig}
\end{figure}

\begin{example}
\label{fluredeq.exm}
Consider the LFSPNs $N$ and $N'$ from Figure \ref{flblfspn.fig}, for which it holds $N\bis_{fl}N'$.

In Figure \ref{flbqdrg.fig}, the quotient discrete reachability graphs $DRG_{\bis_{fl}}(N)$ and $DRG_{\bis_{fl}}(N')$
are depicted, for which we have $DRG_{\bis_{fl}}(N)\simeq DRG_{\bis_{fl}}(N')$. In Figure \ref{flbqctmc.fig}, the
quotient underlying CTMCs $CTMC_{\bis_{fl}}(N)$ and $CTMC_{\bis_{fl}}(N')$ are drawn, for which it holds
$CTMC_{\bis_{fl}}(N)\simeq CTMC_{\bis_{fl}}(N')\simeq CTMC(N)$.

We have ${\bf Q}_{\bis_{fl}}={\bf Q}_{\bis_{fl}}'={\bf Q},\ {\bf R}_{\bis_{fl}}={\bf R}_{\bis_{fl}}'={\bf R}$ and
$SJ_{\bis_{fl}}=SJ_{\bis_{fl}}'=SJ,\ VAR_{\bis_{fl}}=VAR_{\bis_{fl}}'=VAR$.

The collector matrix ${\bf V}$ for the largest fluid autobisimulation ${\cal R}_{fl}(N)$ on $N$ and the distributor
matrix ${\bf W}$ for ${\bf V}$ are

$$\begin{array}{cc}
{\bf V}=\left(\begin{array}{cc}
1 & 0\\
0 & 1\\
0 & 1\\
\end{array}\right),
&
{\bf W}=\left(\begin{array}{ccc}
1 & 0 & 0\\
0 & \frac{1}{2} & \frac{1}{2}
\end{array}\right).
\end{array}$$

Then it is easy to check that

$${\bf W}{\bf Q}'{\bf V}={\bf Q},\ {\bf W}{\bf R}'{\bf V}={\bf R}.$$

Hence, it holds that

$${\bf 1}{\bf W}Diag(SJ'){\bf V}=SJ,\ {\bf 1}{\bf W}Diag(VAR'){\bf V}=VAR,$$

$$\mbox{where }\begin{array}{ccc}
{\bf 1}=(1,1),
&
Diag(SJ')=\left(\begin{array}{ccc}
\frac{1}{2} & 0 & 0\\
0 & \frac{1}{2} & 0\\
0 & 0 & \frac{1}{2}
\end{array}\right),
&
Diag(VAR')=\left(\begin{array}{ccc}
\frac{1}{4} & 0 & 0\\
0 & \frac{1}{4} & 0\\
0 & 0 & \frac{1}{4}
\end{array}\right).
\end{array}$$

\end{example}

\section{Logical characterization}
\label{flulogs.sec}

In this section, a logical characterization of fluid trace and bisimulation equivalences is accomplished via formulas
of the novel fluid modal logics. The results obtained could be interpreted as an operational characterization of the
corresponding logical equivalences.

\subsection{Logic $HML_{flt}$}

The modal logic $HML_{NPMTr}$ has been introduced in \cite{BBo06,Bern07a,BBo08} (called $HML_{MTr}$ in
\cite{BBo06,Bern07a}) on (sequential) and concurrent Markovian process calculi SMPC (called MPC in \cite{BBo06,BBo08})
and CMPC for logical interpretation of Markovian trace equivalence. $HML_{NPMTr}$ is based on the logic HML
\cite{HM85}, to which a new interpretation function has been added that takes as its arguments a process state and a
sum or a sequence of the average sojourn times.

We now propose a novel fluid modal logic $HML_{flt}$ for the characterization of fluid trace equivalence. For this, we
extend the interpretation function of $HML_{NPMTr}$ with an additional argument, which is the sequence of the potential
fluid flow rates for the single continuous place of an LFSPN (remember that in the {\em standard} definition of fluid
trace equivalence we compare only LFSPNs, each having exactly one continuous place).

Note that Markovian trace equivalence and the corresponding interpretation function for $HML_{MTr}$ in \cite{BBo06} are
defined by summing up the average sojourn times in the process states. In our definition of fluid trace equivalence, we
consider sequences of the average sojourn times in the discrete markings of LFSPNs. Hence, our fluid extension of
$HML_{NPMTr}$ is based rather on the definitions from \cite{Bern07a,BBo08}, where the latter approach (i.e. the
sequences instead of sums) has been presented.

\begin{definition}
Let $\top$ denote the truth and $a\in Act$. A {\em formula} of $HML_{flt}$ is defined as follows:

$$\Phi ::=\ \top\mid\langle a\rangle\Phi .$$

\end{definition}

${\bf HML}_{flt}$ denotes the set of {\em all formulas of the logic} $HML_{flt}$.

\begin{definition}
Let $N$ be an LFSPN and $M\in DRS(N)$. The {\em interpretation function} $[\![\,]\!]_{flt}:DRS(N)\times{\mathbb
R}_{>0}^*\times{\mathbb R}^*\rightarrow{\bf HML}_{flt}$ is defined as follows:
\begin{enumerate}

\item $[\![\top]\!]_{flt}(M,\varsigma ,\varrho )=\left\{
\begin{array}{ll}
0, & (\varsigma\neq SJ(M))\vee(\varrho\neq RP(M));\\
1, & (\varsigma =SJ(M))\wedge (\varrho =RP(M));
\end{array}
\right.$

\item $[\![\langle a\rangle\Phi]\!]_{flt}(M,\varsigma ,\varrho )=\left\{
\begin{array}{ll}
0, & (\varsigma =\varepsilon )\vee (\varrho =\varepsilon)\vee\\
 & ((\varsigma =s\circ\hat{\varsigma})\wedge (SJ(M)\neq s))\vee\\
 & ((\varrho =r\circ\hat{\varrho})\wedge (RP(M)\neq r));\\
\sum_{\{t\mid M\stackrel{t}{\rightarrow}\widetilde{M},\ L_N(t)=a\}}PT(t,M)
[\![\Phi]\!]_{flt}(\widetilde{M},\hat{\varsigma},\hat{\varrho}), &
(\varsigma =s\circ\hat{\varsigma})\wedge (SJ(M)=s)\wedge\\
 & (\varrho =r\circ\hat{\varrho})\wedge (RP(M)=r).
\end{array}
\right.$

\end{enumerate}
\end{definition}

Note that the item $1$ in the definition above describes the situation when only the empty transition sequence should
start in the discrete marking $M$ to reach the state (which is $M$ itself), described by the identically true formula.
Since we have just a single (mentioned) true state, it remains to check that second and third arguments of the
interpretation function are the sequences of length one, as well as that they are equal to the average sojourn time and
fluid flow rate in $M$, respectively.

\begin{definition}
Let $N$ be an LFSPN. Then we define $[\![\Phi]\!]_{flt}(N,\varsigma ,\varrho )=[\![\Phi]\!]_{flt}(M_N,\varsigma
,\varrho )$. Two LFSPNs $N$ and $N'$ are {\em logically equivalent} in $HML_{flt}$, denoted by $N=_{HML_{flt}}N'$, if
$\forall\Phi\in{\bf HML}_{flt}\ \forall\varsigma\in{\mathbb R}_{>0}^*\ \forall\varrho\in{\mathbb R}^*\
[\![\Phi]\!]_{flt}(N,\varsigma ,\varrho )=[\![\Phi]\!]_{flt}(N',\varsigma ,\varrho )$.
\end{definition}

Let $N$ be an LFSPN and $M\in DRS(N),\ a\in Act$. The set of discrete markings reached from $M$ by execution of action
$a$, called the {\em image set}, is defined as $Image(M,a)=\{\widetilde{M}\mid
M\stackrel{t}{\rightarrow}\widetilde{M},\ L_N(t)=a\}$. An LFSPN $N$ is an {\em image-finite} one, if $\forall M\in
DRS(N)\ \forall a\in Act\ |Image(M,a)|<\infty$.

In order to get the intended logical characterization, we need in some auxiliary definitions considering the transition
sequences starting not just in the initial discrete marking of an LFSPN, but in any reachable one.

Let $N$ be an LFSPN and $M\in DRS(N)$. The set of {\em all (finite) transition sequences} in $N$ {\em starting in the
discrete marking} $M$ is defined as

$$TranSeq(N,M)=\{\vartheta\mid\vartheta =\varepsilon\mbox{ or }\vartheta =t_1\cdots t_n,\ M=M_0
\stackrel{t_1}{\rightarrow}M_1\stackrel{t_2}{\rightarrow}\cdots\stackrel{t_n}{\rightarrow}M_n\}.$$

Let $\vartheta =t_1\cdots t_n\in TranSeq(N,M)$ and $M=M_0\stackrel{t_1}{\rightarrow}M_1\stackrel{t_2}{\rightarrow}\cdots
\stackrel{t_n}{\rightarrow}M_n$. The {\em probability to execute the transition sequence $\vartheta$} is

$$PT(\vartheta )=\prod_{i=1}^n PT(t_i,M_{i-1}).$$

For $\vartheta =\varepsilon$ we define $PT(\varepsilon )=1$.

Let $(\sigma ,\varsigma ,\varrho )\in Act^*\times{\mathbb R}_{>0}^*\times{\mathbb R}^*$. The set of {\em $(\sigma
,\varsigma ,\varrho )$-selected (finite) transition sequences} in $N$ {\em starting in the discrete marking} $M$ is
defined as

$$TranSeq(N,M,\sigma ,\varsigma ,\varrho )=\{\vartheta\in TranSeq(N,M)\mid L_N(\vartheta )=\sigma ,\ SJ(\vartheta
)=\varsigma ,\ RP(\vartheta )=\varrho\}.$$

The (cumulative) {\em probability to execute $(\sigma ,\varsigma ,\varrho )$-selected transition sequences starting in
the discrete marking} $M$ is

$$PT(M,\sigma ,\varsigma ,\varrho )=\sum_{\vartheta\in TranSeq(N,M,\sigma ,\varsigma ,\varrho )}PT(\vartheta ).$$

The following lemma provides a recursive definition of $PT(M,\sigma ,\varsigma ,\varrho )$ that will be used later in
the proofs.

\begin{lemma}
Let $N$ be an LFSPN and $M\in DRS(N)$. Then for all $(\sigma ,\varsigma ,\varrho )\in Act^*\times{\mathbb
R}_{>0}^*\times{\mathbb R}^*$ such that $\sigma =a\cdot\hat{\sigma},\ \varsigma =s\circ\hat{\varsigma},\ \varrho
=r\circ\hat{\varrho}$, where $a\in Act,\ s\in{\mathbb R}_{>0},\ r\in{\mathbb R}$, we have

$$PT(M,\sigma ,\varsigma ,\varrho )=\sum_{\{t\mid M\stackrel{t}{\rightarrow}\widetilde{M},\ \sigma =a\cdot\hat{\sigma},\
L_N(t)=a,\ \varsigma =s\circ\hat{\varsigma},\ SJ(M)=s,\ \varrho =r\circ\hat{\varrho},\ RP(M)=r\}}PT(t,M)
PT(\widetilde{M},\hat{\sigma},\hat{\varsigma},\hat{\varrho}).$$

\label{ptmsivsvr.lem}
\end{lemma}
{\em Proof.}
It holds that $PT(M,\sigma ,\varsigma ,\varrho )=
\sum_{\vartheta\in TranSeq(N,M,\sigma ,\varsigma ,\varrho )}PT(\vartheta )=\\
\sum_{\{t_1,\ldots ,t_n\mid M=M_0\stackrel{t_1}{\rightarrow}M_1\stackrel{t_2}{\rightarrow}\cdots
\stackrel{t_n}{\rightarrow}M_n,\ L_N(t_1\cdots t_n)=\sigma ,\ SJ(M_0\cdots M_n)=\varsigma ,\ RP(M_0\cdots
M_n)=\varrho\}}\prod_{i=1}^n PT(t_i,M_{i-1})=\\
\sum_{\{t_1\mid M=M_0\stackrel{t_1}{\rightarrow}M_1,\ L_N(t_1)=a,\ SJ(M_0)=s,\ RP(M_0)=r\}}\\
\sum_{\{t_2,\ldots ,t_n\mid M_1\stackrel{t_2}{\rightarrow}M_2\stackrel{t_3}{\rightarrow}\cdots
\stackrel{t_n}{\rightarrow}M_n,\ L_N(t_2\cdots t_n)=\hat{\sigma},\ SJ(M_1\cdots M_n)=\hat{\varsigma},\ RP(M_1\cdots
M_n)=\hat{\varrho}\}}PT(t_1,M_0)\prod_{i=2}^n PT(t_i,M_{i-1})=\\
\sum_{\{t_1\mid M=M_0\stackrel{t_1}{\rightarrow}M_1,\ L_N(t_1)=a,\ SJ(M_0)=s,\ RP(M_0)=r\}}\\
PT(t_1,M_0)\left(\sum_{\{t_2,\ldots ,t_n\mid M_1\stackrel{t_2}{\rightarrow}M_2\stackrel{t_3}{\rightarrow}\cdots
\stackrel{t_n}{\rightarrow}M_n,\ L_N(t_2\cdots t_n)=\hat{\sigma},\ SJ(M_1\cdots M_n)=\hat{\varsigma},\ RP(M_1\cdots
M_n)=\hat{\varrho}\}}\prod_{i=2}^n PT(t_i,M_{i-1})\right)=\\
\sum_{\{t_1\mid M=M_0\stackrel{t_1}{\rightarrow}M_1,\ L_N(t_1)=a,\ SJ(M_0)=s,\ RP(M_0)=r\}}PT(t_1,M_0)
PT(M_1,\hat{\sigma},\hat{\varsigma},\hat{\varrho})$.
%
Let us now $t=t_1$ and $\widetilde{M}=M_1$. \hfill $\eop$

The following propositions demonstrate that there exists a bijective correspondence between fluid stochastic traces of
LFSPNs and formulas of ${\bf HML}_{flt}$, by proving that the probabilities of the triples $(\sigma ,\varsigma ,\varrho
)\in Act^*\times{\mathbb R}_{>0}^*\times{\mathbb R}^*$ coincide in the net and logical frameworks.

\begin{proposition}
Let $N$ be an image-finite LFSPN. Then for each $\sigma\in Act^*$ there exists $\Phi_{\sigma}\in{\bf HML}_{flt}$ such
that $\forall M\in DRS(N)\ \forall\varsigma\in{\mathbb R}_{>0}^*\ \forall\varrho\in{\mathbb R}^*$

$$[\![\Phi_{\sigma}]\!]_{flt}(M,\varsigma ,\varrho )=PT(M,\sigma ,\varsigma ,\varrho ).$$

\label{fltrform.pro}
\end{proposition}
{\em Proof.} We prove by induction on the length $n$ of the action sequence $\sigma$.
\begin{itemize}

\item $n=0$

We have $|\sigma |=0$, hence, $\sigma =\varepsilon$. In this case, we take $\Phi_{\sigma}=\top$. Let $M\in DRS(N),\
\varsigma\in{\mathbb R}_{>0}^*,\ \varrho\in{\mathbb R}^*$.

If
$(\varsigma\neq SJ(M))\vee(\varrho\neq RP(M))$ then $TranSeq(N,M,\sigma ,\varsigma ,\varrho )=\emptyset$ and

$$[\![\Phi_{\sigma}]\!]_{flt}(M,\varsigma ,\varrho )=0=PT(M,\sigma ,\varsigma ,\varrho ).$$

Otherwise, if $(\varsigma =SJ(M))\wedge (\varrho =RP(M))$ then $TranSeq(N,M,\sigma ,\varsigma ,\varrho
)=\{\varepsilon\}$ and

$$[\![\Phi_{\sigma}]\!]_{flt}(M,\varsigma ,\varrho )=1=PT(M,\sigma ,\varsigma ,\varrho ).$$

\item $n\rightarrow n+1$

We have $|\sigma |=n+1$, hence, $\sigma =a\cdot\hat{\sigma}$, where $a\in Act$ and $|\hat{\sigma}|=n$. In this
case, we take $\Phi_{\sigma}=\langle a\rangle\Phi_{\hat{\sigma}}$, where the indiction hypothesis holds for
$\hat{\sigma}$ and $\Phi_{\hat{\sigma}}$. Let $M\in DRS(N),\ \varsigma\in{\mathbb R}_{>0}^*,\ \varrho\in{\mathbb
R}^*$.

If no transition labeled with action $a$ is enabled in $M$ or $(\varsigma =\varepsilon )\vee (\varrho
=\varepsilon)\vee ((\varsigma =s\circ\hat{\varsigma})\wedge (SJ(M)\neq s))\vee ((\varrho
=r\circ\hat{\varrho})\wedge (RP(M)\neq r))$ then $TranSeq(N,M,\sigma ,\varsigma ,\varrho )=\emptyset$ and

$$[\![\Phi_{\sigma}]\!]_{flt}(M,\varsigma ,\varrho )=0=PT(M,\sigma ,\varsigma ,\varrho ).$$

Otherwise, if transitions labeled with action $a$ are enabled in $M$ and $(\varsigma =s\circ\hat{\varsigma})\wedge
(SJ(M)=s)\wedge (\varrho =r\circ\hat{\varrho})\wedge (RP(M)=r)$ then $TranSeq(N,M,\sigma ,\varsigma ,\varrho
)\neq\emptyset$ and

$$[\![\Phi_{\sigma}]\!]_{flt}(M,\varsigma ,\varrho )=\sum_{\{t\mid M\stackrel{t}{\rightarrow}\widetilde{M},\
L_N(t)=a\}}PT(t,M)[\![\Phi_{\hat{\sigma}}]\!]_{flt}(\widetilde{M},\hat{\varsigma},\hat{\varrho}),$$

as well as

$$PT(M,\sigma ,\varsigma ,\varrho )=\sum_{\{t\mid M\stackrel{t}{\rightarrow}\widetilde{M},\ L_N(t)=a\}}PT(t,M)
PT(\widetilde{M},\hat{\sigma},\hat{\varsigma},\hat{\varrho}).$$

By the induction hypothesis, for all discrete markings $\widetilde{M}$ reachable from $M$ by firing transitions
labeled with action $a$ we have

$$[\![\Phi_{\hat{\sigma}}]\!]_{flt}(\widetilde{M},\hat{\varsigma},\hat{\varrho})=
PT(\widetilde{M},\hat{\sigma},\hat{\varsigma},\hat{\varrho}),$$

thus, we have proven. \hfill $\eop$

\end{itemize}

\begin{proposition}
Let $N$ be an image-finite LFSPN. Then for each $\Phi\in{\bf HML}_{flt}$ there exists $\sigma_{\Phi}\in Act^*$ such
that $\forall M\in DRS(N)\ \forall\varsigma\in{\mathbb R}_{>0}^*\ \forall\varrho\in{\mathbb R}^*$

$$PT(M,\sigma_{\Phi},\varsigma ,\varrho )=[\![\Phi]\!]_{flt}(M,\varsigma ,\varrho ).$$

\label{formfltr.pro}
\end{proposition}
{\em Proof.} We prove by induction on the syntactical structure of the logical formula $\Phi$.
\begin{itemize}

\item $\Phi =\top$

In this case, we take $\sigma_{\Phi}=\varepsilon$. Let $M\in DRS(N),\ \varsigma\in{\mathbb R}_{>0}^*,\
\varrho\in{\mathbb R}^*$.

If
$(\varsigma\neq SJ(M))\vee(\varrho\neq RP(M))$ then $TranSeq(N,M,\sigma ,\varsigma ,\varrho )=\emptyset$ and

$$PT(M,\sigma ,\varsigma ,\varrho )=0=[\![\Phi_{\sigma}]\!]_{flt}(M,\varsigma ,\varrho ).$$

Otherwise, if $(\varsigma =SJ(M))\wedge (\varrho =RP(M))$ then $TranSeq(N,M,\sigma ,\varsigma ,\varrho
)=\{\varepsilon\}$ and

$$PT(M,\sigma ,\varsigma ,\varrho )=1=[\![\Phi_{\sigma}]\!]_{flt}(M,\varsigma ,\varrho ).$$

\item $\Phi =\langle a\rangle\Phi$

In this case, we take $\sigma_{\Phi}=a\cdot\sigma_{\widehat{\Phi}}$, where the indiction hypothesis holds for
$\widehat{\Phi}$ and $\sigma_{\widehat{\Phi}}$. Let $M\in DRS(N),\ \varsigma\in{\mathbb R}_{>0}^*,\
\varrho\in{\mathbb R}^*$.

If no transition labeled with action $a$ is enabled in $M$ or $(\varsigma =\varepsilon )\vee (\varrho
=\varepsilon)\vee ((\varsigma =s\circ\hat{\varsigma})\wedge (SJ(M)\neq s))\vee ((\varrho
=r\circ\hat{\varrho})\wedge (RP(M)\neq r))$ then $TranSeq(N,M,\sigma ,\varsigma ,\varrho )=\emptyset$ and

$$PT(M,\sigma ,\varsigma ,\varrho )=0=[\![\Phi_{\sigma}]\!]_{flt}(M,\varsigma ,\varrho ).$$

Otherwise, if transitions labeled with action $a$ are enabled in $M$ and $(\varsigma =s\circ\hat{\varsigma})\wedge
(SJ(M)=s)\wedge (\varrho =r\circ\hat{\varrho})\wedge (RP(M)=r)$ then $TranSeq(N,M,\sigma ,\varsigma ,\varrho
)\neq\emptyset$ and

$$PT(M,\sigma_{\Phi},\varsigma ,\varrho )=\sum_{\{t\mid M\stackrel{t}{\rightarrow}\widetilde{M},\ L_N(t)=a\}}PT(t,M)
PT(\widetilde{M},\sigma_{\widehat{\Phi}},\hat{\varsigma},\hat{\varrho}),$$

as well as

$$[\![\Phi ]\!]_{flt}(M,\varsigma ,\varrho )=\sum_{\{t\mid M\stackrel{t}{\rightarrow}\widetilde{M},\ L_N(t)=a\}}PT(t,M)
[\![\widehat{\Phi}]\!]_{flt}(\widetilde{M},\hat{\varsigma},\hat{\varrho}).$$

By the induction hypothesis, for all discrete markings $\widetilde{M}$ reachable from $M$ by firing transitions
labeled with action $a$ we have

$$PT(\widetilde{M},\sigma_{\widehat{\Phi}},\hat{\varsigma},\hat{\varrho})=
[\![\widehat{\Phi}]\!]_{flt}(\widetilde{M},\hat{\varsigma},\hat{\varrho}),$$

thus, we have proven. \hfill $\eop$

\end{itemize}

\begin{theorem}
For image-finite LFSPNs $N$ and $N'$

$$N\equiv_{fl}N'\ \Leftrightarrow\ N=_{HML_{flt}}N'.$$

\label{HML-flt.the}
\end{theorem}
{\em Proof.} The result follows from Proposition \ref{fltrform.pro} and Proposition \ref{formfltr.pro}, which establish
a bijective correspondence between fluid stochastic traces of LFSPNs and formulas of ${\bf HML}_{flt}$. \hfill $\eop$

Thus, in the trace semantics, we obtained a logical characterization of the fluid behavioural equivalence or,
symmetrically, an operational characterization of the fluid modal logic equivalence.

\begin{example}
\label{HML-flt.exm} Consider the LFSPNs $N$ and $N'$ from Figure \ref{fltlfspn.fig}, for which it holds
$N\equiv_{fl}N'$, hence, $N=_{HML_{flt}}N'$. In particular, for $\Phi =\langle\{a\}\rangle\langle\{b\}\rangle\top$ we
have $\sigma_{\Phi}=a\cdot b$ and $[\![\Phi]\!]_{flt}(M_N,\frac{1}{2}\circ\frac{1}{2}\circ\frac{1}{2},1\circ (-2)\circ
1)=PT(t_1 t_2)=1\cdot\frac{1}{2}=\frac{1}{2}=1\cdot\frac{1}{2}=PT(t_1't_3')=
[\![\Phi]\!]_{flt}(M_{N'},\frac{1}{2}\circ\frac{1}{2}\circ\frac{1}{2},1\circ (-2)\circ 1)$.
\end{example}

\subsection{Logic $HML_{flb}$}

The modal logic $HML_{MB}$ has been introduced in \cite{Bern07a,BBo08} on sequential and concurrent Markovian process
calculi SMPC (called MPC in \cite{BBo08}) and CPMC for logical interpretation of Markovian bisimulation equivalence.
$HML_{MB}$ is based on the logic HML \cite{HM85}, in which the diamond operator was decorated with the rate lower
bound. Hence, $HML_{MB}$ can be also seen as a modification of the logic $PML$ \cite{LS91}, where the probability lower
bound that decorates the diamond operator was replaced with the rate lower bound.

We now propose a novel fluid modal logic $HML_{flb}$ for the characterization of fluid bisimulation equivalence. For
this, we
add to $HML_{MB}$ a new modality $\wr_r$, where $r\in{\mathbb R}$ is the potential fluid flow rate value for the
single continuous place of an LFSPN (remember that in the {\em standard} definition of fluid bisimulation equivalence
we compare only LFSPNs, each having exactly one continuous place). The formula $\wr_r$ is used to check whether the
potential fluid flow rate in a discrete marking of an LFSPN
equals $r$, the fact that
refers to a particular condition from the fluid bisimulation definition. Thus, $\wr_r$ can be seen as a supplement to
the PML and $HML_{MB}$ formula $\nabla_a$, where $a\in Act$, since $\nabla_a$ is used to check whether the transitions
labeled with the action $a$ cannot be fired in a state (discrete marking), the fact violating the bisimulation transfer
property.

\begin{definition}
Let $\top$ denote the truth and $a\in Act,\ r\in{\mathbb R},\ \lambda\in{\mathbb R}_{>0}$. A {\em formula} of
$HML_{flb}$ is defined as follows:

$$\Phi ::=\ \top\mid\neg\Phi\mid\Phi\wedge\Phi\mid\nabla_a\mid\wr_r\mid\langle a\rangle_{\lambda}\Phi .$$

\end{definition}

We define $\langle a\rangle\Phi=\exists\lambda\ \langle a\rangle_{\lambda}\Phi$ and $\Phi\vee\Psi
=\neg(\neg\Phi\wedge\neg\Psi )$.

${\bf HML}_{flb}$ denotes the set of {\em all formulas of the logic} $HML_{flb}$.

\begin{definition}
Let $N$ be a LFSPN and $M\in DRS(N)$. The {\em satisfaction relation} $\models_{flb}\subseteq DRS(N)\times{\bf
HML}_{flb}$ is defined as follows:
\begin{enumerate}

\item $M\models_{flb}\top$ --- always;

\item $M\models_{flb}\neg\Phi$, if $M\not\models_N\Phi$;

\item $M\models_{flb}\Phi\wedge\Psi$, if $M\models_N\Phi$ and $M\models_N\Psi$;

\item $M\models_{flb}\nabla_a$, if it does not hold that $M\stackrel{a}{\rightarrow}DRS(N)$;

\item $M\models_{flb}\wr_r$, if $RP(M)=r$;

\item $M\models_{flb}\langle a\rangle_{\lambda}\Phi$, if
$\exists{\cal H}\subseteq DRS(N)\ M\stackrel{a}{\rightarrow}_{\mu}{\cal H},\ \mu\geq\lambda$ and
$\forall\widetilde{M}\in{\cal H}\ \widetilde{M}\models_{flb}\Phi$.

\end{enumerate}
\end{definition}

Note that $\langle a\rangle_{\mu}\Phi$ implies $\langle a\rangle_{\lambda}\Phi$, if $\mu\geq\lambda$.

\begin{definition}
Let $N$ be an LFSPN. Then we write $N\models_{flb}\Phi$, if $M_N\models_{flb}\Phi$. LFSPNs $N$ and $N'$ are {\em
logically equivalent} in $HML_{flb}$, denoted by $N=_{HML_{flb}}N'$, if $\forall\Phi\in{\bf HML}_{flb}\
N\models_{flb}\Phi\ \Leftrightarrow\ N'\models_{flb}\Phi$.
\end{definition}

Let $N$ be an LFSPN and $M\in DRS(N),\ a\in Act$. The set of discrete markings reached from $M$ by execution of action
$a$, called the {\em image set}, is defined as $Image(M,a)=\{\widetilde{M}\mid
M\stackrel{t}{\rightarrow}\widetilde{M},\ L_N(t)=a\}$. An LFSPN $N$ is an {\em image-finite} one, if $\forall M\in
DRS(N)\ \forall a\in Act\ |Image(M,a)|<\infty$.

\begin{theorem}
For image-finite LFSPNs $N$ and $N'$

$$N\bis_{fl}N'\ \Leftrightarrow\ N=_{HML_{flb}}N'.$$

\label{HML-flb.the}
\end{theorem}
{\em Proof.} Our reasoning is based on the proofs of Theorem 6.4 from \cite{LS91} about characterization of
probabilistic bisimulation equivalence for probabilistic transition systems and Theorem 1 from \cite{CGH99} about
characterization of strong equivalence for PEPA. The differences are the LFSPNs context, and what we also respect the
fluid flow rates in the discrete markings with the satisfaction check for the formulas $\wr_r,\ r\in{\mathbb R}$, as
presented below.

($\Leftarrow$) Let us define the equivalence relation ${\cal R}=\{(M_1,M_2)\in (DRS(N)\cup DRS(N'))^2\mid
\forall\Phi\in{\bf HML}_{flb}\\
M_1\models_{flb}\Phi\ \Leftrightarrow\ M_2\models_{flb}\Phi\}$. We have $(M_N,M_{N'})\in{\cal R}$. Let us prove that
${\cal R}$ is a fluid bisimulation.

Assume that $M_N\stackrel{a}{\rightarrow}_{\lambda}{\cal H}\in (DRS(N)\cup DRS(N'))/_{\cal R}$. Let
$M_{N'}\stackrel{a}{\rightarrow}_{\lambda_1'}M_1',\ldots ,M_{N'}\stackrel{a}{\rightarrow}_{\lambda_i'}M_i',
M_{N'}\stackrel{a}{\rightarrow}_{\lambda_{i+1}'}M_{i+1}',\ldots ,M_{N'}\stackrel{a}{\rightarrow}_{\lambda_n'}M_n'$ be
the changes of the discrete marking $M_{N'}$ as a result of executing the action $a$. Since the LFSPN $N'$ is
image-finite one, the number of such
changes is finite. The discrete marking changes are ordered so that $M_1',\ldots ,M_i'\in{\cal H}$ and $M_{i+1}',\ldots
,M_n'\not\in{\cal H}$.

Then $\exists\Phi_{i+1},\ldots ,\Phi_n\in{\bf HML}_{flb}$ such that $\forall j\ (i+1\leq j\leq n)\ \forall M\in{\cal
H}\ M\models_{flb}\Phi_j$, but $M_j'\not\models_{flb}\Phi_j$. We have $M_N\models_{flb}\langle
a\rangle_{\lambda}(\wedge_{j=i+1}^n\Phi_j)$ and $M_{N'}\models_{flb}\langle
a\rangle_{\lambda '}(\wedge_{j=i+1}^n\Phi_j)$, where $\lambda '=\sum_{j=1}^i\lambda_j'$.

Assume that $\lambda >\lambda '$. Then $M_{N'}\not\models_{flb}\langle a\rangle_{\lambda}(\wedge_{j=i+1}^n\Phi_j)$,
which contradicts to $(M_N,M_{N'})\in{\cal R}$. Hence, $\lambda\leq\lambda '$. Consequently,
$M_{N'}\stackrel{a}{\rightarrow}_{\lambda '}{\cal H}$, where $\lambda\leq\lambda '$. By symmetry of ${\cal R}$, we have
$\lambda\geq\lambda '$. Thus, $\lambda =\lambda '$, and ${\cal R}$ is a fluid bisimulation.

($\Rightarrow$) Let for LFSPNs $N$ and $N'$ we have $N\bis_{fl}N'$. Then $\exists{\cal R}:N\bis_{fl}N'$ and
$(M_N,M_{N'})\in{\cal R}$. It is sufficient to consider only the cases $\nabla_a,\wr_r$ and $\langle
a\rangle_{\lambda}\Phi$, since the remaining cases are trivial.

{\bf The case} $\nabla_a$.

Assume that $M_N\models_{flb}\nabla_a$. Then it does not hold that $M_N\stackrel{a}{\rightarrow}DRS(N)$. Hence, there
exist no $t$ and $\widetilde{M}$ such that $M_N\stackrel{t}{\rightarrow}\widetilde{M}$ and $L_N(t)=a$. Since summing by
the empty index set produces zero, the transitions from each discrete marking always lead to the discrete markings of
the discrete reachability set to which that discrete marking belongs and $(M_N,M_{N'})\in{\cal R}$, we get
$0=\sum_{\{t\mid\exists\widetilde{M}\in DRS(N)\ M_N\stackrel{t}{\rightarrow}\widetilde{M},\
L_N(t)=a\}}\Omega_N(t,M_N)=\\
RM_a(M_N,DRS(N))=RM_a(M_N,DRS(N)\cup DRS(N'))=\sum_{{\cal H}\in (DRS(N)\cup DRS(N'))/_{\cal R}}RM_a(M_N,{\cal H})=\\
\sum_{{\cal H}\in (DRS(N)\cup DRS(N'))/_{\cal R}}RM_a(M_{N'},{\cal H})=RM_a(M_{N'},DRS(N)\cup DRS(N'))=
RM_a(M_{N'},DRS(N'))=\\
\sum_{\{t'\mid\exists\widetilde{M}'\in DRS(N')\ M_{N'}\stackrel{t'}{\rightarrow}\widetilde{M}',\
L_{N'}(t')=a\}}\Omega_{N'}(t',M_{N'})$. Hence, there exist no $t'$ and $\widetilde{M}'$ such that
$M_{N'}\stackrel{t'}{\rightarrow}\widetilde{M}'$ and $L_{N'}(t')=a$. Thus, it does not hold that
$M_{N'}\stackrel{a}{\rightarrow}DRS(N')$ and we have $M_{N'}\models_{flb}\nabla_a$.

{\bf The case} $\wr_r$.

Assume that $M_N\models_{flb}\wr_r$. Then, respecting that $(M_N,M_{N'})\in{\cal R}$, we get $r=RP(M_N)=RP(M_{N'})$,
hence, $M_{N'}\models_{flb}\wr_r$.

{\bf The case} $\langle a\rangle_{\lambda}\Phi$.

Assume that $M_N\models_{flb}\langle a\rangle_{\lambda}\Phi$. Then $\exists{\cal H}\subseteq DRS(N)$ such that
$M_N\stackrel{a}{\rightarrow}_{\mu}{\cal H},\ \mu\geq\lambda$ and $\forall M\in{\cal H}\ M\models_{flb}\Phi$. Let us
define $\widetilde{\cal H}=\bigcup\{\overline{\cal H}\in (DRS(N)\cup DRS(N'))/_{\cal R}\mid\overline{\cal H}\cap{\cal
H}\neq\emptyset\}$. Then $\forall\widetilde{M}\in\widetilde{\cal H}\ \exists M\in{\cal H}\ (M,\widetilde{M})\in{\cal
R}$. Since $\forall M\in{\cal H}\ M\models_{flb}\Phi$, we have $\forall\widetilde{M}\in\widetilde{\cal H}\
\widetilde{M}\models_{flb}\Phi$ by the induction hypothesis.

Since ${\cal H}\subseteq\widetilde{\cal H}$, we get $M_N\stackrel{a}{\rightarrow}_{\tilde{\mu}}\widetilde{\cal H},\
\tilde{\mu}\geq\mu$. Since $\widetilde{\cal H}$ is the union of the equivalence classes with respect to ${\cal R}$, we
have $(M_N,M_{N'})\in{\cal R}$ implies $M_{N'}\stackrel{a}{\rightarrow}_{\tilde{\mu}}\widetilde{\cal H}$. Since
$\tilde{\mu}\geq\mu\geq\lambda$, we get $M_{N'}\models_{flb}\langle a\rangle_\lambda\Phi$. Therefore, $N'$ satisfies
all the formulas which $N$ does. By symmetry of ${\cal R},\ N$ satisfies all the formulas which $N'$ does. Thus, the
sets of satisfiable formulas for $N$ and $N'$ coincide. \hfill $\eop$
%

Thus, in the bisimulation semantics, we obtained a logical characterization of the fluid behavioural equivalence or,
symmetrically, an operational characterization of the fluid modal logic equivalence.

\begin{example}
\label{HML-flb.exm}
Consider the LFSPNs $N$ and $N'$ from Figure \ref{fltlfspn.fig}, for which it holds $N\nbis_{fl}N'$, hence,
$N\neq_{HML_{flb}}N'$. Indeed, for $\Phi =\langle a\rangle_2\langle b\rangle_1\top$ we have $N\models_{flb}\Phi$, but
$N'\not\models_{flb}\Phi$, since only in the LFSPN $N'$ action $a$ can occur so that action $b$ cannot occur
afterwards.

Let us now take the LFSPNs $N$ and $N'$ from Figure \ref{flblfspn.fig}, for which it holds $N\bis_{fl}N'$, hence,
$N=_{HML_{flb}}N'$. In particular, for $\Psi =\wr_1\wedge\langle a\rangle_2(\wr_{-2}\wedge\langle b\rangle_2\top )$ we
have $N\models_{flb}\Psi$ and $N'\models_{flb}\Psi$.
\end{example}

Table \ref{flumodint.tab} demonstrates how the modalities and interpretation functions of the logics $HML_{flt}$ and
$HML_{flb}$ respect the following behavioural aspects of LFSPNs: {\em semantics type} (linear or branching time), {\em
functional activity} (consisting in the action occurrences), {\em stochastic timing} (specified by the transition
rates) and {\em fluid flow} (defined by the fluid rates). In case of the composite constructions, the variables
describing particular aspects of behaviour are presented nearby in parentheses.

\begin{table}
\caption{Behavioural aspects of LFSPNs in the logical modalities and interpretations}
\vspace{-3mm}
\label{flumodint.tab}
\begin{center}
\begin{tabular}{|c||c|c|c|c|}
\hline
Fluid & Semantics type & Functional activity & Stochastic timing & Fluid flow\\
modal logic & (linear/branching time) & (action occurrences) & (transition rates) & (fluid rates)\\
\hline\hline
$HML_{flt}$ & $\top$ & $\langle a\rangle$ & $[\![\cdot ]\!]_{flt}(M,\varsigma ,\varrho )$ (by $\varsigma$) &
$[\![\cdot ]\!]_{flt}(M,\varsigma ,\varrho )$ (by $\varrho$)\\
\hline
$HML_{flb}$ & $\top$, $\neg$, $\wedge$ & $\nabla_a$, $\langle a\rangle_\lambda$ (by $a$) & $\langle a\rangle_\lambda$
(by $\lambda$) & $\wr_r$\\
\hline
\end{tabular}
\end{center}
\end{table}

\section{Preservation of the quantitative behaviour}
\label{quantbeh.sec}

It is clear that the proposed fluid bisimulation equivalence of LFSPNs preserves their qualitative (functional)
behaviour which is based on the actions assigned to the fired transitions. Let us examine if fluid bisimulation
equivalence also preserves the quantitative (performance) behaviour of LFSPNs, taken for the steady states of their
underlying CTMCs and associated SFMs.
The quantitative behaviour takes into account the values of the rates and probabilities, as well as those of the
related probability mass, distribution, density and mass at lower boundary functions. Then we shall define the
quotients of the mentioned probability functions by fluid bisimulation equivalence with a goal to describe the quotient
(by $\bis_{fl}$) associated SFMs.

The following proposition demonstrates that for two LFSPNs related by $\bis_{fl}$ their aggregate steady-state
probabilities coincide for each equivalence class of discrete markings.

\begin{proposition}
Let $N,N'$ be LFSPNs with ${\cal R}:N\bis_{fl}N'$ and $\varphi =(\varphi_1,\ldots ,\varphi_n),\ n=|DRS(N)|$, be the
steady-state PMF for $CTMC(N)$ and $\varphi '=(\varphi_1',\ldots ,\varphi_m'),\ m=|DRS(N')|$, be the steady-state PMF
for $CTMC(N')$. Then for all ${\cal H}\in (DRS(N)\cup DRS(N'))/_{\cal R}$ we have

$$\sum_{\{i\mid M_i\in{\cal H}\cap DRS(N)\}}\varphi_i=\sum_{\{j\mid M_j'\in{\cal H}\cap DRS(N')\}}\varphi_j'.$$

\label{statprob.pro}
\end{proposition}
{\em Proof.} The steady-state PMF $\varphi =(\varphi_1,\ldots ,\varphi_n)$ for $CTMC(N)$ is a solution of the linear
equation system

$$\left\{
\begin{array}{l}
\varphi{\bf Q}={\bf 0}\\
\varphi{\bf 1}^T=1
\end{array}
\right..$$

Then for all $i\ (1\leq i\leq n)$ we have

$$\left\{
\begin{array}{l}
\sum_{j=1}^n{\cal Q}_{ji}\varphi_j=0\\
\sum_{j=1}^n\varphi_j=1
\end{array}
\right..$$

By definition of ${\cal Q}_{ij}\ (1\leq i,j\leq n)$ we have

$$\left\{
\begin{array}{l}
\sum_{j=1}^n RM(M_j,M_i)\varphi_j=0\\
\sum_{j=1}^n\varphi_j=1
\end{array}
\right..$$

Let ${\cal H}\in (DRS(N)\cup DRS(N'))/_{\cal R}$. We sum the left and right sides of the first equation from the system
above for all $i$ such that $M_i\in{\cal H}\cap DRS(N)$. The resulting equation is

$$\sum_{\{i\mid M_i\in{\cal H}\cap DRS(N)\}}\sum_{j=1}^n RM(M_j,M_i)\varphi_j=0.$$

Let us denote the aggregate steady-state PMF for $CTMC(N)$ by $\varphi_{{\cal H}\cap DRS(N)}=\sum_{\{i\mid M_i\in{\cal
H}\cap DRS(N)\}}\varphi_i$. Then, by Remark 2
from Section \ref{flubiseq.sec}, for the left-hand side of the equation above, we get\\
$\sum_{\{i\mid M_i\in{\cal H}\cap DRS(N)\}}\sum_{j=1}^n RM(M_j,M_i)\varphi_j=
\sum_{j=1}^n\varphi_j\sum_{\{i\mid M_i\in{\cal H}\cap DRS(N)\}}RM(M_j,M_i)=\\
\sum_{j=1}^n RM(M_j,{\cal H})\varphi_j=\sum_{\widetilde{\cal H}\in (DRS(N)\cup DRS(N'))/_{\cal R}}
\sum_{\{j\mid M_j\in\widetilde{\cal H}\cap DRS(N)\}}RM(M_j,{\cal H})\varphi_j=\\
\sum_{\widetilde{\cal H}\in (DRS(N)\cup DRS(N'))/_{\cal R}}
\sum_{\{j\mid M_j\in\widetilde{\cal H}\cap DRS(N)\}}RM(\widetilde{\cal H},{\cal H})\varphi_j=\\
\sum_{\widetilde{\cal H}\in (DRS(N)\cup DRS(N'))/_{\cal R}}RM(\widetilde{\cal H},{\cal H})
\sum_{\{j\mid M_j\in\widetilde{\cal H}\cap DRS(N)\}}\varphi_j\!=\!\!
\sum_{\widetilde{\cal H}\in (DRS(N)\cup DRS(N'))/_{\cal R}}RM(\widetilde{\cal H},{\cal H})
\varphi_{\widetilde{\cal H}\cap DRS(N)}$.

For the left-hand side of the second equation from the system above, we have\\
$\sum_{j=1}^n\varphi_j=\sum_{\widetilde{\cal H}\in (DRS(N)\cup DRS(N'))/_{\cal R}}
\sum_{\{j\mid M_j\in\widetilde{\cal H}\cap DRS(N)\}}\varphi_j=
\sum_{\widetilde{\cal H}\in (DRS(N)\cup DRS(N'))/_{\cal R}}\varphi_{{\cal H}\cap DRS(N)}$.

Thus, the aggregate linear equation system for $CTMC(N)$ is

$$\left\{
\begin{array}{l}
\sum_{\widetilde{\cal H}\in (DRS(N)\cup DRS(N'))/_{\cal R}}RM(\widetilde{\cal H},{\cal H})
\varphi_{\widetilde{\cal H}\cap DRS(N)}=0\\
\sum_{\widetilde{\cal H}\in (DRS(N)\cup DRS(N'))/_{\cal R}}\varphi_{{\cal H}\cap DRS(N)}=1
\end{array}
\right..$$

Let us denote the aggregate steady-state PMF for $CTMC(N')$ by $\varphi_{{\cal H}\cap DRS(N')}'=\sum_{\{j\mid
M_j'\in{\cal H}\cap DRS(N')\}}\varphi_j'$. Then, in a similar way, the aggregate linear equation system for $CTMC(N')$
is

$$\left\{
\begin{array}{l}
\sum_{\widetilde{\cal H}\in (DRS(N)\cup DRS(N'))/_{\cal R}}RM(\widetilde{\cal H},{\cal H})
\varphi_{\widetilde{\cal H}\cap DRS(N')}'=0\\
\sum_{\widetilde{\cal H}\in (DRS(N)\cup DRS(N'))/_{\cal R}}\varphi_{{\cal H}\cap DRS(N')}'=1
\end{array}
\right..$$

Let $(DRS(N)\cup DRS(N'))/_{\cal R}=\{{\cal H}_1,\ldots ,{\cal H}_l\}$. Then the aggregate steady-state PMFs
$\varphi_{{\cal H}_k\cap DRS(N)}$ and $\varphi_{{\cal H}_k\cap DRS(N')}'\ (1\leq k\leq l)$ satisfy the same aggregate
system of $l+1$ linear equations with $l$ independent equations and $l$ unknowns.
The aggregate linear equation system has a unique solution when a single aggregate steady-state PMF exists, which is
the case here. Hence, $\varphi_{{\cal H}_k\cap DRS(N)}=\varphi_{{\cal H}_k\cap DRS(N')}'\ (1\leq k\leq l)$.
\hfill$\eop$

Let $N$ be an LFSPN and $\varphi$ be the steady-state PMF for $CTMC(N)$. Let $\varphi_{\cal K},\ {\cal K}\in
DRS(N)/_{{\cal R}_{fl}}(N)$, be the elements of the steady-state PMF for $CTMC_{\bis_{fl}}(N)$, denoted by
$\varphi_{\bis_{fl}}$. By (the proof of) Proposition \ref{statprob.pro}, for all ${\cal K}\in DRS(N)/_{{\cal
R}_{fl}(N)}$ we have

$$\varphi_{\cal K}=\sum_{\{i\mid M_i\in{\cal K}\}}\varphi_i.$$

Let ${\bf V}$ be the collector matrix for the largest fluid autobisimulation ${\cal R}_{fl}(N)$ on $N$. One can see
that

$$\varphi{\bf V}=\varphi_{\bis_{fl}}.$$

We have $\left\{
\begin{array}{l}
\varphi{\bf Q}={\bf 0}\\
\varphi{\bf 1}^T=1
\end{array}
\right..$ After right-multiplying both sides of the first equation by ${\bf V}$ and since
${\bf V}{\bf 1}^T={\bf 1}^T$, we get $\left\{
\begin{array}{l}
\varphi{\bf Q}{\bf V}={\bf 0}\\
\varphi{\bf V}{\bf 1}^T=1
\end{array}
\right..$ Since ${\bf Q}{\bf V}={\bf V}{\bf Q}_{\bis_{fl}}$, we obtain
$\left\{
\begin{array}{l}
\varphi{\bf V}{\bf Q}_{\bis_{fl}}={\bf 0}\\
\varphi{\bf V}{\bf 1}^T=1
\end{array}
\right..$ Since $\varphi{\bf V}=\varphi_{\bis_{fl}}$, we conclude that $\varphi_{\bis_{fl}}$ is a solution of the
linear equation system

$$\left\{
\begin{array}{l}
\varphi_{\bis_{fl}}{\bf Q}_{\bis_{fl}}={\bf 0}\\
\varphi_{\bis_{fl}}{\bf 1}^T=1
\end{array}
\right..$$

Thus, the treatment of $CTMC_{\bis_{fl}}(N)$ instead of $CTMC(N)$ simplifies the analytical solution, since we have
less states, but constructing the TRM ${\bf Q}_{\bis_{fl}}$ for $CTMC_{\bis_{fl}}(N)$
also requires some efforts, including determining ${\cal R}_{fl}(N)$ and calculating the rates to move from one
equivalence class to another. The behaviour of $CTMC_{\bis_{fl}}(N)$ stabilizes quicker than that of $CTMC(N)$ (if each
of them has a single steady state), since ${\bf Q}_{\bis_{fl}}$ is denser matrix than ${\bf Q}$ (the TRM for $CTMC(N)$)
due to the fact that the former matrix is smaller and the transitions between the equivalence classes ``include'' all
the transitions between the discrete markings belonging to these equivalence classes.

The following proposition demonstrates that for two LFSPNs related by $\bis_{fl}$ their aggregate steady-state fluid
PDFs coincide for each equivalence class of discrete markings.

\begin{proposition}
Let $N,N'$ be LFSPNs with ${\cal R}:N\bis_{fl}N'$ and $F(x)=(F_1(x),\ldots ,F_n(x)),\ n=|DRS(N)|$, be the steady-state
fluid PDF for the SFM of $N$ and $F'(x)=(F_1'(x),\ldots ,F_m'(x)),\ m=|DRS(N')|$, be the steady-state fluid PDF for the
SFM of $N'$. Then for all ${\cal H}\in (DRS(N)\cup DRS(N'))/_{\cal R}$ we have

$$\sum_{\{i\mid M_i\in{\cal H}\cap DRS(N)\}}F_i(x)=\sum_{\{j\mid M_j'\in{\cal H}\cap DRS(N')\}}F_j'(x),\ x>0.$$

\label{statpdf.pro}
\end{proposition}
{\em Proof.} The ordinary differential equation characterizing the steady-state PDF for the SFM of $N$ is

$$\frac{dF(x)}{dx}{\bf R}=F(x){\bf Q},\ x>0.$$

The upper boundary constraint is
$F(\infty )=\varphi$, where $\varphi$ is the steady-state PMF for $CTMC(N)$.

Then for all $i\ (1\leq i\leq n)$ we have

$${\cal R}_{ii}\frac{dF_i(x)}{dx}=\sum_{j=1}^n{\cal Q}_{ji}F_j(x),\ x>0.$$

The upper boundary constraints are
$\forall i\ (1\leq i\leq n)\ F_i(\infty )=\varphi_i$, where $\varphi =(\varphi_1,\ldots ,\varphi_n)$ is the
steady-state PMF for $CTMC(N)$.

By definition of ${\cal R}_{ij}$ and ${\cal Q}_{ij}\ (1\leq i,j\leq n)$ we have

$$RP(M_i)\frac{dF_i(x)}{dx}=\sum_{j=1}^n RM(M_j,M_i)F_j(x),\ x>0.$$

Let ${\cal H}\in (DRS(N)\cup DRS(N'))/_{\cal R}$. We sum the left and right sides of the equation above for all $i$
such that $M_i\in{\cal H}\cap DRS(N)$. The resulting equation is

$$\sum_{\{i\mid M_i\in{\cal H}\cap DRS(N)\}}RP(M_i)\frac{dF_i(x)}{dx}=\sum_{\{i\mid M_i\in{\cal H}\cap DRS(N)\}}
\sum_{j=1}^n RM(M_j,M_i)F_j(x),\ x>0.$$

\noindent Let us denote the aggregate fluid flow PDF for the SFM of $N$ by $F_{{\cal H}\cap DRS(N)}(x)=\sum_{\{i\mid
M_i\in{\cal H}\cap DRS(N)\}}F_i(x)$. Then, by Remark 3
from Section \ref{flubiseq.sec}, for the left-hand side of the equation above, we get\\
$\sum_{\{i\mid M_i\in{\cal H}\cap DRS(N)\}}RP(M_i)\frac{dF_i(x)}{dx}=
\sum_{\{i\mid M_i\in{\cal H}\cap DRS(N)\}}RP({\cal H})\frac{dF_i(x)}{dx}=
RP({\cal H})\sum_{\{i\mid M_i\in{\cal H}\cap DRS(N)\}}\frac{dF_i(x)}{dx}=\\
RP({\cal H})\frac{d}{dx}\left(\sum_{\{i\mid M_i\in{\cal H}\cap DRS(N)\}}F_i(x)\right)=
RP({\cal H})\frac{dF_{{\cal H}\cap DRS(N)}(x)}{dx}$.

Analogously, for the right-hand side of the equation above, we get\\
$\sum_{\{i\mid M_i\in{\cal H}\cap DRS(N)\}}\sum_{j=1}^n RM(M_j,M_i)F_j(x)=
\sum_{j=1}^n F_j(x)\sum_{\{i\mid M_i\in{\cal H}\cap DRS(N)\}}RM(M_j,M_i)=\\
\sum_{j=1}^n RM(M_j,{\cal H})F_j(x)=
\sum_{\widetilde{\cal H}\in (DRS(N)\cup DRS(N'))/_{\cal R}}
\sum_{\{j\mid M_j\in\widetilde{\cal H}\cap DRS(N)\}}RM(M_j,{\cal H})F_j(x)=\\
\sum_{\widetilde{\cal H}\in (DRS(N)\cup DRS(N'))/_{\cal R}}
\sum_{\{j\mid M_j\in\widetilde{\cal H}\cap DRS(N)\}}RM(\widetilde{\cal H},{\cal H})F_j(x)=\\
\sum_{\widetilde{\cal H}\in (DRS(N)\cup DRS(N'))/_{\cal R}}RM(\widetilde{\cal H},{\cal H})
\sum_{\{j\mid M_j\in\widetilde{\cal H}\cap DRS(N)\}}F_j(x)=\\
\sum_{\widetilde{\cal H}\in (DRS(N)\cup DRS(N'))/_{\cal R}}RM(\widetilde{\cal H},{\cal H})
F_{\widetilde{\cal H}\cap DRS(N)}(x)$.

By combining both the resulting sides of the differential equation, we get the aggregate differential equation system
for the SFM of $N$:

$$RP({\cal H})\frac{dF_{{\cal H}\cap DRS(N)}(x)}{dx}=
\sum_{\widetilde{\cal H}\in (DRS(N)\cup DRS(N'))/_{\cal R}}RM(\widetilde{\cal H},{\cal H})
F_{\widetilde{\cal H}\cap DRS(N)}(x),\ x>0.$$

\noindent Let us denote the aggregate fluid flow PDF for the SFM of $N'$ by $F_{{\cal H}\cap DRS(N')}'(x)=\sum_{\{j\mid
M_j'\in{\cal H}\cap DRS(N')\}}F_j'(x)$. Then, in a similar way, we get the aggregate differential equation system for
the SFM of $N'$:

$$RP({\cal H})\frac{dF_{{\cal H}\cap DRS(N')}'(x)}{dx}=
\sum_{\widetilde{\cal H}\in (DRS(N)\cup DRS(N'))/_{\cal R}}RM(\widetilde{\cal H},{\cal H})
F_{\widetilde{\cal H}\cap DRS(N')}'(x),\ x>0.$$

By Proposition \ref{statprob.pro}, the upper boundary constraints associated with the aggregate differential equation
systems for the SFMs of $N$ and $N'$ coincide:
$F_{{\cal H}\cap DRS(N)}(\infty )=\sum_{\{i\mid M_i\in{\cal H}\cap DRS(N)\}}F_i(\infty )=\\
\sum_{\{i\mid M_i\in{\cal H}\cap DRS(N)\}}\varphi_i=\sum_{\{j\mid M_j'\in{\cal H}\cap DRS(N')\}}\varphi_i'=
\sum_{\{j\mid M_j'\in{\cal H}\cap DRS(N')\}}F_j'(\infty )=F_{{\cal H}\cap DRS(N')}'(\infty )$.

Let $(DRS(N)\cup DRS(N'))/_{\cal R}=\{{\cal H}_1,\ldots ,{\cal H}_l\}$. By analogy with the above results for ${\cal
H}\in (DRS(N)\cup DRS(N'))/_{\cal R}$, we can demonstrate that for each ${\cal H}_k\ (1\leq k\leq l)$ the aggregate
differential equation systems for the SFMs of $N$ and $N'$ and the associated upper boundary constraints coincide.

For each ${\cal H}_k\ (1\leq k\leq l)$, the lower boundary constraints are $\exists M_i\in{\cal H}_k\cap DRS(N)\
RP(M_i)>0\ \Rightarrow\ F_i(0)=0$ and $\exists M_j'\in{\cal H}_k\cap DRS(N')\ RP(M_j')>0\ \Rightarrow\ F_j'(0)=0$.
Since $\forall M_i\in{\cal H}_k\cap DRS(N)\ \forall M_j'\in{\cal H}_k\cap DRS(N')\ RP(M_i)=RP({\cal H}_k\cap
DRS(N))=RP({\cal H}_k)=RP({\cal H}_k\cap DRS(N'))=RP(M_j')$, we have $F_{{\cal H}_k\cap DRS(N)}(0)=0\ \Leftarrow\
RP({\cal H}_k)>0\ \Rightarrow\ F_{{\cal H}_k\cap DRS(N')}'(0)=0\ (1\leq k\leq l)$.

Then the aggregate fluid flow PDFs $F_{{\cal H}_k\cap DRS(N)}(x)$ and $F_{{\cal H}_k\cap DRS(N')}'(x)\ (1\leq k\leq l)$
satisfy the same aggregate system of $l$ differential equations with $l$ unknowns and the same upper and lower boundary
constraints. The spectral decomposition method, described in Section \ref{fspncont.sec}, provides such an aggregate
differential equation system with a unique solution. Hence, $F_{{\cal H}_k\cap DRS(N)}(x)=F_{{\cal H}_k\cap
DRS(N')}'(x)\ (1\leq k\leq l)$. \hfill$\eop$

Let $N$ be an LFSPN and $F(x)$ be the steady-state fluid PDF for the SFM of $N$. Let $F_{\cal K}(x),\ {\cal K}\in
DRS(N)/_{{\cal R}_{fl}(N)}$, be the elements of the steady-state fluid PDF for the quotient (by $\bis_{fl}$) SFM of
$N$, denoted by $F_{\bis_{fl}}(x)$. By (the proof of) Proposition \ref{statpdf.pro}, for all ${\cal K}\in
DRS(N)/_{{\cal R}_{fl}(N)}$ we have

$$F_{\cal K}(x)=\sum_{\{i\mid M_i\in{\cal K}\}}F_i(x),\ x>0.$$

Let ${\bf V}$ be the collector matrix for the largest fluid autobisimulation ${\cal R}_{fl}(N)$ on $N$. One can see
that

$$F(x){\bf V}=F_{\bis_{fl}}(x),\ x>0.$$

We have $\frac{dF(x)}{dx}{\bf R}=F(x){\bf Q},\ x>0$. After right-multiplying both sides of the above equation by ${\bf
V}$, we get $\frac{dF(x)}{dx}{\bf R}{\bf V}=F(x){\bf Q}{\bf V},\ x>0$. Since ${\bf R}{\bf V}={\bf V}{\bf
R}_{\bis_{fl}}$ and ${\bf Q}{\bf V}={\bf V}{\bf Q}_{\bis_{fl}}$, we obtain $\frac{dF(x)}{dx}{\bf V}{\bf
R}_{\bis_{fl}}=F(x){\bf V}{\bf Q}_{\bis_{fl}},\\
x>0$. By linearity of differentiation operator, we have $\frac{d}{dx}(F(x){\bf V}){\bf R}_{\bis_{fl}}=F(x){\bf V}{\bf
Q}_{\bis_{fl}},\ x>0$. Since $F(x){\bf V}=F_{\bis_{fl}}(x)$, we conclude that $F_{\bis_{fl}}(x)$ is a solution of the
system of ordinary differential equations

$$\frac{dF_{\bis_{fl}}(x)}{dx}{\bf R}_{\bis_{fl}}=F_{\bis_{fl}}(x){\bf Q}_{\bis_{fl}},\ x>0.$$

Thus, the treatment of the quotient (by $\bis_{fl}$) SFM of $N$ instead of SFM of $N$ simplifies the analytical
solution.

The following proposition demonstrates that for two LFSPNs related by $\bis_{fl}$ their aggregate steady-state fluid
probability density functions coincide for each equivalence class of discrete markings.

\begin{proposition}
Let $N,N'$ be LFSPNs with ${\cal R}:N\bis_{fl}N'$ and $f(x)=(f_1(x),\ldots ,f_n(x)),\ n=|DRS(N)|$, be the steady-state
fluid probability density function
for the SFM of $N$ and $f'(x)=(f_1'(x),\ldots ,f_m'(x)),\ m=|DRS(N')|$, be the steady-state fluid probability density
function
for the SFM of $N'$. Then for all ${\cal H}\in (DRS(N)\cup DRS(N'))/_{\cal R}$ we have

$$\sum_{\{i\mid M_i\in{\cal H}\cap DRS(N)\}}f_i(x)=\sum_{\{j\mid M_j'\in{\cal H}\cap DRS(N')\}}f_j'(x),\ x>0.$$

\label{statpdsf.pro}
\end{proposition}
{\em Proof.} Remember that $f_i(x)=\frac{dF_i(x)}{dx}\ (1\leq i\leq n)$ and $f_j'(x)=\frac{dF_j'(x)}{dx}\ (1\leq j\leq
m)$. Let ${\cal H}\in (DRS(N)\cup DRS(N'))/_{\cal R}$. By Proposition \ref{statpdf.pro}, we have

$$\sum_{\{i\mid M_i\in{\cal H}\cap DRS(N)\}}F_i(x)=\sum_{\{j\mid M_j'\in{\cal H}\cap DRS(N')\}}F_j'(x),\ x>0.$$

By differentiating both sides of this equation by $x$ and applying the property for differentiating a sum, we get

$$\sum_{\{i\mid M_i\in{\cal H}\cap DRS(N)\}}f_i(x)=\!\sum_{\{i\mid M_i\in{\cal H}\cap DRS(N)\}}\frac{dF_i(x)}{dx}=\!
\sum_{\{j\mid M_j'\in{\cal H}\cap DRS(N')\}}\frac{dF_j'(x)}{dx}=\!\sum_{\{j\mid M_j'\in{\cal H}\cap DRS(N')\}}f_j'(x),\
x>0.$$

\hfill$\eop$

Let $N$ be an LFSPN and $f(x)$ be the steady-state fluid probability density function for the SFM of $N$. Let $f_{\cal
K}(x),\ {\cal K}\in DRS(N)/_{{\cal R}_{fl}(N)}$, be the elements of the steady-state fluid probability density function
for the quotient (by $\bis_{fl}$) SFM of $N$, denoted by $f_{\bis_{fl}}(x)$. By (the proof of) Proposition
\ref{statpdsf.pro}, for all ${\cal K}\in DRS(N)/_{{\cal R}_{fl}(N)}$ we have

$$f_{\cal K}(x)=\sum_{\{i\mid M_i\in{\cal K}\}}f_i(x),\ x>0.$$

Let ${\bf V}$ be the collector matrix for the largest fluid autobisimulation ${\cal R}_{fl}(N)$ on $N$. One can see
that

$$f(x){\bf V}=f_{\bis_{fl}}(x),\ x>0.$$

We have $\frac{df(x)}{dx}{\bf R}=f(x){\bf Q},\ x>0$. Like it has been done after Proposition \ref{statpdf.pro}, we can
prove that $f_{\bis_{fl}}(x)$ is a solution of the system of ordinary differential equations

$$\frac{df_{\bis_{fl}}(x)}{dx}{\bf R}_{\bis_{fl}}=f_{\bis_{fl}}(x){\bf Q}_{\bis_{fl}},\ x>0.$$

Alternatively, we can use the fact $f(x)=\frac{dF(x)}{dx}$. Since $f(x){\bf V}=f_{\bis_{fl}}(x),\ x>0$, and $F(x){\bf
V}=F_{\bis_{fl}}(x),\\
x>0$, by linearity of differentiation operator, we have $f_{\bis_{fl}}(x)=f(x){\bf V}=\frac{dF(x)}{dx}{\bf
V}=\frac{d}{dx}(F(x){\bf V})=\frac{dF_{\bis_{fl}}(x)}{dx}$.

We also have $\frac{dF_{\bis_{fl}}(x)}{dx}{\bf R}_{\bis_{fl}}=F_{\bis_{fl}}(x){\bf Q}_{\bis_{fl}},\ x>0$. Since
$f_{\bis_{fl}}(x)=\frac{dF_{\bis_{fl}}(x)}{dx}$, by differentiating both sides of the previous equation, we get
$\frac{d}{dx}\left(f_{\bis_{fl}}(x){\bf R}_{\bis_{fl}}\right)=\frac{d}{dx}\left(F_{\bis_{fl}}(x){\bf
Q}_{\bis_{fl}}\right),\ x>0$. By linearity of differentiation operator and since
$f_{\bis_{fl}}(x)=\frac{dF_{\bis_{fl}}(x)}{dx}$, we conclude that $f_{\bis_{fl}}(x)$ is a solution of the system of
ordinary differential equations

$$\frac{df_{\bis_{fl}}(x)}{dx}{\bf R}_{\bis_{fl}}=f_{\bis_{fl}}(x){\bf Q}_{\bis_{fl}},\ x>0.$$

The following proposition demonstrates that for two LFSPNs related by $\bis_{fl}$ their aggregate steady-state buffer
empty probabilities coincide for each equivalence class of discrete markings.

\begin{proposition}
Let $N,N'$ be LFSPNs with ${\cal R}:N\bis_{fl}N'$ and $\ell =(\ell_1,\ldots ,\ell_n),\ n=|DRS(N)|$, be the steady-state
buffer empty probability for the SFM of $N$ and $\ell '(x)=(\ell_1',\ldots ,\ell_m'),\ m=|DRS(N')|$, be the
steady-state buffer empty probability for the SFM of $N'$. Then for all ${\cal H}\in (DRS(N)\cup DRS(N'))/_{\cal R}$ we
have

$$\sum_{\{i\mid M_i\in{\cal H}\cap DRS(N)\}}\ell_i=\sum_{\{j\mid M_j'\in{\cal H}\cap DRS(N')\}}\ell_j'.$$

\label{statbufemp.pro}
\end{proposition}
{\em Proof.} Remember that by the total probability law for the stationary behaviour for the SFM of $N$, we have

$$\ell =\varphi -\int_{0+}^{\infty}f(x)dx.$$

Then for all $i\ (1\leq i\leq n)$ we have

$$\ell_i=\varphi_i-\int_{0+}^{\infty}f_i(x)dx.$$

Let ${\cal H}\in (DRS(N)\cup DRS(N'))/_{\cal R}$. We sum the left and right sides of the equation above for all $i$ such
that $M_i\in{\cal H}\cap DRS(N)$. The resulting equation is

$$\sum_{\{i\mid M_i\in{\cal H}\cap DRS(N)\}}\ell_i=\sum_{\{i\mid M_i\in{\cal H}\cap DRS(N)\}}\varphi_i-
\sum_{\{i\mid M_i\in{\cal H}\cap DRS(N)\}}\int_{0+}^{\infty}f_i(x)dx.$$

Consider the right-hand side of the equation above. We apply to it the property for integrating a sum, then Proposition
\ref{statprob.pro} and Proposition \ref{statpdsf.pro}, finally, the total probability law for the stationary behaviour
for the SFM of $N$. Then we get
$\sum_{\{i\mid M_i\in{\cal H}\cap DRS(N)\}}\ell_i=
\sum_{\{i\mid M_i\in{\cal H}\cap DRS(N)\}}\varphi_i-
\sum_{\{i\mid M_i\in{\cal H}\cap DRS(N)\}}\int_{0+}^{\infty}f_i(x)dx=\\
\sum_{\{i\mid M_i\in{\cal H}\cap DRS(N)\}}\varphi_i-
\int_{0+}^{\infty}\sum_{\{i\mid M_i\in{\cal H}\cap DRS(N)\}}f_i(x)dx=
\sum_{\{j\mid M_j'\in{\cal H}\cap DRS(N')\}}\varphi_i'-\\
\int_{0+}^{\infty}\sum_{\{j\mid M_j'\in{\cal H}\cap DRS(N')\}}f_i'(x)dx=
\sum_{\{j\mid M_j'\in{\cal H}\cap DRS(N')\}}\varphi_i'-
\sum_{\{j\mid M_j'\in{\cal H}\cap DRS(N')\}}\int_{0+}^{\infty}f_i'(x)dx=\\
\sum_{\{j\mid M_j'\in{\cal H}\cap DRS(N')\}}\ell_j'$. \hfill$\eop$

Let $N$ be an LFSPN and $\ell$ be the steady-state buffer empty probability for the SFM of $N$. Let $\ell_{\cal K},\
{\cal K}\in DRS(N)/_{{\cal R}_{fl}(N)}$, be the elements of the steady-state buffer empty probability for the quotient
(by $\bis_{fl}$) SFM of $N$, denoted by $\ell_{\bis_{fl}}$. By (the proof of) Proposition \ref{statbufemp.pro}, for all
${\cal K}\in DRS(N)/_{{\cal R}_{fl}(N)}$ we have

$$\ell_{\cal K}=\sum_{\{i\mid M_i\in{\cal K}\}}\ell_i.$$

Let ${\bf V}$ be the collector matrix for the largest fluid autobisimulation ${\cal R}_{fl}(N)$ on $N$. One can see
that

$$\ell{\bf V}=\ell_{\bis_{fl}}.$$

We have $\ell =\varphi -\int_{0+}^{\infty}f(x)dx$. After right-multiplying both sides of the equation by ${\bf V}$, we
get $\ell{\bf V}=\varphi{\bf V}-\left(\int_{0+}^{\infty}f(x)dx\right){\bf V}$. Since $\ell{\bf V}=\ell_{\bis_{fl}}$ and
$\varphi{\bf V}=\varphi_{\bis_{fl}}$, by linearity of integration operator, we obtain
$\ell_{\bis_{fl}}=\varphi_{\bis_{fl}}-\int_{0+}^{\infty}f(x){\bf V}dx$. Since $f(x){\bf V}=f_{\bis_{fl}}(x),\ x>0$, we
conclude that $\ell_{\bis_{fl}}$ is a solution of the linear equation system

$$\ell_{\bis_{fl}}=\varphi_{\bis_{fl}}-\int_{0+}^{\infty}f_{\bis_{fl}}(x)dx.$$

Thus, the proposed quotients of the probability functions describe the behaviour of the quotient (by $\bis_{fl}$)
associated SFMs of LFSPNs.

\begin{example}
\label{fluquaeq.exm}
Consider the LFSPNs $N$ and $N'$ from Figure \ref{flblfspn.fig}, for which it holds $N\bis_{fl}N'$.

We have $DRS^-(N)=\{M_2\},\ DRS^0(N)=\emptyset$ and $DRS^+(N)=\{M_1\}$.

The steady-state PMF for $CTMC(N)$ is

$$\varphi =\left(\frac{1}{2},\frac{1}{2}\right).$$

Then the {\em stability condition} for the SFM of $N$ is fulfilled: $FluidFlow(q)=\sum_{i=1}^2\varphi_i RP(M_i)=
\frac{1}{2}\cdot 1+\frac{1}{2}(-2)=-\frac{1}{2}<0$.

For each eigenvalue $\gamma$ we must have $|\gamma{\bf R}-{\bf Q}|=
\left|\begin{array}{cc}
\gamma +2 & -2\\
-2 & -2\gamma +2
\end{array}\right|=-2\gamma (1+\gamma )=0$; hence, $\gamma_1=0$ and $\gamma_2=-1$.

The corresponding eigenvectors are the solutions of

$$\begin{array}{cc}
v_1\left(\begin{array}{cc}
2 & -2\\
-2 & 2
\end{array}\right)=0,
&
v_2\left(\begin{array}{cc}
1 & -2\\
-2 & 4
\end{array}\right)=0.
\end{array}$$

Then the (normalized) eigenvectors are $v_1=\left(\frac{1}{2},\frac{1}{2}\right)$ and
$v_2=\left(\frac{2}{3},\frac{1}{3}\right)$.

Since $\varphi =F(\infty )=a_1 v_1$, we have $F(x)=\varphi +a_2 e^{\gamma_2 x}v_2$ and $a_1=1$. Since $\forall M_l\in
DRS^+(N)\\
F_l(0)=\varphi_l+a_2 v_{2l}=0$ and $DRS^+(N)=\{M_1\}$, we have $\varphi_1+a_2 v_{21}=\frac{1}{2}+a_2\frac{2}{3}=0$;
hence, $a_2=-\frac{3}{4}$.

Then the steady-state fluid PDF for the SFM of $N$ is

$$F(x)=\left(\frac{1}{2}-\frac{1}{2}e^{-x},\frac{1}{2}-\frac{1}{4}e^{-x}\right).$$

The steady-state fluid probability density function for the SFM of $N$ is

$$f(x)=\frac{dF(x)}{dx}=\left(\frac{1}{2}e^{-x},\frac{1}{4}e^{-x}\right).$$

The steady-state buffer empty probability for the SFM of $N$ is

$$\ell =F(0)=\left(0,\frac{1}{4}\right).$$

We have $DRS^-(N')=\{M_2',M_3'\},\ DRS^0(N')=\emptyset$ and $DRS^+(N')=\{M_1'\}$.

The steady-state PMF for $CTMC(N')$ is

$$\varphi '=\left(\frac{1}{2},\frac{1}{4},\frac{1}{4}\right).$$

Then the {\em stability condition} for the SFM of $N'$ is fulfilled: $FluidFlow(q')=\sum_{j=1}^3\varphi_j'RP(M_j')=
\frac{1}{2}\cdot 1+\frac{1}{4}(-2)+\frac{1}{4}(-2)=-\frac{1}{2}<0$.

For each eigenvalue $\gamma '$ we must have $|\gamma '{\bf R}'-{\bf Q}'|=
\left|\begin{array}{ccc}
\gamma '+2 & -1 & -1\\
-2 & -2\gamma '+2 & 0\\
-2 & 0 & -2\gamma '+2
\end{array}\right|=-2\gamma '(1+\gamma ')(1-\gamma ')=0$; hence, $\gamma_1'=0,\ \gamma_2'=-1$ and $\gamma_3'=1$.

By the boundedness condition, the positive eigenvalue $\gamma_3'$ and the corresponding eigenvector $v_3'$ should be
excluded from the solution.

The remaining corresponding eigenvectors are the solutions of

$$\begin{array}{cc}
v_1'\left(\begin{array}{ccc}
2 & -1 & -1\\
-2 & 2 & 0\\
-2 & 0 & 2
\end{array}\right)=0,
&
v_2'\left(\begin{array}{ccc}
1 & -1 & -1\\
-2 & 4 & 0\\
-2 & 0 & 4
\end{array}\right)=0.
\end{array}$$

Then the remaining (normalized) eigenvectors are $v_1'=\left(\frac{1}{2},\frac{1}{4},\frac{1}{4}\right)$ and
$v_2'=\left(\frac{2}{3},\frac{1}{6},\frac{1}{6}\right)$.

Since $\varphi '=F'(\infty )=a_1'v_1'$, we have $F'(x)=\varphi '+a_2'e^{\gamma_2'x}v_2'$ and $a_1'=1$. Since
$\forall M_l'\in DRS^+(N')\\
F_l'(0)=\varphi_l'+a_2'v_{2l}'=0$ and $DRS^+(N')=\{M_1'\}$, we have $\varphi_1'+a_2'v_{21}'=\frac{1}{2}+
a_2'\frac{2}{3}=0$; hence, $a_2=-\frac{3}{4}$.

Then the steady-state fluid PDF for the SFM of $N'$ is

$$F'(x)=\left(\frac{1}{2}-\frac{1}{2}e^{-x},\frac{1}{4}-\frac{1}{8}e^{-x},\frac{1}{4}-\frac{1}{8}e^{-x}\right).$$

The steady-state fluid probability density function for the SFM of $N'$ is

$$f'(x)=\frac{dF'(x)}{dx}=\left(\frac{1}{2}e^{-x},\frac{1}{8}e^{-x},\frac{1}{8}e^{-x}\right).$$

The steady-state buffer empty probability for the SFM of $N'$ is

$$\ell '=F'(0)=\left(0,\frac{1}{8},\frac{1}{8}\right).$$

In Figure \ref{flupdfs.fig}, the plots of the elements $F_1,F_2,F_2'$ of the steady-state fluid PDFs $F=(F_1,F_2)$ and
$F'=(F_1',F_2',F_3')$ for the SFMs of $N$ and $N'$ as functions of $x$ are depicted. It is sufficient to consider the
functions $F_1(x)=\frac{1}{2}-\frac{1}{2}e^{-x},\ F_2(x)=\frac{1}{2}-\frac{1}{4}e^{-x},\
F_2'(x)=\frac{1}{4}-\frac{1}{8}e^{-x}$ only, since $F_1=F_1'$ and $F_2'=F_3'$.

\begin{figure}
\begin{center}
\includegraphics[width=\textwidth]{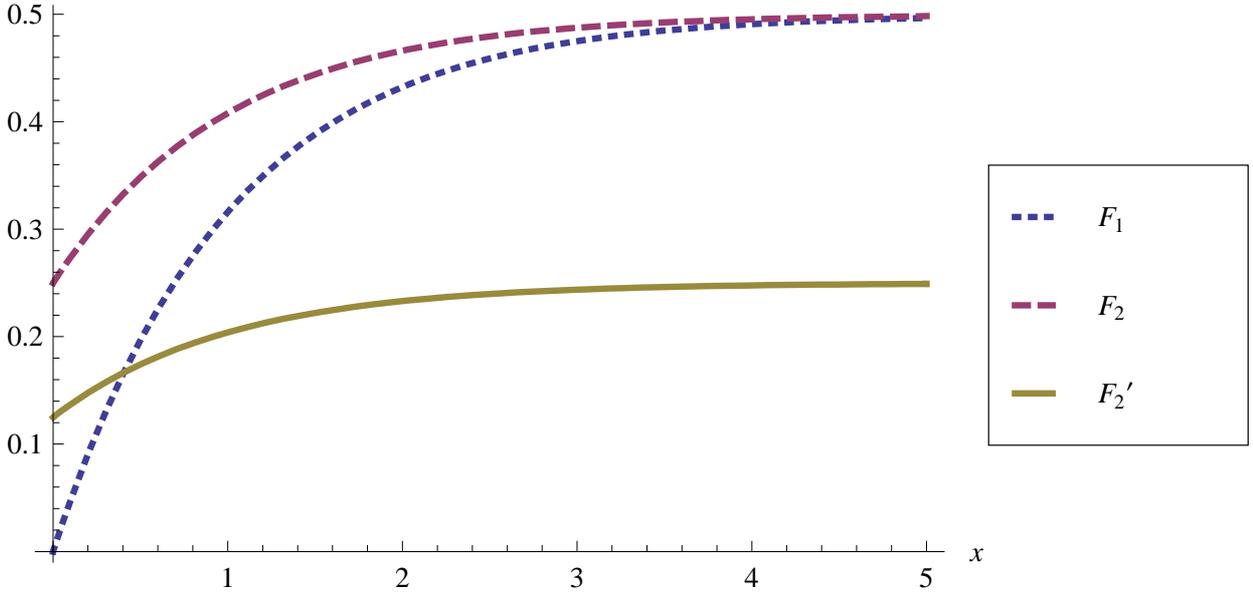}
\end{center}
\vspace{-10mm}
\caption{The elements of the steady-state fluid PDFs for the SFMs of $N$ and $N'$ as functions of $x$}
\label{flupdfs.fig}
\end{figure}

We have $(DRS(N)\cup DRS(N'))/_{{\cal R}_{fl}(N,N')}=\{{\cal H}_1,{\cal H}_2\}$, where ${\cal H}_1=\{M_1,M_1'\}$ and
${\cal H}_2=\{M_2,M_2',M_3'\}$.

First, consider the equivalence class
${\cal H}_1$.
\begin{itemize}

\item The aggregate steady-state probabilities for ${\cal H}_1$ coincide: $\varphi_{{\cal H}_1\cap DRS(N)}=
\sum_{\{i\mid M_i\in{\cal H}_1\cap DRS(N)\}}\varphi_i=\varphi_1=\frac{1}{2}=\varphi_1'=
\sum_{\{j\mid M_j'\in{\cal H}_1\cap DRS(N')\}}\varphi_j'=\varphi_{{\cal H}_1\cap DRS(N')}'$.

\item The aggregate steady-state buffer empty probabilities for ${\cal H}_1$ coincide: $\ell_{{\cal H}_1\cap
DRS(N)}=\\
\sum_{\{i\mid M_i\in{\cal H}_1\cap DRS(N)\}}\ell_i=\ell_1=0=\ell_1'=
\sum_{\{j\mid M_j'\in{\cal H}_1\cap DRS(N')\}}\ell_j'=\ell_{{\cal H}_1\cap DRS(N')}'$.

\item The aggregate steady-state fluid PDFs for ${\cal H}_1$ coincide: $F_{{\cal H}_1\cap DRS(N)}(x)=
\sum_{\{i\mid M_i\in{\cal H}_1\cap DRS(N)\}}F_i(x)=F_1(x)=\frac{1}{2}-\frac{1}{2}e^{-x}=F_1'(x)=
\sum_{\{j\mid M_j'\in{\cal H}_1\cap DRS(N')\}}F_j'(x)=F_{{\cal H}_1\cap DRS(N')}'(x)$, where $x>0$.

\item The aggregate steady-state fluid probability density functions for ${\cal H}_1$ coincide: $f_{{\cal H}_1\cap
DRS(N)}(x)=\\
\sum_{\{i\mid M_i\in{\cal H}_1\cap DRS(N)\}}f_i(x)=f_1(x)=\frac{1}{2}e^{-x}=f_1'(x)=
\sum_{\{j\mid M_j'\in{\cal H}_1\cap DRS(N')\}}f_j'(x)=f_{{\cal H}_1\cap DRS(N')}'(x)$, where $x>0$.

\end{itemize}

Second, consider the equivalence class
${\cal H}_2$.
\begin{itemize}

\item The aggregate steady-state probabilities for ${\cal H}_2$ coincide: $\varphi_{{\cal H}_2\cap
DRS(N)}=\sum_{\{i\mid M_i\in{\cal H}_2\cap DRS(N)\}}\varphi_i=\varphi_2=\\
\frac{1}{2}=\frac{1}{4}+\frac{1}{4}=\varphi_2'+\varphi_3'=\sum_{\{j\mid M_j'\in{\cal H}_2\cap DRS(N')\}}\varphi_j'=
\varphi_{{\cal H}_2\cap DRS(N')}'$.

\item The aggregate steady-state buffer empty probabilities for ${\cal H}_2$ coincide: $\ell_{{\cal H}_2\cap
DRS(N)}=\\
\sum_{\{i\mid M_i\in{\cal H}_2\cap DRS(N)\}}\ell_i=\ell_2=\frac{1}{4}=\frac{1}{8}+\frac{1}{8}=\ell_2'+\ell_3'=
\sum_{\{j\mid M_j'\in{\cal H}_2\cap DRS(N')\}}\ell_j'=\ell_{{\cal H}_2\cap DRS(N')}'$.

\item The aggregate steady-state fluid PDFs for ${\cal H}_2$ coincide: $F_{{\cal H}_2\cap DRS(N)}(x)=
\sum_{\{i\mid M_i\in{\cal H}_2\cap DRS(N)\}}F_i(x)=\\
F_2(x)=\frac{1}{2}-\frac{1}{4}e^{-x}=\frac{1}{4}-\frac{1}{8}e^{-x}+\frac{1}{4}-\frac{1}{8}e^{-x}=F_2'(x)+F_3'(x)=
\sum_{\{j\mid M_j'\in{\cal H}_2\cap DRS(N')\}}F_j'(x)=F_{{\cal H}_2\cap DRS(N')}'(x)$, where $x>0$.

\item The aggregate steady-state fluid probability density functions for ${\cal H}_2$ coincide: $f_{{\cal H}_2\cap
DRS(N)}(x)=\\
\sum_{\{i\mid M_i\in{\cal H}_2\cap DRS(N)\}}f_i(x)=f_2(x)=\frac{1}{4}e^{-x}=\frac{1}{8}e^{-x}+
\frac{1}{8}e^{-x}=f_2'(x)+f_3'(x)=\sum_{\{j\mid M_j'\in{\cal H}_2\cap DRS(N')\}}f_j'(x)=f_{{\cal H}_2\cap
DRS(N')}'(x)$, where $x>0$.

\end{itemize}

One can also see that $\varphi_{\bis_{fl}}=\varphi_{\bis_{fl}}'=\varphi ,\ \ell_{\bis_{fl}}=\ell_{\bis_{fl}}'=\ell ,\
F_{\bis_{fl}}(x)=F_{\bis_{fl}}'(x)=F(x),\ x>0$, and $f_{\bis_{fl}}(x)=f_{\bis_{fl}}'(x)=f(x),\ x>0$.

\end{example}

\section{Preservation of the functionality and performance}
\label{funcperf.sec}

In this section we
demonstrate how fluid bisimulation equivalence preserves
the functionality and performance of the equivalent LFSPNs.

Consider the LFSPNs $N$ and $N'$ from Figure \ref{flblfspn.fig}, for which it holds $N\bis_{fl}N'$.

Many steady-state hybrid performance indices may be aggregated to make them consistent with fluid bisimulation, as well
as with the quotienting of the discrete reachability graphs and underlying CTMCs, and with the induced lumping of the
discrete markings into the equivalence classes. Thus, the aggregate (up to $\bis_{fl}$) steady-state performance
measures of $N$ based on the probability functions
$\varphi ,\ell ,F(x)$ and $f(x)$ should coincide with those of $N'$ based on
$\varphi ',\ell ',F'(x)$ and $f'(x)$, respectively. Let us check this for the equivalence class ${\cal H}_2$.
\begin{itemize}

%
%
\item The {\em aggregate fraction (proportion) of time spent in the set of discrete markings} ${\cal H}_2\cap DRS(N)$
is\\
$TimeFract({\cal H}_2\cap DRS(N))=\sum_{\{i\mid M_i\in{\cal H}_2\cap DRS(N)\}}\varphi_i=\varphi_2=\frac{1}{2}$.

The {\em aggregate fraction (proportion) of time spent in the set of discrete markings} ${\cal H}_2\cap DRS(N')$ is\\
$TimeFract({\cal H}_2\cap DRS(N'))=\sum_{\{j\mid M_j'\in{\cal H}_2\cap DRS(N')\}}\varphi_i'=\varphi_2'+\varphi_3'=
\frac{1}{4}+\frac{1}{4}=\frac{1}{2}$.

\item The {\em aggregate firing frequency (throughput) of the transitions} enabled in the discrete markings from\\
${\cal H}_2\cap DRS(N)$ is $FiringFreq_{{\cal H}_2\cap DRS(N)}=
\sum_{t\in T_N}FiringFreq_{{\cal H}_2\cap DRS(N)}(t)=\\
\sum_{t\in T_N}\sum_{\{i\mid t\in Ena(M_i),\ M_i\in{\cal H}_2\cap DRS(N)\}}\varphi_i\Omega_N(t,M_i)=
\varphi_2\Omega_N(t_2,M_2)+\varphi_2\Omega_N(t_3,M_2)=\frac{1}{2}\cdot 1+\frac{1}{2}\cdot 1=\frac{1}{2}+\frac{1}{2}=1$.

The {\em aggregate firing frequency (throughput) of the transitions} enabled in the discrete markings from\\
${\cal H}_2\cap DRS(N')$ is $FiringFreq_{{\cal H}_2\cap DRS(N')}=
\sum_{t'\in T_{N'}}FiringFreq_{{\cal H}_2\cap DRS(N')}(t')=\\
\sum_{t'\in T_{N'}}\sum_{\{j\mid t'\in Ena(M_j'),\ M_j'\in{\cal H}_2\cap DRS(N')\}}\varphi_j'\Omega_{N'}(t',M_j')=
\varphi_2'\Omega_{N'}(t_3',M_2')+\varphi_3'\Omega_{N'}(t_4',M_3')=\frac{1}{4}\cdot 2+\frac{1}{4}\cdot 2=
\frac{1}{2}+\frac{1}{2}=1$.

\item The {\em aggregate exit frequency of the discrete markings} from ${\cal H}_2\cap DRS(N)$ is
$ExitFreq({\cal H}_2\cap DRS(N))=\frac{\sum_{\{i\mid M_i\in{\cal H}_2\cap DRS(N)\}}\varphi_i}{SJ({\cal H}_2\cap
DRS(N))}=\frac{\varphi_2}{SJ(M_2)}=\frac{1}{2}\cdot\frac{2}{1}=1$.

The {\em aggregate exit frequency of the discrete markings} from ${\cal H}_2\cap DRS(N')$ is
$ExitFreq({\cal H}_2\cap DRS(N'))=\frac{\sum_{\{j\mid M_j'\in{\cal H}_2\cap DRS(N')\}}\varphi_j'}{SJ({\cal H}_2\cap
DRS(N'))}=\frac{\varphi_2'+\varphi_3'}{SJ(M_2')}=\frac{\varphi_2'+\varphi_3'}{SJ(M_3')}=
\left(\frac{1}{4}+\frac{1}{4}\right)\frac{2}{1}=1$.

\item The {\em aggregate mean potential fluid flow rate for the continuous place} $q$ in the discrete markings from
${\cal H}_2\cap DRS(N)$ is $FluidFlow_{{\cal H}_2\cap DRS(N)}(q)=
\sum_{\{i\mid M_i\in{\cal H}_2\cap DRS(N)\}}\varphi_i RP({\cal H}_2\cap DRS(N))=\varphi_2 RP(M_2)=\frac{1}{2}(-2)=
-1$.

The {\em aggregate mean potential fluid flow rate for the continuous place} $q$ in the discrete markings from
${\cal H}_2\cap DRS(N')$ is $FluidFlow_{{\cal H}_2\cap DRS(N')}(q)=
\sum_{\{j\mid M_j'\in{\cal H}_2\cap DRS(N')\}}\varphi_j'RP({\cal H}_2\cap DRS(N'))=(\varphi_2'+\varphi_3')RP(M_2')=
(\varphi_2'+\varphi_3')RP(M_3')=\left(\frac{1}{4}+\frac{1}{4}\right)(-2)=-1$.

\item The {\em aggregate traversal frequency of the move from the discrete markings} from ${\cal H}_2\cap DRS(N)$ {\em
to the discrete markings} from ${\cal H}_1\cap DRS(N)$ is $TravFreq({\cal H}_2\cap DRS(N),{\cal H}_1\cap DRS(N))=\\
\sum_{\{i\mid M_i\in{\cal H}_2\cap DRS(N)\}}\varphi_i RM({\cal H}_2\cap DRS(N),{\cal H}_1\cap DRS(N))=
\varphi_2 RM(M_2,M_1)=\frac{1}{2}\cdot 2=1$.

The {\em aggregate traversal frequency of the move from the discrete markings} from ${\cal H}_2\cap DRS(N')$ {\em to
the discrete markings} from ${\cal H}_1\cap DRS(N')$ is $TravFreq({\cal H}_2\cap DRS(N'),{\cal H}_1\cap DRS(N'))=\\
\sum_{\{j\mid M_j'\in{\cal H}_2\cap DRS(N')\}}\varphi_j'RM({\cal H}_2\cap DRS(N'),{\cal H}_1\cap DRS(N'))=
(\varphi_2'+\varphi_3')RM(M_2',M_1')=\\
(\varphi_2'+\varphi_3')RM(M_3',M_1')=\left(\frac{1}{4}+\frac{1}{4}\right)2=1$.

\item The {\em aggregate probability
of the
positive fluid level in the continuous place} $q$ in the discrete markings from ${\cal H}_2\cap DRS(N)$ is
$FluidLevel_{{\cal H}_2\cap DRS(N)}(q)=\sum_{\{i\mid M_i\in{\cal H}_2\cap DRS(N)\}}(\varphi_i-\ell_i)=\varphi_2-\ell_2=
\frac{1}{2}-\frac{1}{4}=\frac{1}{4}$.

The {\em aggregate probability
of the
positive fluid level in the continuous place} $q'$ in the discrete markings from ${\cal H}_2\cap DRS(N')$ is
$FluidLevel_{{\cal H}_2\cap DRS(N')}(q')=\sum_{\{j\mid M_j'\in{\cal H}_2\cap DRS(N')\}}(\varphi_j'-\ell_j')=
(\varphi_2'-\ell_2')+(\varphi_3'-\ell_3')=\left(\frac{1}{4}-\frac{1}{8}\right)+\left(\frac{1}{4}-\frac{1}{8}\right)=
\frac{1}{8}+\frac{1}{8}=\frac{1}{4}$.

\end{itemize}

The following aggregate steady-state performance measures of $N$ do not coincide with those of $N'$ for the equivalence
class ${\cal H}_2$, since this index is based on the flow rates of continuous arcs {\em from} or {\em to} a continuous
place. However, fluid bisimulation equivalence respects only the {\em total difference between} the flow rates of all
the continuous arcs {\em from} a continuous place and the flow rates of all continuous arcs {\em to} the continuous
place, and this difference is calculated only for a {\em single discrete marking} among several bisimilar ones.
Nevertheless, we present these performance indices below with a goal to illustrate their calculation.
\begin{itemize}

\item The {\em aggregate mean proportional flow rate across the continuous arcs} from the continuous place $q$ to the
transitions enabled in the discrete markings from ${\cal H}_2\cap DRS(N)$ is\\
$FluidFlowOut_{{\cal H}_2\cap DRS(N)}(q)=\sum_{t\in T_N}FluidFlow_{{\cal H}_2\cap DRS(N)}(q,t)=\\
\sum_{t\in T_N}\sum_{\{i\mid t\in Ena(M_i),\ M_i\in{\cal H}_2\cap DRS(N)\}}
\left(\ell_i\left(\frac{\sum_{u\in Ena(M)}R_N((u,q),M)}{\sum_{v\in Ena(M)}R_N((q,v),M)}-1\right)+
\varphi_i\right)R_N((q,t),M)=\\
\left(\ell_2\left(\frac{\sum_{u\in Ena(M_2)}R_N((u,q),M_2)}{\sum_{v\in Ena(M_2)}R_N((q,v),M_2)}-1\right)+
\varphi_2\right)R_N((q,t_2),M_2)+\\
\left(\ell_2\left(\frac{\sum_{u\in Ena(M_2)}R_N((u,q),M_2)}{\sum_{v\in Ena(M_2)}R_N((q,v),M_2)}-1\right)+
\varphi_2\right)R_N((q,t_3),M_2)$.

We have $\frac{\sum_{u\in Ena(M_2)}R_N((u,q),M_2)}{\sum_{v\in Ena(M_2)}R_N((q,v),M_2)}-1=
\frac{R_N((t_2,q),M_2)+R_N((t_3,q),M_2)}{R_N((q,t_2),M_2)+R_N((q,t_3),M_2)}-1=\frac{1+2}{2+3}-1=-\frac{2}{5}$.
%

Thus, $\left(\ell_2\left(\frac{\sum_{u\in Ena(M_2)}R_N((u,q),M_2)}{\sum_{v\in Ena(M_2)}R_N((q,v),M_2)}-1\right)+
\varphi_2\right)R_N((q,t_2),M_2)+\\
\left(\ell_2\left(\frac{\sum_{u\in Ena(M_2)}R_N((u,q),M_2)}{\sum_{v\in Ena(M_2)}R_N((q,v),M_2)}-1\right)+
\varphi_2\right)R_N((q,t_3),M_2)=\left(\frac{1}{4}\left(-\frac{2}{5}\right)+\frac{1}{2}\right)2+
\left(\frac{1}{4}\left(-\frac{2}{5}\right)+\frac{1}{2}\right)3=\frac{4}{5}+\frac{6}{5}=2$.
\item The {\em aggregate mean proportional flow rate across the continuous arcs} to the continuous place $q$ from the
transitions enabled in the discrete markings from ${\cal H}_2\cap DRS(N)$ is\\
$FluidFlowIn_{{\cal H}_2\cap DRS(N)}(q)=\sum_{t\in T_N}FluidFlow_{{\cal H}_2\cap DRS(N)}(t,q)=\\
\sum_{t\in T_N}\sum_{\{i\mid t\in Ena(M_i),\ M_i\in{\cal H}_2\cap DRS(N)\}}
\left(\ell_i\left(\frac{\sum_{v\in Ena(M)}R_N((q,v),M)}{\sum_{u\in Ena(M)}R_N((u,q),M)}-1\right)+
\varphi_i\right)R_N(((t,q),M)=\\
\left(\ell_2\left(\frac{\sum_{v\in Ena(M_2)}R_N((q,v),M_2)}{\sum_{u\in Ena(M_2)}R_N((u,q),M_2)}-1\right)+
\varphi_2\right)R_N((t_2,q),M_2)+\\
\left(\ell_2\left(\frac{\sum_{v\in Ena(M_2)}R_N((q,v),M_2)}{\sum_{u\in Ena(M_2)}R_N((u,q),M_2)}-1\right)+
\varphi_2\right)R_N((t_3,q),M_2)$.

We have $\frac{\sum_{v\in Ena(M_2)}R_N((q,v),M_2)}{\sum_{u\in Ena(M_2)}R_N((u,q),M_2)}-1=
\frac{R_N((q,t_2),M_2)+R_N((q,t_3),M_2)}{R_N((t_2,q),M_2)+R_N((t_3,q),M_2)}-1=\frac{2+3}{1+2}-1=\frac{2}{3}$.
%

Thus, $\left(\ell_2\left(\frac{\sum_{v\in Ena(M_2)}R_N((q,v),M_2)}{\sum_{u\in Ena(M_2)}R_N((u,q),M_2)}-1\right)+
\varphi_2\right)R_N((t_2,q),M_2)+\\
\left(\ell_2\left(\frac{\sum_{v\in Ena(M_2)}R_N((q,v),M_2)}{\sum_{u\in Ena(M_2)}R_N((u,q),M_2)}-1\right)+
\varphi_2\right)R_N((t_3,q),M_2)=\left(\frac{1}{4}\cdot\frac{2}{3}+\frac{1}{2}\right)1+
\left(\frac{1}{4}\cdot\frac{2}{3}+\frac{1}{2}\right)2=\frac{2}{3}+\frac{4}{3}=2$.
\end{itemize}

\section{Document preparation system}
\label{docprsys.sec}

\begin{figure}
\begin{center}
\includegraphics{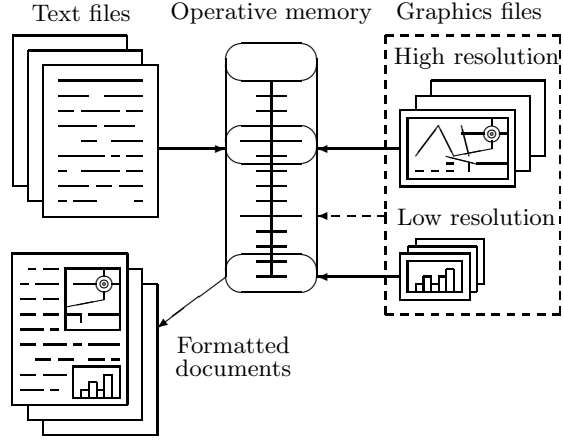}
\end{center}
\vspace{-7mm}
\caption{The diagram of the document preparation system}
\label{dpsdiagr.fig}
\end{figure}

Let us consider an application example describing three different models of a document preparation system. The system
receives (in an arbitrary order or in parallel) the collections of the text and graphics files as its inputs and writes
them into the operative memory of a computer. The system then reads the (mixed)
data from there and produces properly formatted output documents consisting of text and images. In general, it is
supposed that the text file collections
are transferred into the operative memory slower, but for longer time than the graphics ones. In detail, the low
resolution graphics is transferred into the operative memory with the same speed as the high resolution one, but it
takes less time than for the latter. The data from the operative memory is consumed for processing
quicker, but for shorter time than the input file collections of any type. The operative memory capacity is supposed to
be unlimited (for example, there exist some special mechanisms to ensure that the memory upper boundary can always be
increased, such as using the page file, stored on a hard drive of the computer). Clearly, the lower boundary of the
operative memory is zero. The diagram of the system is depicted in Figure \ref{dpsdiagr.fig}.

The meaning of the actions that label the transitions of the LFSPNs which will specify the three models of the document
preparation system is as follows. The action $tx$ represents writing the text files into the operative memory. The
action $gr$ represents putting the graphics files into the operative memory. Particularly, the action $gl$ corresponds
to writing the low resolution graphics while $gh$ specifies writing the high resolution graphics. The action $dt$
represents reading the data (consisting of the portions of the input text and images) from the operative memory. In
each LFSPN, a single continuous place containing fluid will represent the operative memory with a data volume stored.

\begin{figure}
\begin{center}
\includegraphics{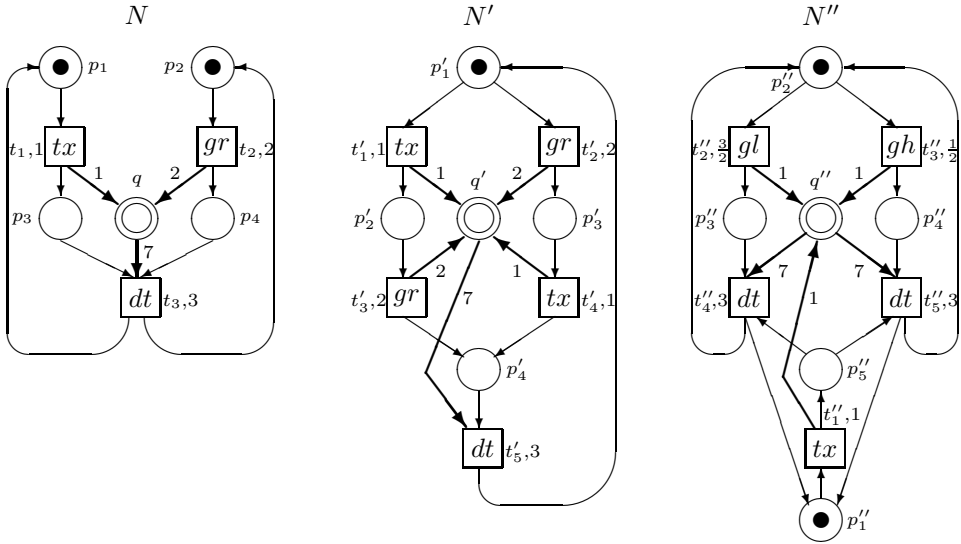}
\end{center}
\vspace{-7mm}
\caption{The LFSPNs of the standard and enhanced document preparation systems}
\label{dpsfspn.fig}
\end{figure}

In Figure \ref{dpsfspn.fig}, the LFSPNs $N$ and $N'$ specifying the {\em standard} document preparation system, as well
as the LFSPN $N''$ representing the {\em enhanced} one that differentiates between the low and high resolution
graphics, are presented. The rate of all transitions labeled with the action $tx$ is $1$, the rate of those labeled
with $gr$ is $2$ and the rate of those labeled with $dt$ is $3$. Further, the rate of the transition with the label
$gl$ is $\frac{3}{2}$ and the rate of that with the label $gh$ is $\frac{1}{2}$. The rate of the fluid flow along the
continuous arcs from the transitions labeled with the action $tx$ is $1$ while that from the transitions labeled with
$gr$ is $2$. Next, the fluid flow rate
from the transitions with the label $gl$ or $gh$ is the same and equals $1$. The rate of the fluid flow along the
continuous arcs to the transitions labeled with the action $dt$ is $7$.

We have $N\bis_{fl}N'$. Since LFSPNs have an {\em interleaving} semantics due to the {\em continuous} time approach and
the {\em race} condition applied to transition firings, the parallel execution of actions (here in $N$) is modeled by
the sequential non-determinism (in $N'$). Fluid bisimulation equivalence is an interleaving relation constructed in
conformance with the LFSPNs semantics. In our application example, one can see that the ``sequential'' LFSPN $N'$ may
be replaced with the fluid bisimulation equivalent and structurally simpler ``concurrent'' LFSPN $N$, the latter having
less transitions and arcs. Thus, the mentioned equivalence can be used not just to reduce behaviour of LFSPNs (as we
have seen in the previous examples), but also to simplify their structure.

We have $DRS(N)=\{M_1,M_2,M_3,M_4\}$, where $M_1=(1,1,0,0),\ M_2=(0,1,1,0),\ M_3=(1,0,0,1),\ M_4=(0,0,1,1)$;
$DRS(N')=\{M_1',M_2',M_3',M_4'\}$, where $M_1'=(1,0,0,0),\ M_2'=(0,1,0,0),\ M_3'=(0,0,1,0),\ M_4'=(0,0,0,1)$; and
$DRS(N'')=\{M_1'',M_2'',M_3'',M_4'',M_5'',M_6''\}$, where $M_1''=(1,1,0,0,0),\ M_2''=(1,0,1,0,0),\ M_3''=(0,1,0,0,1),\
M_4''=(1,0,0,1,0),\ M_5''=(0,0,1,0,1),\ M_6''=(0,0,0,1,1)$.

\begin{figure}
\begin{center}
\includegraphics{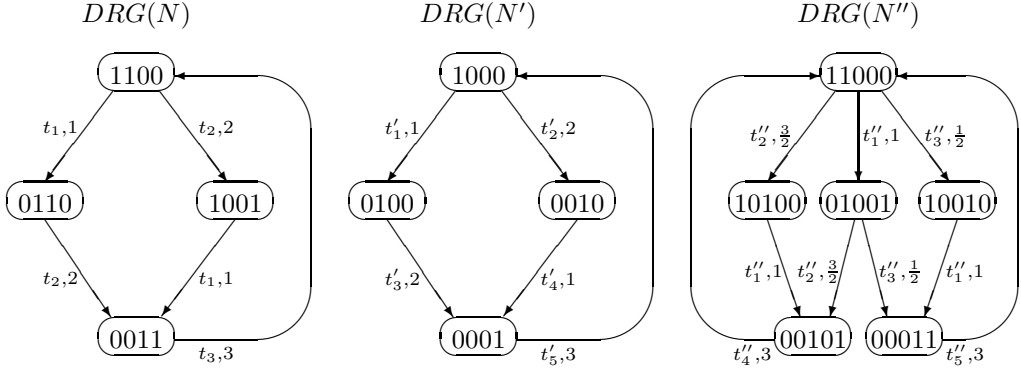}
\end{center}
\vspace{-7mm}
\caption{The discrete reachability graphs of the LFSPNs of the standard and enhanced document preparation systems}
\label{dpsdrg.fig}
\end{figure}

In Figure \ref{dpsdrg.fig}, the discrete reachability graphs $DRG(N),\ DRG(N')$ and $DRG(N'')$ are depicted. Then it is
clear that the discrete parts of the LFSPNs $N$ and $N'$ have the same behaviour.

Let $N'''$ is an abstraction of $N''$ by assuming that the actions $gl$ and $gh$ coincide with the action $gr$. Then it
holds $N\bis_{fl}N'\bis_{fl}N'''$.
In such a case, $DRS(N''')=\{M_1''',M_2''',M_3''',M_4''',M_5''',M_6'''\}$ coincides with $DRS(N'')$ up to the trivial
renaming bijection on the places. Further, $DRG(N''')$ coincides with $DRG(N'')$ up to the analogous renaming the
transitions.

\begin{figure}
\begin{center}
\includegraphics{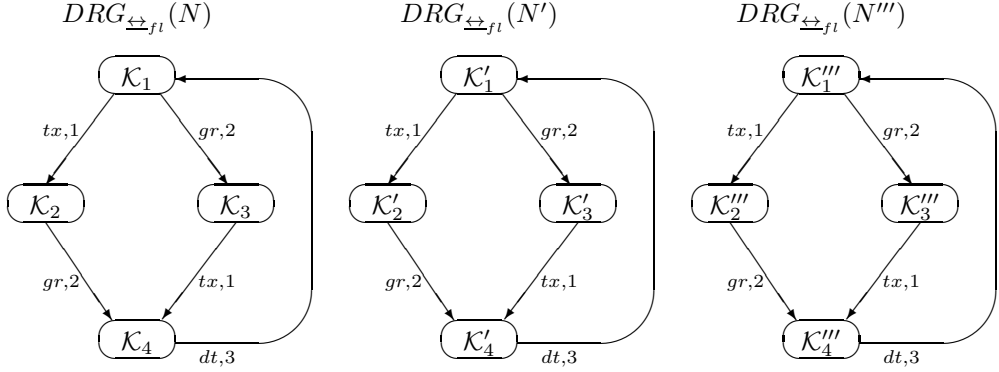}
\end{center}
\vspace{-7mm}
\caption{The quotient discrete reachability graphs of the LFSPNs of the standard document preparation system and that
of the abstract LFSPN of the enhanced document preparation system}
\label{dpsqdrg.fig}
\end{figure}

Let ${\cal K}_1=\{M_1\},\ {\cal K}_2=\{M_2\},\ {\cal K}_3=\{M_3\},\ {\cal K}_4=\{M_4\}$ and ${\cal K}_1'=\{M_1'\},\
{\cal K}_2'=\{M_2'\},\ {\cal K}_3'=\{M_3'\},\ {\cal K}_4'=\{M_4'\}$, as well as ${\cal K}_1'''=\{M_1'''\},\ {\cal
K}_2'''=\{M_2''',M_4'''\},\ {\cal K}_3'''=\{M_3'''\},\ {\cal K}_4'''=\{M_5''',M_6'''\}$. In Figure \ref{dpsqdrg.fig},
the quotient (by $\bis_{fl}$) discrete reachability graphs $DRG_{\bis_{fl}}(N),\ DRG_{\bis_{fl}}(N')$ and
$DRG_{\bis_{fl}}(N''')$ are depicted. Obviously, $DRG_{\bis_{fl}}(N)\simeq DRG_{\bis_{fl}}(N')\simeq
DRG_{\bis_{fl}}(N''')$. Then it is clear that the discrete parts of the LFSPNs $N,\ N'$ and $N'''$ have the same {\em
quotient} behaviour. Thus, quotienting by fluid bisimulation equivalence can be used to substantially reduce behaviour
of LFSPNs. It is also clear that the discrete parts of the LFSPNs $N$ and $N'$ have the same {\em complete} and {\em
quotient} behaviour.

The sojourn time average and variance vectors of $N'''$ are

$$SJ'''=\left(\frac{1}{3},1,\frac{1}{2},1,\frac{1}{3},\frac{1}{3}\right),\
VAR'''=\left(\frac{1}{9},1,\frac{1}{4},1,\frac{1}{9},\frac{1}{9}\right).$$

The complete and quotient sojourn time average and variance vectors of $N$ and $N'$, as well as the quotient
corresponding vectors of $N'''$, are

$$SJ=SJ_{\bis_{fl}}=SJ'=SJ_{\bis_{fl}}'=SJ_{\bis_{fl}}'''=\left(\frac{1}{3},\frac{1}{2},1,\frac{1}{3}\right),$$

$$VAR=VAR_{\bis_{fl}}=VAR'=VAR_{\bis_{fl}}'=VAR_{\bis_{fl}}'''=\left(\frac{1}{9},\frac{1}{4},1,\frac{1}{9}\right).$$

The TRM ${\bf Q}'''$ for $CTMC(N''')$, TPM ${\bf P}'''$ for $EDTMC(N''')$ and FRM ${\bf R}'''$ for the SFM of $N'''$
are

$$\footnotesize\begin{array}{ccc}
{\bf Q}'''=\left(\begin{array}{cccccc}
-3 & \frac{3}{2} & 1 & \frac{1}{2} & 0 & 0\\
0 & -1 & 0 & 0 & 1 & 0\\
0 & 0 & -2 & 0 & \frac{3}{2} & \frac{1}{2}\\
0 & 0 & 0 & -1 & 0 & 1\\
3 & 0 & 0 & 0 & -3 & 0\\
3 & 0 & 0 & 0 & 0 & -3
\end{array}\right),
&
{\bf P}'''=\left(\begin{array}{cccccc}
0 & \frac{1}{2} & \frac{1}{3} & \frac{1}{6} & 0 & 0\\
0 & 0 & 0 & 0 & 1 & 0\\
0 & 0 & 0 & 0 & \frac{3}{4} & \frac{1}{4}\\
0 & 0 & 0 & 0 & 0 & 1\\
1 & 0 & 0 & 0 & 0 & 0\\
1 & 0 & 0 & 0 & 0 & 0
\end{array}\right),
&
{\bf R}'''=\left(\begin{array}{cccccc}
3 & 0 & 0 & 0 & 0 & 0\\
0 & 1 & 0 & 0 & 0 & 0\\
0 & 0 & 2 & 0 & 0 & 0\\
0 & 0 & 0 & 1 & 0 & 0\\
0 & 0 & 0 & 0 & -7 & 0\\
0 & 0 & 0 & 0 & 0 & -7
\end{array}\right).
\end{array}$$

The TRMs ${\bf Q},\ {\bf Q}_{\bis_{fl}},\ {\bf Q}',\ {\bf Q}_{\bis_{fl}}'$ and ${\bf Q}_{\bis_{fl}}'''$ for $CTMC(N),\
CTMC_{\bis_{fl}}(N),\ CTMC(N'),\ CTMC_{\bis_{fl}}(N')$\\
and $CTMC_{\bis_{fl}}(N''')$; TPMs ${\bf P},\ {\bf P}_{\bis_{fl}},\ {\bf P}',\ {\bf P}_{\bis_{fl}}'$ and ${\bf
P}_{\bis_{fl}}'''$ for $EDTMC(N),\ EDTMC_{\bis_{fl}}(N),\\
EDTMC(N'),\ EDTMC_{\bis_{fl}}(N')$ and $EDTMC_{\bis_{fl}}(N''')$; as well as FRMs ${\bf R},\ {\bf R}_{\bis_{fl}},\ {\bf
R}',\ {\bf R}_{\bis_{fl}}'$ and ${\bf R}_{\bis_{fl}}'''$ for the complete and quotient SFMs of $N,\ N'$ and for the
quotient SFM of $N'''$ are

$$\begin{array}{c}
{\bf Q}={\bf Q}_{\bis_{fl}}={\bf Q}'={\bf Q}_{\bis_{fl}}'={\bf Q}_{\bis_{fl}}'''=
\left(\begin{array}{cccc}
-3 & 1 & 2 & 0\\
0 & -2 & 0 & 2\\
0 & 0 & -1 & 1\\
3 & 0 & 0 & -3
\end{array}\right),\\[7mm]
{\bf P}={\bf P}_{\bis_{fl}}={\bf P}'={\bf P}_{\bis_{fl}}'={\bf P}_{\bis_{fl}}'''=
\left(\begin{array}{cccc}
0 & \frac{1}{3} & \frac{2}{3} & 0\\
0 & 0 & 0 & 1\\
0 & 0 & 0 & 1\\
1 & 0 & 0 & 0
\end{array}\right),\\[7mm]
{\bf R}={\bf R}_{\bis_{fl}}={\bf R}'={\bf R}_{\bis_{fl}}'={\bf R}_{\bis_{fl}}'''=
\left(\begin{array}{cccc}
3 & 0 & 0 & 0\\
0 & 2 & 0 & 0\\
0 & 0 & 1 & 0\\
0 & 0 & 0 & -7
\end{array}\right).
\end{array}$$

Thus, the respective discrete and continuous parts of the LFSPNs $N$ and $N'$ have the same {\em complete} and {\em
quotient} behaviour while $N'''$ has the same {\em quotient} one.
Then it is enough to consider only LFSPN~$N$~from~now~on.

The discrete markings of LFSPN $N$ are interpreted as follows: $M_1$: both the text and graphics file collections are
written to the memory, $M_2$: the text file collection is resided in the memory and the graphics one is written to the
memory, $M_3$: the graphics file collection is resided in the memory and the text one is written to the memory, $M_4$:
the text and graphics file collections are resided in the memory and the data is read from there (if it is not empty).

We have $DRS^-(N)=\{M_4\},\ DRS^0(N)=\emptyset$ and $DRS^+(N)=\{M_1,M_2,M_3\}$.

The steady-state PMF for $CTMC(N)$ is

$$\varphi =
\left(\frac{2}{9},\frac{1}{9},\frac{4}{9},\frac{2}{9}\right).$$

Then the {\em stability condition} for the SFM of $N$
is fulfilled: $FluidFlow(q)=\sum_{i=1}^4\varphi_i RP(M_i)=
\frac{2}{9}\cdot 3+\frac{1}{9}\cdot 2+\frac{4}{9}\cdot 1+\frac{2}{9}(-7)=-\frac{2}{9}<0$.

For each eigenvalue $\gamma$ we must have $|\gamma{\bf R}-{\bf Q}|=\left|\begin{array}{cccc}
3(\gamma +1) & -1 & -2 & 0\\
0 & 2(\gamma +1) & 0 & -2\\
0 & 0 & \gamma +1 & -1\\
-3  & 0 & 0 & -7\gamma +3
\end{array}\right|=-42\gamma^4-108\gamma^3-72\gamma^2-6\gamma =0$; hence, $\gamma_1=0,\ \gamma_2=-1,\
\gamma_3=-\frac{1}{14}(11+\sqrt{93}),\ \gamma_4=-\frac{1}{14}(11-\sqrt{93})$.

The corresponding eigenvectors are the solutions of

$$\begin{array}{cc}
v_1\left(\begin{array}{cccc}
3 & -1 & -2 & 0\\
0 & 2 & 0 & -2\\
0 & 0 & 1 & -1\\
-3  & 0 & 0 & 3
\end{array}\right)=0,
&
v_2\left(\begin{array}{cccc}
0 & -1 & -2 & 0\\
0 & 0 & 0 & -2\\
0 & 0 & 0 & -1\\
-3  & 0 & 0 & 10
\end{array}\right)=0,\\
\multicolumn{2}{c}{v_3\left(\begin{array}{cccc}
\frac{3}{14}(3-\sqrt{93}) & -1 & -2 & 0\\
0 & \frac{1}{7}(3-\sqrt{93}) & 0 & -2\\
0 & 0 & \frac{1}{14}(3-\sqrt{93}) & -1\\
-3  & 0 & 0 & \frac{1}{2}(17+\sqrt{93})
\end{array}\right)=0,}\\
\multicolumn{2}{c}{v_4\left(\begin{array}{cccc}
\frac{3}{14}(3+\sqrt{93}) & -1 & -2 & 0\\
0 & \frac{1}{7}(3+\sqrt{93}) & 0 & -2\\
0 & 0 & \frac{1}{14}(3+\sqrt{93}) & -1\\
-3  & 0 & 0 & \frac{1}{2}(17-\sqrt{93})
\end{array}\right)=0.}
\end{array}$$

Then the
eigenvectors are $v_1=\left(\frac{2}{9},\frac{1}{9},\frac{4}{9},\frac{2}{9}\right),\ v_2=(0,-1,2,0),\
v_3=\left(\frac{14}{3-\sqrt{93}},\frac{98}{(3-\sqrt{93})^2},\frac{392}{(3-\sqrt{93})^2},1\right),\\
v4=\left(\frac{14}{3+\sqrt{93}},\frac{98}{(3+\sqrt{93})^2},\frac{392}{(3+\sqrt{93})^2},1\right)$.

Since $\varphi =F(\infty )=a_1 v_1$, we have $F(x)=\varphi +a_2 e^{\gamma_2 x}v_2+a_3 e^{\gamma_3 x}v_3+
a_4 e^{\gamma_4 x}v_4$ and $a_1=1$. Since $\forall M_l\in DRS^+(N)\ F_l(0)=\varphi_l+a_2 v_{2l}+a_3 v_{3l}+
a_4 v_{4l}=0$ and $DRS^+(N)=\{M_1,M_2,M_3\}$, we have the following linear equation system:
$\left\{\begin{array}{l}
\varphi_1+a_2 v_{21}+a_3 v_{31}+a_4 v_{41}=\frac{2}{9}+\frac{14}{3-\sqrt{93}}a_3+\frac{14}{3+\sqrt{93}}a_4=0\\
\varphi_2+a_2 v_{22}+a_3 v_{32}+a_4 v_{42}=\frac{1}{9}-a_2+\frac{98}{(3-\sqrt{93})^2}a_3+
\frac{98}{(3+\sqrt{93})^2}a_4=0\\
\varphi_3+a_2 v_{23}+a_3 v_{33}+a_4 v_{43}=\frac{4}{9}+2a_2+\frac{392}{(3-\sqrt{93})^2}a_3+
\frac{392}{(3+\sqrt{93})^2}a_4=0
\end{array}\right..$

By solving the system, we get $a_2=0,\ a_3=\frac{2(31-3\sqrt{93})}{93(3+\sqrt{93})},\
a_4=-\frac{2(10+\sqrt{93})}{21\sqrt{93}}$.
Thus, $F(x)=\left(\frac{2}{9},\frac{1}{9},\frac{4}{9},\frac{2}{9}\right)+
\frac{2(31-3\sqrt{93})}{93(3+\sqrt{93})}e^{-\frac{1}{14}(11+\sqrt{93})x}
\left(\frac{14}{3-\sqrt{93}},\frac{98}{(3-\sqrt{93})^2},\frac{392}{(3-\sqrt{93})^2},1\right)\!-\!
\frac{2(10+\sqrt{93})}{21\sqrt{93}}e^{-\frac{1}{14}(11-\sqrt{93})x}
\left(\frac{14}{3+\sqrt{93}},\frac{98}{(3+\sqrt{93})^2},\frac{392}{(3+\sqrt{93})^2},1\right)$.

Then the steady-state fluid PDF for the SFM of $N$ is

%
$$\begin{array}{c}
F(x)=\left(\frac{2}{9}-\frac{(31-3\sqrt{93})}{279}e^{-\frac{1}{14}(11+\sqrt{93})x}+
\frac{4(10+\sqrt{93})}{3\sqrt{93}(3+\sqrt{93})}e^{-\frac{1}{14}(11-\sqrt{93})x},\right.\\
\frac{1}{9}-\frac{7(31-3\sqrt{93})}{279(3-\sqrt{93})}e^{-\frac{1}{14}(11+\sqrt{93})x}+
\frac{28(10+\sqrt{93})}{3\sqrt{93}(3+\sqrt{93})^2}e^{-\frac{1}{14}(11-\sqrt{93})x},\\
\frac{4}{9}-\frac{28(31-3\sqrt{93})}{279(3-\sqrt{93})}e^{-\frac{1}{14}(11+\sqrt{93})x}+
\frac{112(10+\sqrt{93})}{3\sqrt{93}(3+\sqrt{93})^2}e^{-\frac{1}{14}(11-\sqrt{93})x},\\
\left.\frac{2}{9}+\frac{2(31-3\sqrt{93})}{93(3+\sqrt{93})}e^{-\frac{1}{14}(11+\sqrt{93})x}+
\frac{2(10+\sqrt{93})}{21\sqrt{93}}e^{-\frac{1}{14}(11-\sqrt{93})x}\right).
\end{array}$$

The steady-state fluid probability density function for the SFM of $N$ is

$$\begin{array}{c}
f(x)=\frac{dF(x)}{dx}=\left(\frac{e^{-\frac{1}{14}(11+\sqrt{93})x}(31-\sqrt{93}+
(31+\sqrt{93})e^{\frac{\sqrt{93}x}{7}})}{1953},
\frac{e^{-\frac{1}{14}(11+\sqrt{93})x}(-1+e^{\frac{\sqrt{93}x}{7}})}{9\sqrt{93}},\right.\\
\left.\frac{e^{-\frac{1}{14}(11+\sqrt{93})x}(-1+e^{\frac{\sqrt{93}x}{7}})}{9\sqrt{93}},
\frac{e^{-\frac{1}{14}(11+\sqrt{93})x}(14(-31+\sqrt{93})+
(620+48\sqrt{93})e^{\frac{\sqrt{93}x}{7}})}{4557(3+\sqrt{93})}\right).
\end{array}$$

The steady-state buffer empty probability for the SFM of $N$ is

$$\ell =F(0)=\left(0,0,0,\frac{2}{63}\right).$$

\begin{figure}
\begin{center}
\includegraphics[width=\textwidth]{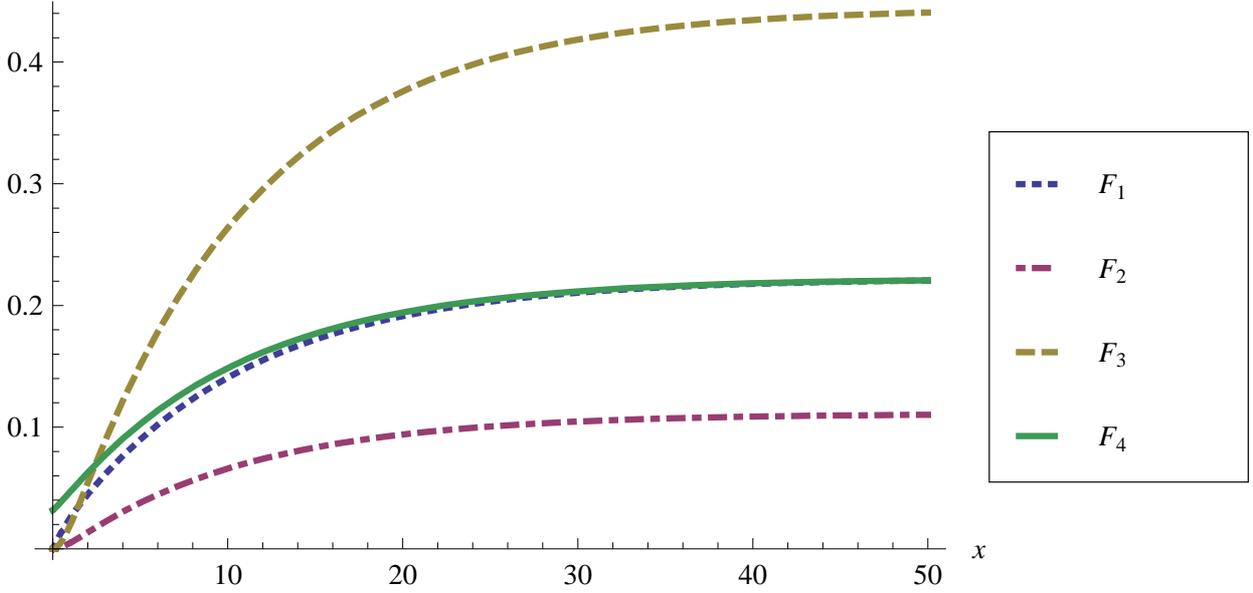}
\end{center}
\vspace{-10mm}
\caption{The elements of the steady-state fluid PDF for the SFM of the concurrent LFSPN of the document preparation
system}
\label{dpsflpdfs.fig}
\end{figure}

In Figure \ref{dpsflpdfs.fig}, the plots of the elements $F_1,F_2,F_3,F_4$ of the steady-state fluid PDF
$F=(F_1,F_2,F_3,F_4)$ for the SFM of $N$, as functions of $x$, are depicted.

We can now calculate some steady-state performance measures for the document preparation system.
\begin{itemize}

\item The {\em fraction of time when both the text and graphics file collections are written to the memory} is

$$TimeFract(\{M_1\})=\varphi_1=\frac{2}{9}.$$

\item The {\em average number of the text file collections received per unit of time} is

$$FiringFreq(t_1)=\sum_{\{i\mid t_1\in Ena(M_i),\ M_i\in DRS(N)\}}\varphi_i\Omega_N(t_1,M_i)=
\varphi_1\Omega_N(t_1,M_1)=\frac{2}{9}\cdot 1=\frac{2}{9}.$$

\item The {\em throughput of the system} is

$$FiringFreq(t_3)=\sum_{\{i\mid t_3\in Ena(M_i),\ M_i\in DRS(N)\}}\varphi_i\Omega_N(t_3,M_i)=
\varphi_4\Omega_N(t_3,M_4)=\frac{2}{9}\cdot 3=\frac{2}{3}.$$

\item The {\em probability that the memory is not empty} is

$$\begin{array}{c}
FluidLevel(q)=1-\displaystyle\sum_{\{i\mid M_i\in DRS(N)\}}\ell_i=1-(\ell_1+\ell_2+\ell_3+\ell_4)=
1-\frac{2}{63}=\frac{61}{63}.
\end{array}$$

\item The {\em probability that the operative memory contains at least $5$ Mb data} is

$$\begin{array}{c}
FluidLevel(q,5)=1-\displaystyle\sum_{\{i\mid M_i\in DRS(N)\}}F_i(5)=1-(F_1(5)+F_2(5)+F_3(5)+F_4(5))=\\
\frac{e^{-\frac{5}{14}(11+\sqrt{93})}(5673-631\sqrt{93}+e^{\frac{5\sqrt{93}}{7}}(5673+631\sqrt{93}))}{11718}\approx
0.6181.
\end{array}$$

\end{itemize}

Since $N\bis_{fl}N'\bis_{fl}N'''$, the LFSPNs $N,\ N'$ and $N'''$ satisfy the same formulas of $HML_{flt}$ (with the
identical interpretation values) and $HML_{flb}$. For instance, consider the following formulas for LFSPN $N$.

We have $[\![\langle tx\rangle\langle gr\rangle\top]\!]_{flt}(M_N,\frac{1}{3}\circ\frac{1}{2}\circ\frac{1}{3},3\circ
2\circ (-7))=PT(t_1 t_2)=\frac{1}{3}\cdot 1=\frac{1}{3}$, i.e. the value $\frac{1}{3}$ is the {\em probability} that
the text files are written into the operative memory with the potential flow rate $3$ during the exponentially
distributed time period with the average $\frac{1}{3}$; {\em then} the graphics files are written into the memory with
the potential flow rate $2$ during the exponentially distributed time period with the average $\frac{1}{2}$; {\em
finally}, the data is read from the memory with the potential flow rate $-7$ for the exponentially distributed time
period with the average $\frac{1}{3}$.

Further, it holds $M_N\models_{flb}\wr_3\wedge (\langle tx\rangle_1\top\vee\langle gr\rangle_2\top )$, i.e. it is {\em
valid} that the text files are written into the operative memory with the potential flow rate $3$ during the
exponentially distributed time period with the rate $1$ {\em or} the graphics files are written into the memory with
the same potential flow rate $3$ during the exponentially distributed time period with the rate $2$.

\section{Conclusion}
\label{conclusion.sec}

In this paper, we have defined two behavioural equivalences that preserve the qualitative and quantitative behavior of
LFSPNs, related to both their discrete part (labeled CTSPNs and the underlying CTMCs) and continuous part (the
associated SFMs). We have proposed on LFSPNs a linear-time relation of fluid trace equivalence and a branching-time
relation of fluid bisimulation equivalence. Both equivalences respect {\em functional activity}, {\em stochastic
timing} and {\em fluid flow} in the behaviour of LFSPNs. We have demonstrated that fluid trace equivalence preserves
average potential fluid change volume for the transition sequences of each given length. We have proven that fluid
bisimulation equivalence implies fluid trace equivalence and the reverse implication does not hold in general. We have
explained how to reduce the discrete reachability graphs and underlying CTMCs of LFSPNs with respect to fluid
bisimulation equivalence by applying the technique that builds the quotients of the respective labeled transition
systems by the largest fluid bisimulation. We have defined the quotients of the probability functions by fluid
bisimulation equivalence to describe the quotient associated SFMs. We have characterized logically fluid trace and
bisimulation equivalences with two novel fluid modal logics $HML_{flt}$ and $HML_{flb}$. The characterizations give
rise to better understanding of basic features of the equivalences. According to \cite{Ace03}, we have demonstrated
that the fluid equivalences are reasonable notions, by constructing their natural and pleasant modal characterizations.
In addition, they offer a possibility for the logical reasoning on resemblance of the fluid behaviour, while before it
was only possible in the operational manner. For example, let $N$ be one of the fluid (trace of bisimulation)
equivalent LFSPNs that model the production line mentioned in Section \ref{introduction.sec}. In the initial discrete
marking $M_N$, we now can specify and verify formally the properties described there: the {\em probability} given by
the interpretation $[\![\langle f_1\rangle\langle f_2\rangle\top]\!]_{flt}(M_N,s_1 s_2 s_3,r_1 r_2 r_3)$ in $HML_{flt}$
and the {\em validity} of the satisfaction $M_N\models_{flb}\wr_{r_1}\wedge (\langle
f_1\rangle_{\lambda_1}\top\vee\langle f_2\rangle_{\lambda_2}\top )$ in $HML_{flb}$. We have proven that fluid
bisimulation equivalence preserves the qualitative and stationary quantitative behaviour, hence, it guarantees that the
functionality and performance measures of the equivalent systems
coincide. We have presented a case study of the three
LFSPNs, all modeling the document preparation system, with intention to show how fluid bisimulation equivalence can be
used to simplify the LFSPNs structure and behaviour.

In the future, we plan to define a fluid place bisimulation relation that connects ``similar'' continuous places of
LFSPNs, like place bisimulations \cite{AS92,APS94,Tar98,Tar00,Tar07} relate discrete places of (standard) Petri nets.
The {\em lifting} of the relation to the discrete-continuous LFSPN markings (with discrete markings treated as the
multisets of places) will respect both the fluid distribution among the related continuous places and the rates of
fluid flow through them. For this purpose, we should introduce a novel notion of the multiset analogue with
non-negative real-valued multiplicities of the elements. While multiset is a mapping from a countable set to all
natural numbers, we need a more sophisticated mapping from the set of continuous places to all non-negative real
numbers, corresponding to the associated fluid levels. Such an extension of the multiset notion may use the results of
\cite{Bli91,Syr01}, concerning hybrid sets (the multiplicities of the elements are arbitrary integers) and fuzzy
multisets (the multiplicities belong to the interval [0;1]). In this way, both the initial amount of fluid and its
transit flow rate in each discrete marking may be distributed among several continuous places of an LFSPN, such that
all of them are bisimilar to a particular continuous place of the equivalent LFSPN. The interesting point here is that
fluid distributed among several bisimilar continuous places should be taken as the fluid contained in a single
continuous place, resulting from aggregating those ``constituent'' continuous places with the use of fluid place
bisimulation. Then the fluid level in the ``aggregate'' continuous place will be a sum of the fluid levels in the
``constituent'' continuous places. The probability density function for the sum of random variables representing the
fluid levels in the ``constituent'' continuous places is defined via {\em convolution} operation. In this approach, we
should avoid or treat correctly the situations when the fluid flow in the ``aggregate'' continuous place becomes
suddenly non-continuous. This happens when some of the ``constituent'' continuous places are emptied while the others
still contain a positive amount of fluid. Obviously, such a discontinuity is a result of applying the aggregation since
it is not caused by either reaching the lower fluid boundary (zero fluid level) or change of the current discrete
marking.

We assume that summation of the fluid levels in the continuous places may be implemented with the constructions
proposed in \cite{Gri02} for {\em extended FSPNs} (EFSPNs). EFSPNs have special deterministic fluid jump arcs that are
used to transfer a deterministic amount of fluid from one continuous place to another via intermediate stochastic
transitions connecting both places (deterministic fluid transfer). Analogously, random fluid jump arcs in EFSPNs are
used to transfer a random amount of fluid from one continuous place to another (random fluid transfer). We can also use
fluid transitions, mentioned in \cite{Gri02} as a direction for future development of the FSPNs formalism. Fluid
transitions that transfer fluid from their input to their output continuous places are used to implement fluid volume
conservation. If one of the input continuous places of a fluid transition becomes empty (i.e. the lower fluid boundary
is reached) then the rate of the transition should change in a certain way. The continuous arcs between continuous
places and fluid transitions may have multiplicities that multiply (change according to a factor) the fluid flow along
the arcs. Fluid transitions may be controlled by a discrete marking, using the guard functions associated with them or
applying the inhibitor and test arcs, i.e. by the constructions that do not affect discrete markings.

Further, we intend to apply to LFSPNs an analogue of the effective reduction technique based on the place bisimulations
of Petri nets \cite{AS92,APS94}. In this way, we shall merge several equivalent continuous places and, in some cases,
the transitions between them. This should result in the significant reductions of LFSPNs. The number of continuous
places in an LFSPN impacts drastically the complexity of its solution. The analytical solution is normally possible for
just a few continuous places (or even only for one). In all other cases, when modeling realistic large and complex
systems, we have to apply numerical techniques to solve systems of partial differential equations, or the method of
simulation. Hence, the reduction of the number of continuous places accomplished with the place bisimulation merging
appears to be even more important for LFSPNs than for Petri nets.

%

\end{document}